\documentclass[final]{IEEEtran}
\ifCLASSINFOpdf
\else
\fi

\usepackage{amsthm,amssymb,graphicx,multirow,amsmath,color,amsfonts}
\usepackage[update,prepend]{epstopdf}
\usepackage[noadjust]{cite}
\usepackage{caption}
\usepackage{tabulary}
\usepackage{bbm} 
\usepackage{multirow}
\usepackage{comment}
\usepackage{tabularx}
\usepackage[caption=false,font=normalsize,labelfont=sf,textfont=sf]{subfig}
\usepackage{balance}
\usepackage{enumitem,kantlipsum}

\usepackage[dvipsnames]{xcolor}

\renewenvironment{IEEEbiography}[1]
  {\IEEEbiographynophoto{#1}}
  {\endIEEEbiographynophoto}

\newcolumntype{s}{>{\hsize=.8\hsize}X}
\newcolumntype{b}{>{\hsize=1.2\hsize}X}

\def\nb0{{\mathbf{0}}}
\def\nb1{{\mathbf{1}}}

\newtheorem{lemma}{Lemma}

\newtheorem{definition}{Definition}

\newtheorem{theorem}{Theorem}
\newtheorem{prop}{Proposition}

\newtheorem{remark}{Remark}

\allowdisplaybreaks 

\usepackage{setspace}	

\begin{document}
\title{
Modeling and Analysis of Dynamic Charging for EVs: A Stochastic Geometry Approach
}
\author{
Duc Minh Nguyen, Mustafa A. Kishk, and Mohamed-Slim Alouini
\thanks{The authors are with Computer, Electrical and Mathematical Science and Engineering Division, King Abdullah University of Science and Technology, Thuwal  23955-6900, Saudi Arabia (email: \{ducminh.nguyen; mustafa.kishk; slim.alouini\}@kaust.edu.sa).} 
}

\maketitle
\begin{abstract}
With the increasing demand for greener and more energy efficient transportation solutions, electric vehicles (EVs) have emerged to be the future of transportation across the globe. However, currently, one of the biggest bottlenecks of EVs is the battery. Small batteries limit the EVs driving range, while big batteries are expensive and not environmentally friendly. One potential solution to this challenge is the deployment of \textit{charging roads}, i.e., dynamic wireless charging systems installed under the roads that enable EVs to be charged while driving. In this paper, we use tools from stochastic geometry to establish a framework that enables evaluating the performance of charging roads deployment in metropolitan cities. We first present the course of actions that a driver should take when driving from a random source to a random destination in order to maximize dynamic charging during the trip. Next, we analyze the distribution of the distance to the nearest charging road. This distribution is vital for studying multiple performance metrics such as the trip efficiency, which we define as the fraction of the total trip spent on charging roads. Next, we derive the probability that a given trip passes through at least one charging road. The derived probability distributions can be used to assist urban planners and policy makers in designing the deployment plans of dynamic wireless charging systems. In addition, they can also be used by drivers and automobile manufacturers in choosing the best driving routes given the road conditions and level of energy of EV battery. 
\end{abstract}
\IEEEpeerreviewmaketitle
\section{Introduction}
\label{intro}
As the global trend towards sustainable energy has significantly transformed several industries, automobile manufacturing is not an exception. Almost every major car manufacturer now has models that run entirely on electric batteries. It is expected that in the next decades, electric vehicles (EVs) will account for a large portion of total car production~\cite{sperling2013future}. 

Although there are several challenges  with EVs, e.g., engines, sensors, one of the biggest bottlenecks of EVs is the battery~\cite{goingelectric}. Ideally, batteries for EVs should last for a comparable distance compared to gasoline tanks. The battery should also be quickly charged and remain in good condition after thousands of charging cycles. Moreover, it should be affordable and finally, be environment-friendly. However, there are a couple of important trade-offs with the current EVs batteries that need to be considered~\cite{areviewonthekeyissuesforbattery}. For example, large-capacity batteries, which are optimized for driving distance, are expensive to make, slow to charge, and not friendly to the environment. On the other hand, small-capacity batteries are more affordable but require more frequent charging. 

There have been steady advances in producing batteries that fit the demand~\cite{rapidlyfallingcostsofbattery}. However, as batteries require charging, we still need solutions on how to charge them effectively and efficiently. In the literature, several works have been presented about optimally deploying charging stations~\cite{pevcharginginfra,asurveyonpevcharginginfra,qosaware,stochasticgeometryplanningofEV}. Although this solution seems to extend the driving range of EVs, one major drawback of charging stations is the waiting time, especially in metropolitan cities. For example, during rush hours, charging stations may not meet the charging demand if each EV needs half an hour or more to charge. As the frequency of people using EVs in their daily commute increases, a better solution that benefits all commuters is dynamic wireless vehicle charging systems~\cite{reviewofstaticanddynamic,areviewoninductivecharging}, i.e, roads that are able to charge EVs while driving without the need to stop. Charging roads are equipped with wireless power transfer technology that enables complete wireless charging~\cite{onroadchargingofev,modernadvancesinwpt}. This technology has been thoroughly researched in several research institutes around the world such as KAIST (Korea)~\cite{advancesinwirelesspowertransfer}, University of Auckland (UoA - New Zealand)~\cite{theinductivepowertransferstory}, and Oak Ridge National Laboratory (ORNL - United States)~\cite{oakridgenationallab}. At KAIST, since 2009, researchers have introduced six generations of dynamic wireless charging systems for both driving vehicles and stationary EVs with improved charging efficiency in each new generation~\cite{systemarchitecture}. UoA has been active in designing coil structure and layout for dynamic charging systems~\cite{theinductivepowertransferstory,developmentofasinglesidedflux}. ORNL has focused on integrated wireless charging systems for vehicles. They have tested their systems in a few popular car models~\cite{oakridgenationallab,rav4} and achieved good power transfer efficiency~\cite{rav4performance}. Samples of dynamic electric vehicular charging systems have also been demonstrated by QualComm Halo~\cite{qualcomm}, and later by WiTricity~\cite{witricity}. Given the prominent future of dynamic charging technology, there is a high demand for efficiently utilizing charging roads and assessing the impact of its deployment at large scale, i.e., city level. While there are several studies on exploiting charging roads, e.g., optimizing routing policies for EVs in a specific city~\cite{optimalrouteplanning,morp}, researches on modeling the impact of dynamic charging on a generic city have not been comprehensively carried out.   

Motivated by the great potential of charging roads, as explained above, we study the system-level modeling and analysis of large-scale deployment of charging roads in metropolitan cities. Our goal is to provide an analytical framework that could be useful to urban planners, city policy makers, car manufacturers, and drivers. To be more precise, we consider a setup in which a fraction of the total number of roads is equipped with wireless charging capabilities. The proposed framework relies on considering a generic city, with respect to the density of roads and the fraction of charging roads. This, in turns, extends the applicability of the proposed framework, compared to a well-planned deployment scenario which only works for specific cities. Given a randomly located source and a randomly located destination, we analyze two metrics: (i) the probability that any given trip passes through at least one charging road, and (ii) the distribution of the distance from the source to the nearest charging road. These two metrics are crucial for city planners and policymakers to evaluate how effective and efficient the deployment of the charging roads is, as demonstrated in more details in Sec.~\ref{performancemetrics}.

To compute the two metrics, we adopt a policy that a driver will take given a source and a destination. Since routing with charging roads is a new research topic with few published studies, we select a general and intuitive driving policy. We assume that the driver will always choose the shortest route, as this assumption has been frequently used in several routing applications on road networks~\cite{heuristicshortestpath,findingshortestpath,probabilisticpathqueries}. In addition, if there are multiple shortest routes, we go on to assume that the driver will always choose the one that maximizes the time spent on charging roads. Note that this policy can be combined with more sophisticated constraints to model different EVs routing scenarios. For example, EVs need to meet an arrival time constraint so that not only the shortest route but the traffic congestion should be also considered. Another example is for a fleet of EVs to choose a driving policy to maintain connectivity constraint. However, since our work is one of the first attempts to study the impact of deploying dynamic charging at a city level, we begin with the most basic policy of taking the shortest route and maximizing the time spent on charging roads.

We make use of stochastic geometry as the main tool for our analysis, as it has been used extensively to model vehicular networks and has been proven useful to study several network-related problems~\cite{geodesicsandflow, meantrafficbehavior,PLCPfoundations}. For example, in~\cite[Chapter~10]{PLCPfoundations}, the charging facilities for EVs are modeled as a Poisson point process on each line of a Poisson line process (i.e., a road), and the distribution of the length of the shortest path between an arbitrary EV and its nearest facility is derived. Stochastic geometry offers a statistical approach to assessing our two proposed metrics. Unlike a deterministic approach, in which analysis is done given a specific road system in a particular city, stochastic geometry let us model the road system in an urban city as a stochastic process. Therefore, it allows us to study the two metrics averaging over all the random sources and destinations~\cite{stochasticgeometryforwireless,PLCPfoundations}. Specifically, we adopt the Manhattan Poisson Line Process (MPLP) to model the street network as a grid-like structure since it resembles the actual system of roads in several modern cities, e.g., New York~\cite{gridasalgorithm}, Chicago~\cite{mamadecitychicago}, Vancouver~\cite{dreamcityvancouver}, Barcelona~\cite{citiesasemergentmodel}. A more comprehensive overview of stochastic geometry, MPLP, and their application in vehicular network modeling is discussed in Sec.~\ref{vehicularnetworkmodeling} and Sec.~\ref{PLP}. The main contributions of this paper are summarized next.

\begin{itemize}
    \item We introduce a routing policy that a driver would take for all situations given a random source and a random destination, assuming that the driver will always choose the shortest route and maximize the time spent on charging roads throughout the trip.
    \item Given the above routing policy, and conditioned on the location of the source and the destination, we derive the distribution of the distance from the source to the nearest charging road.
    \item We derive the probability that a given trip passes through at least one charging road.
    \item We rigorously verify our analytical results for the two performance metrics through Monte-Carlo simulations.
\end{itemize}

To the best of our knowledge, this paper is the first one to incorporate stochastic geometry into the performance analysis of charging road deployment in metropolitan cities. Thus, it sheds light on how analytical tools such as stochastic geometry can be used to assess the performance of charging roads deployment in a generic urban city.

Our paper is organized as follows. We first review the previous related works Sec.~\ref{prevwork}. Sec.~\ref{analytical_frame} describes our analytical framework. Then, the routing policy and distribution of the distance to the nearest charging road are elaborated in Sec.~\ref{policy}. Sec.~\ref{atleast1charging} introduces the probability that a trip passes through at least one charging road. In Sec.~\ref{analytical_frame}, we demonstrate our analytical results verified by Monte-Carlo simulations. Lastly, Sec.~\ref{conclusion} concludes the paper with some final remarks.

A summary of the notations used in the paper is given in Table \ref{notation}. Some notations will be defined in more details as they appear in later sections of the paper. 

\begin{table*}[h]
\centering
\captionsetup{font=normalsize}
\caption{Summary of notations}
\label{notation}
\begin{tabular}{p{1.9cm}| p{8.8cm}}
\hline
\textbf{Notation} & \textbf{Description} \\ \hline
$p$                              & ratio of the number of charging roads to the total number of roads\\ 
$\lambda$                   & density of the 1D Poisson Point Process that generates the horizontal or vertical lines \\
$d_h$                 & horizontal distance between a source and a destination \\\
$d_v$               & vertical distance between a source and a destination\\
$D_n$               & distance from a source to the nearest charging road\\
$D_\mathrm{N-HC}$          & distance from a source to the nearest horizontal charging road\\
$D_\mathrm{N-VC}$          & distance from a source to the nearest vertical charging road\\
$D_\mathrm{N-HNC}$         & distance from a source to the nearest horizontal non-charging road\\
$D_\mathrm{N-VNC}$         & distance from a source to the nearest vertical non-charging road\\
$X_1$                       & distance between the nearest vertical non-charging road and the nearest vertical charging road from source \\
$X_2$                       & distance between the nearest horizontal non-charging road and the nearest horizontal charging road from source \\
$d_L$                       & distance from source to the nearest horizontal road in the opposite direction of the destination \\
$T_c$               & event that a given trip passes through at least one charging road\\
$\overline{T_c}$    & event that a given trip passes through no charging roads\\\hline

\end{tabular}
\end{table*}

\section{Related Work}
\label{prevwork}

\subsection{Wireless Charging for Electric Vehicles}
The wireless charging technology for EVs can be categorized into two major branches: capacitive power transfer and inductive power transfer. Capacitive power transfer utilizes the electric field interaction between coupled capacitor. Hence, it is only viable to transfer energy through a short air gap between $10^{-4}$ and $10^{-3}$ meters~\cite{asurveyofwirelesspowertransfer}, which is not suitable for charging EVs while running. Thus, we mainly focus on the inductive wireless charging system, in which there can be an air gap up to a few meters between a power transmitter in the roads and a receiver in the vehicles~\cite{acriticalreview}. It can transfer power electro-magnetically to EVs while driving and its main components consist of long primary windings (installed under the road) and secondary pick-up windings (installed in the EV). There are several other components that go into the wireless charging system. Optimizing the design of those components is an active research field on its own~\cite{generaldesignrequirement}. For example, the relationship among the length of winding tracks, the speed of vehicles, and the efficiency of dynamic charging systems is studied in~\cite{anoptimizedtracklength}, in which optimal track lengths for different vehicular speeds are presented. In~~\cite{sizingofinductivepowerpads}, the impact of sizing inductive power transfer power pads on the resulting power profile of a dynamic charging system is explored using Gaussian modeling and phase analysis. Fundamental principles of wireless charging using magnetic field resonance, designs of resonant magnetic coils, and electromagnetic field noise suppression methods are introduced in~\cite{coildesign}. The problem of allocating power from charging lanes to in-motion EVs is initially studied in~\cite{anefficientwptsystem} by balancing the state of charge of EVs. Later, a solution to allocating power to EVs, enabling them to arrive at destinations while achieving goals such as balancing the state of charge and power stored in EVs, or minimizing the total power charged, is addressed using a greedy approach in~\cite{powerdistributionscheduling}.

Since an inductive wireless power transfer system can power EVs while driving, it significantly increases the driving range of EVs without the need to stop and charge at stationary charging stations~\cite{drivingrangeextension}. Furthermore, the current design and cost of deployment of charging roads suggest that it is most suitable to deploy charging roads in metropolitan cities. Since the total energy transferred to an EV is the power of the charging system multiplied by the time that the vehicle spends on the charging road, it is desirable to maximize the time vehicles spend on charging roads, given a fixed power of the charging system. However, longer charging roads directly increase the cost of deployment~\cite{acriticalreview}. Also, it is preferred to have a high density of traffic travelled on the charging roads to fully utilize the charging system and reduce waste of energy~\cite{doublecouplesystems}. Thus, the suitable place to install a charging road system is in an urban setting since it has a high density of transportation, slower vehicle driving speed compared to highways, and shorter driving trajectory compared to highways~\cite{catcharger}. Hence, we can maximize charging performance while minimizing deployment cost. Indeed, in the literature, several researches have been proposed to optimally utilize the wireless charging systems in an urban road network~\cite{gameapproachonmodeling}. For example, a stationary wireless charging stations deployment scheme for taxicabs that optimizes the idle time and continuous operability is introduced in~\cite{employingopportunisticcharging}. A charging scheduling system that targets to reduce the charging and operating costs for large-scale electric bus fleet is presented in~\cite{bCharge}. Given the need for dynamic charging systems in urban areas, our work aims to assess the impact of deploying charging roads in metropolitan cities.

\subsection{Vehicular Network Modeling}
\label{vehicularnetworkmodeling}
In the literature, several models have been proposed for vehicular networks~\cite{sotaofvehiculartraffic}. The classic Erdos-Renyo (ER) graph model proposes that a graph of $n$ nodes is constructed by connecting those $n$ nodes randomly, i.e., each edge has an equal probability $p$ of being included in the graph~\cite{ontheevolutionofrandom}. However, ER graphs do not closely represent several real-world networks since they have low clustering coefficients and do not account for the formation of hubs. Watts-Strogatz small-world network models~\cite{collectivedynamicsof} address the first limitation of ER graph by accounting for clustering while maintaining the average path length as the ER graphs. Hammersley graphs~\cite{connectedspatialnetworksover} define a vertex with exactly four edges, while all vertices in the network follows an infinite Poisson Point Process. However, all of these network models do not correctly reflect the road systems in metropolitan cities since they do not capture the continuity of streets. To alleviate this problem, a good alternative is to model the streets in vehicular networks as a set of random lines, which collectively forms a \textit{line process}~\cite{stochasticgeometryandarchitecture, analysisofshortestpaths}. A well-known model for line processes is the Poisson Line Process (PLP)~\cite{stochasticgeometryandits}. Several modern cities in the world, e.g., New York, have a grid-like street network that can be closely modeled with a special case of PLP named \textit{Manhattan Poisson Line Process (MPLP)}. Several properties of PLP and MPLP that are useful for modeling vehicular networks is discussed in~\cite{poissoncoxpointprocess}. A method to analyze the coverage of wireless signals propagating through the streets modeled with MPLP is introduced in~\cite{mmwave}. In this paper, given the goal to assess the deployment of dynamic charging roads, we choose to model vehicular networks in a metropolitan settings using a MPLP. Details about MPLP and our network model are elaborated in Sec.~\ref{PLP} and Sec.~\ref{systemmodel}, respectively. 

\subsection{Charging Lanes Deployment}
As the importance of charging roads are realized by researchers and companies around the world, some researches have been presented on the deployment of charging lanes for EVs. For example, a plan to support electric buses running on a pre-defined route to minimize cost of deployment is introduced in~\cite{systemarchitecture}. A categorization and clustering method to choose the landmarks to deploy charging lanes in metropolitan cities is presented in~\cite{catcharger}. An integer programming approach to modeling the charging lanes installation based on geospatial data is demonstrated in~\cite{optimalinstallation}. Unlike those studies, our work aims to provide a general analytical framework to assess the deployment of charging road in metropolitan cities, and thus can be applied to several big cities and benefit various groups from city planners to EV manufacturers.  

\section{Analytical Framework}
\label{analytical_frame}

\subsection{Poisson Point Process and Poisson Line Process Preliminaries}
\label{PLP}
Since our vehicular network model in this paper is based on Poisson point process and Poisson line process, we briefly review the essence of those processes in this section. For a more detailed discussion regarding this topic, we refer the reader to sources such as~\cite{stochasticgeometryforwireless,PLCPfoundations,stochasticgeometryandits, aprimeroncellularnetwork,stochasticgeometryformodeling,modelingandanalysisofcellular, coverageanalysisofavehicular}.

\textit{Poisson Point Process}. Intuitively, a point process is a random collection of points in some spaces. Let $N(B)$ denote the number of points in a Borel set $B$. A point process is a homogeneous Poisson Point Process (PPP) with intensity parameter $\beta > 0 $ if:
\begin{itemize}
    \item $N(B)$ $\sim$ Poisson($\beta m(B)$), where $m(B)$ is the measure of set B, i.e., $\mathbb{P}(N(B)=v) = \frac{(\beta m(B))^v}{v!}e^{-\beta m(B)}$.
    \item For Borel sets $B_1$ and $B_2$ such that $B_1$ and $B_2$ are mutually exclusive, $N(B_1)$ and $N(B_2)$ are independent.
\end{itemize}

One important property of PPP is that given $N(B)=v$, the locations of those $v$ points are independent and identically distributed and uniform in $B$.

\textit{Poisson Line Process}. Similar to a point process, a line process is a random collection of lines in a 2D plane. A random undirected line in this plane can be fully characterized by a set of two parameters ($\rho$, $\theta$), where $\rho \in \mathbb{R}$ is a real number denoting the perpendicular distance from the origin $o \equiv (0,0)$ and $\theta$ is the angle between the positive $x$-axis and the line, i.e., $\theta \in [0,\pi)$. It is worth noting that $\rho$ is positive if the line is above or to the right of the origin, and negative otherwise. We can represent all the possible values of ($\rho$, $\theta$) in a plane denoting $\mathcal{C} \equiv [0,\pi) \times \mathbb{R}$. Since the mapping between the set of points in $\mathcal{C}$ and the set of lines in $\mathbb{R}^2$ is one-to-one, one can generate a line process in $\mathbb{R}^2$ by generating a point process in $\mathcal{C}$. For example, a set of lines generated by a PPP on $\mathcal{C}$ is a Poisson Line Process (PLP).

In this paper, we will focus on a special instance of PLP, namely the Manhattan Poisson line processes (MPLP) \cite{mmwave}. A MPLP has $\rho \in \mathbb{R}$ and $\theta \in \{0,\frac{\pi}{2}\}$, i.e., the set of lines has a grid-like shape. 

\subsection{System Model}
\label{systemmodel}
We model the system of roads in metropolitan cities by an MPLP in $\mathbb{R}^2$, where an overview about line process and MPLP are given in Sec.~\ref{PLP}. The MPLP is characterized by the parameters $\lambda$ and $p$, where $\lambda$ is the density of the 1D PPP that generates the horizontal or vertical lines, and $p$ is the ratio of the number of charging roads to the total number of roads. Then, we consider random positions for the source and the destination of a trip in this road system. There are two main possibilities. One is when the source and the destination are on two parallel roads, the other is when the source and the destination are on two perpendicular roads. 

\subsection{Performance Metrics}
\label{performancemetrics}

In this paper, our goal is to assess the deployment of charging roads in a metropolitan city. To this end, we first give a definition of the measure for the trip distance.

\begin{definition}[Manhattan distance]
In a two-dimensional plane, the Manhattan distance between a point $A(x_1,y_1)$ and a point $B(x_2,y_2)$  is the sum of the vertical distance and the horizontal distance between $A$ and $B$, i.e., $\mid x_1-x_2 \mid + \mid y_1-y_2 \mid$.
\end{definition}

Based on the distance measure, we propose two performance metrics whose definitions are given as follows:

\begin{definition}[Probability distribution of the distance to the nearest charging road $\mathbb{P}(D_n <x)$]
It is the probability that, given the locations of the source and the destination, the travel distance, i.e. Manhattan distance, from the source to the nearest charging road is less than a positive real number $x$.
\end{definition}

\begin{definition}[Probability that a trip passes through at least one charging road $\mathbb{P}(T_c)$]
It is the probability that, given the locations of the source and the destination, a driver travels on at least one charging road.
\end{definition}

These metrics have practical significance in understanding how dynamic wireless charging systems can serve the needs of commuters. For example, urban planners and city policy makers can use these metrics to determine how densely charging roads should be deployed so that 80\% of the time a driver will pass through at least one charging road in his or her trip. Another example would be for car manufacturers to see, based on the distance to the nearest charging road, how big the battery should be designed to fit urban design in a particular city. Given the promising future of electric vehicles and the need for charging roads as illustrated in Section \ref{intro}, our metrics provide useful insights to a diverse group of people about the deployment of dynamic wireless charging systems in metropolitan cities.

\section{Routing Policy \& Distribution of the distance to the nearest charging road}
\label{policy}

We denote the horizontal and vertical distances between source and destination as $d_h$ and $d_v$, respectively. In this section, we analyze the probability that the distance from the source to the nearest charging road, i.e., $D_n$, is less than a positive real number $x$. The distribution of $D_n$ depends on the source, the destination, and the route that a driver will take. Thus, to calculate $\mathbb{P}(D_n < x)$, we break it down into eight sub-events of two groups, i.e., 
\begin{itemize}
    \item When the source and the destination are on two parallel roads and
    \begin{itemize}
        \item Both source and destination roads are charging (Event $E_1$)
        \item Only the source road is charging (Event $E_2$)
        \item Only the destination road is charging (Event $E_3$)
        \item Both source and destination roads are not charging (Event $E_4$)
    \end{itemize}
    \item When the source and the destination are on two perpendicular roads and
    \begin{itemize}
        \item Both source and destination roads are charging (Event $E_5$)
        \item Only the source road is charging (Event $E_6$)
        \item Only the destination road is charging (Event $E_7$)
        \item Both source and destination roads are not charging (Event $E_8$),
    \end{itemize}
\end{itemize}
each of which will be discussed starting from Sec.~\ref{dnxsub1} to Sec.~\ref{dnxsub8}. In particular, each of the eight events $E_i$ represent a specific scenario for the relation between the location of the source and the location of the destination. In each case, $D_n$ is calculated based on an assumption that a driver always chooses the shortest route from the source to the destination. If there are multiple routes with the same minimum distance, priority is given to the routes containing the largest portion of charging roads.
\subsection{Summary of important distributions}
\label{appendx}

In this subsection, we first provide some propositions that appear frequently in the later proofs. 

\begin{prop}
Let $D_\mathrm{N-HC}$ be the distance from source to the nearest horizontal charging road. The CDF of $D_\mathrm{N-HC}$ is 
\begin{equation}
\label{FNHC}
    \mathbb{P}(D_\mathrm{N-HC} < x) = 1-e^{-\lambda p x}.
\end{equation}
The PDF of $D_\mathrm{N-HC}$ is 
\begin{equation}
    \label{fNHC}
    f_{D_\mathrm{N-HC}}(x) = \lambda p e^{-\lambda p x}.
\end{equation}
\end{prop}

\begin{prop}
Let $D_\mathrm{N-VC}$ be the distance from source to the nearest vertical charging road. The CDF of $D_\mathrm{N-VC}$ is 
\begin{equation}
    \label{FNVC}
    \mathbb{P}(D_\mathrm{N-VC} < x) = 1-e^{-\lambda p x}.
\end{equation}
The PDF of $D_\mathrm{N-VC}$ is
\begin{equation}
    \label{fNVC}
    f_{D_\mathrm{N-VC}}(x) = \lambda p e^{-\lambda p x}.
\end{equation}
\end{prop}

\begin{prop}
Let $D_\mathrm{N-HNC}$ be the distance from source to the nearest horizontal non-charging road. The CDF of $D_\mathrm{N-HNC}$ is 
\begin{equation}
    \label{FHNC}
    \mathbb{P}(D_\mathrm{N-HNC} < x) = 1-e^{-\lambda (1-p) x}.
\end{equation}
The PDF of $D_\mathrm{N-HNC}$ is
\begin{equation}
    \label{fHNC}
    f_{D_\mathrm{N-HNC}}(x) = \lambda (1-p) e^{-\lambda (1-p) x}.
\end{equation}
\end{prop}

\begin{prop}
Let $D_\mathrm{N-VNC}$ be the distance from source to the nearest vertical non-charging road. The CDF of $D_\mathrm{N-VNC}$ is 
\begin{equation}
    \label{FVNC}
    \mathbb{P}(D_\mathrm{N-VNC} < x) = 1-e^{-\lambda (1-p) x}.
\end{equation}
The PDF of $D_\mathrm{N-VNC}$ is 
\begin{equation}
    \label{fVNC}
    f_{D_\mathrm{N-VNC}}(x) = \lambda (1-p) e^{-\lambda (1-p) x}.
\end{equation}
\end{prop}

\begin{prop}
Let $d_L$ be the distance from source to the nearest horizontal road in the opposite direction of the destination. The CDF of $d_L$ is 
\begin{equation}
    \label{FDL}
    \mathbb{P}(d_L < x) = 1-e^{-\lambda x}.
\end{equation}
The PDF of $d_L$ is
\begin{equation}
    \label{fDL}
    f_{d_L}(x) = \lambda e^{-\lambda x}.
\end{equation}
\end{prop}

\begin{prop}
\label{distx1}
Let $X_1$ be the distance between the nearest vertical non-charging road and the nearest vertical charging road from source, given that they exist between the source and the destination. The CDF of $X_1$ is given by 

\begin{equation}
\label{FX1}
\begin{split}
    F_{X_1}(x) &= \mathbb{P}(X_1 < x) = 1- \int_{x}^{d_h} \frac{1-e^{-\lambda (1-p) (t-x)}}{1-e^{-\lambda (1-p) t}} \times
    \\&\frac{\lambda p e^{-\lambda p t}}{1-e^{-\lambda p d_h}} {\rm d} t.
\end{split}
\end{equation}
    
The PDF of $X_1$ is given by

\begin{equation}
\label{fx1}
  f_{X_1}(x) = \int_{x}^{d_h} \frac{\lambda^2 (1-p) p e^{-\lambda p t - \lambda (1-p) (t-x)}}{(1-e^{-\lambda p d_h})(1-e^{-\lambda (1-p) t})}  {\rm d} t.
\end{equation}
\end{prop}
\begin{IEEEproof}
    See Appendix \ref{appendxx1x2}
\end{IEEEproof}

\begin{prop}
\label{distx2}
Let $X_2$ be the distance between the nearest horizontal non-charging road and the nearest horizontal charging road from source, given that they exist between the source and the destination. The CDF of $X_2$ is given by 
\begin{equation}
\label{FX2}
\begin{split}
     F_{X_2}(x) &= \mathbb{P}(X_2 < x)  = 1- \int_{x}^{d_v} \frac{1-e^{-\lambda (1-p) (t-x)}}{1-e^{-\lambda (1-p) t}} \times
    \\&\frac{\lambda p e^{-\lambda p t}}{1-e^{-\lambda p d_v}} {\rm d} t.
\end{split}
\end{equation}

The PDF of $X_2$ is given by

\begin{equation}
\label{fx2}
  f_{X_2}(x) =  \int_{x}^{d_v} \frac{\lambda^2 (1-p) p e^{-\lambda p t - \lambda (1-p) (t-x)}}{(1-e^{-\lambda p d_v})(1-e^{-\lambda (1-p) t})}  {\rm d} t.
\end{equation}
\end{prop}
\begin{IEEEproof}
    See Appendix \ref{appendxx1x2}
\end{IEEEproof}

\subsection{Case a: When source (S) and destination (D) are on two parallel roads} \label{dnxsuba}
Let $A$ denote the case when S and D are on two parallel roads. The probability of case $A$ is $\mathbb{P}(A) = \frac{1}{2}$. We consider four scenarios:
\begin{itemize}
    \item Both source and destination roads are charging,
    \item Only source road is charging,
    \item Only destination road is charging,
    \item Both source and destination roads are not charging.
\end{itemize}

In each scenario, we first describe its probability, then present the distribution of $D_n$ given the scenario as a lemma. 

\subsubsection{Both source and destination roads are charging} \label{dnxsub1}

Let $E_1$ denote the case when both source and destination roads are on two parallel roads and are charging. The probability of event $E_1$ is $\frac{p^2}{2} $. 
\begin{lemma}
\label{DnE1}
The distribution of $D_n$ given $E_1$ is given as follows:
\begin{align*}
\mathbb{P}(D_n < x | E_1)\mathbb{P}(E_1) = \frac{p^2}{2}.
\end{align*} 
\end{lemma}
\begin{IEEEproof}
    See Appendix \ref{appendxe1}.
\end{IEEEproof}
 
\subsubsection{Only source road is charging}

Let $E_2$ denote the case when both source and destination roads are on two parallel roads and only the source road is charging. The probability of event $E_2$ is $\frac{p(1-p)}{2} $. 
\begin{lemma}
\label{DnE2}
The distribution of $D_n$ given $E_2$ is given as follows:
\begin{align*}
\mathbb{P}(D_n < x | E_2)\mathbb{P}(E_2) = \frac{p(1-p)}{2}.
\end{align*} 
\end{lemma}
\begin{IEEEproof}
    See Appendix \ref{appendxe2}.
\end{IEEEproof}
 
\subsubsection{Only destination road is charging}

Let $E_3$ denote the case when both source and destination roads are on two parallel roads and only the destination road is charging. The probability of event $E_3$ is $\frac{p(1-p)}{2}$. 
\begin{lemma}
\label{DnE3}
The distribution of $D_n$ given $E_3$ is given as follows:
\begin{align*}
&\mathbb{P}(D_n < x | E_3)\mathbb{P}(E_3) = \Psi_1 (p, \lambda, d_h, d_v, x),
\end{align*}
where $\Psi_1()$ is a function of the (charging) road density (i.e., $p$ and $\lambda$) and the trip information (i.e., $d_h, d_v,$ and $x$).
\end{lemma}
The complete form of $\Psi_1()$ is given in (\ref{lemma3eqn}) in Appendix~\ref{appendxb}.
\subsubsection{Both source and destination roads are not charging}

Let $E_4$ denote the case when both source and destination roads are on two parallel roads and are not charging. The probability of event $E_4$ is $\frac{(1-p)^2}{2}$. 

\begin{lemma}
\label{DnE4}
The distribution of $D_n$ given $E_4$ is given as follows:
\begin{align*}
&\mathbb{P}(D_n < x | E_4)\mathbb{P}(E_4) = \Psi_2 (p, \lambda, d_h, d_v, x),
\end{align*}
where $\Psi_2()$ is a function of the (charging) road density (i.e., $p$ and $\lambda$) and the trip information (i.e., $d_h, d_v,$ and $x$).
\end{lemma}
The complete form of $\Psi_2()$ is given in (\ref{lemma4eqn}) in Appendix \ref{appendxc}.

\subsection{Case b: When source (S) and destination (D) are on two perpendicular roads}
\label{dnxsubb}
Let $B$ denote the case when S and D are on two perpendicular roads. $\mathbb{P}(B) = \frac{1}{2}$. We consider four scenarios: \begin{itemize}
    \item Both source and destination roads are charging,
    \item Only source road is charging,
    \item Only destination road is charging,
    \item Both source and destination roads are not charging.
\end{itemize}

In each scenario, we first describe its probability, then present the distribution of $D_n$ given the scenario as a lemma. 

\subsubsection{Both source and destination roads are charging}

Let $E_5$ denote the case when both source and destination roads are on two perpendicular roads and are charging. The probability of event $E_5$ is $\frac{p^2}{2}$. 

\begin{lemma}
The distribution of $D_n$ given $E_6$ is given as follows:

\begin{align*}
\mathbb{P}(D_n < x | E_5)\mathbb{P}(E_5) = \frac{p^2}{2}.
\end{align*} 
\end{lemma}
\begin{IEEEproof}
    Since both the source road and the destination road are charging, the optimal driving route is always taking the source road and the destination road. Hence, $\mathbb{P}(D_n < x | E_{5})$ is always $1$.
\end{IEEEproof}
 
\subsubsection{Only source road is charging}

Let $E_6$ denote the case when source and destination roads are on two perpendicular roads and only the source road is charging. The probability of event $E_6$ is $\frac{p(1-p)}{2}$. 
\begin{lemma}
\label{DnE6}
The distribution of $D_n$ given $E_6$ is given as follows:

\begin{align*}
\mathbb{P}(D_n < x | E_6)\mathbb{P}(E_6) = \frac{p(1-p)}{2}.
\end{align*} 
\end{lemma}
\begin{IEEEproof}
    See Appendix \ref{appendxe6}.
\end{IEEEproof}
 
\subsubsection{Only destination road is charging}

Let $E_7$ denote the case when source and destination roads are on two perpendicular roads and the destination road is charging. The probability of event $E_7$ is $\frac{p(1-p)}{2}$. 
\begin{lemma}
\label{DnE7}
The distribution of $D_n$ given $E_7$ is given as follows:
\begin{align*}
&\mathbb{P}(D_n < x | E_7)\mathbb{P}(E_7) = \Psi_3 (p, \lambda, d_h, d_v, x),
\end{align*}
where $\Psi_3()$ is a function of the (charging) road density (i.e., $p$ and $\lambda$) and the trip information (i.e., $d_h, d_v,$ and $x$).
\end{lemma}
The complete form of $\Psi_3()$ is given in (\ref{lemmae7eqn}) in Appendix \ref{appendxd}.

\subsubsection{Both source and destination roads are not charging.} \label{dnxsub8}

Let $E_8$ denote the case when source and destination roads are on two perpendicular roads and both are not charging. The probability of event $E_8$ is $\frac{(1-p)^2}{2}$. 

\begin{lemma}
\label{DnE8}
The distribution of $D_n$ given $E_8$ is given as follows:
\begin{align*}
&\mathbb{P}(D_n < x | E_8)\mathbb{P}(E_8) = \Psi_4 (p, \lambda, d_h, d_v, x),
\end{align*}
where $\Psi_4()$ is a function of the (charging) road density (i.e., $p$ and $\lambda$) and the trip information (i.e., $d_h, d_v,$ and $x$).
\end{lemma}
The complete form of $\Psi_4()$ is given in (\ref{lemmae8eqn}) in Appendix \ref{appendxe}.

\subsection{Distribution of the distance to the nearest charging road}
Having derived the distribution of the distance to the nearest charging road, i.e., $D_n$, given eight cases in Lemmas \ref{DnE1}-\ref{DnE8}, we are ready to present the distribution of $D_n$, which is given in the following Theorem.

\begin{theorem}
The probability that the distance from the source to the nearest charging road, i.e. $D_n$, is less than a positive real number $x$ is given by
$$\mathbb{P}(D_n < x) = \sum_{i=1}^{8} \mathbb{P}(D_n < x|E_i)\mathbb{P}(E_i).$$
\end{theorem}

\begin{IEEEproof}
    This result follows directly by substituting $\mathbb{P}(D_n < x|E_i)\mathbb{P}(E_i), i\in[1,8]$ from Lemma \ref{DnE1}-\ref{DnE8}, respectively.
\end{IEEEproof}

\begin{remark}
\label{chap4ext}

The distribution of the distance to the nearest charging road can be used to study multiple important performance metrics. For instance,  it can be used to study a lower bound on the fraction of the total trip spent on non-charging roads, $\frac{D_n}{d_h + d_v}$. Hence, it can also be used to study an upper bound on the fraction of the trip spent on charging roads, $1-\frac{D_n}{d_h + d_v}$. These two extensions provide a relative view on the utilization of charging roads with respective to the total distance traveled in a trip. Since we already discussed the distribution of $D_n$, the distributions of $\frac{D_n}{d_h + d_v}$ and $1-\frac{D_n}{d_h + d_v}$ can be obtained using simple random variable transformation.
\end{remark}

\section{Probability of passing through at least one charging road}
\label{atleast1charging}

We denote the event that any given trip passes through at least one charging road by $T_c$, and the event that any given trip passes through no charging road by $\overline{T_c}$. In this section, we calculate the probability $\mathbb{P}(T_c)$ based on the routing policy explained in Sec.~\ref{policy}. The probability of $T_c$ can be derived as follows. 

\begin{equation*}
\begin{split}
    \mathbb{P}(T_c) = 1 - \mathbb{P}(\overline{T_c})  = 1 - \sum_{i=1}^{8} \mathbb{P}(\overline{T_c} | E_i)\mathbb{P}(E_i),
\end{split}
\end{equation*}

where $E_i$'s are defined in Sec.~\ref{policy}. It is apparent that $\mathbb{P}(\overline{T_c} | E_i) = 0$ for $i \in \{1,2,3,5,6,7\}$ since at least one of the source and destination roads is already a charging road in those cases. Hence, the probability of interest is reduced as the following Theorem.

\begin{theorem}
\label{theorem2}
The probability of passing through at least one charging road $\mathbb{P}(T_c)$ is given by
\begin{equation*}
\begin{split}
    \mathbb{P}(T_c) = 1 - \sum_{i=4,8} \mathbb{P}(\overline{T_c} | E_i)\mathbb{P}(E_i),
\end{split}
\end{equation*}
where
\begin{equation*}
\begin{split}
    &\mathbb{P}(\overline{T_c} | E_4)\mathbb{P}(E_4)
    =[e^{-\lambda d_v}(1-p) \nonumber\\&+ \lambda (1-p) d_v e^{-\lambda (1-p) d_v} e^{-\lambda p d_v} \\&+ e^{-\lambda p d_v}(1-e^{-\lambda (1-p) d_v} - \lambda (1-p) d_v e^{- \lambda (1-p) d_v}) e^{-\lambda p d_h}] \nonumber\\& \times(1-p)^2,
\end{split}
\end{equation*}
\begin{equation*}
\begin{split}
    &\mathbb{P}(\overline{T_c} | E_8)\mathbb{P}(E_8) 
    =[e^{-\lambda d_v} + e^{-\lambda p d_v}(1-e^{-\lambda (1-p) d_v})e^{-\lambda p d_h} \nonumber\\&+ (1-e^{-\lambda p d_v})e^{-\lambda d_h}](1-p)^2.
\end{split}
\end{equation*}
\end{theorem}
\begin{IEEEproof}
    See Appendix \ref{appendxf}.
    \end{IEEEproof}

\section{Numerical Results}
\label{analyticalResult}

In this section, we present the analytical and simulation results of the two performance metrics with various values of $p>0$. 

\subsection{Distribution of the distance to the nearest charging road}



\begin{figure}[!t]
\centering
\captionsetup[subfigure]{font=scriptsize,labelfont=normalsize}
\subfloat[Manhattan\label{sfig:man}]{%
  \includegraphics[width=0.8\linewidth,keepaspectratio]{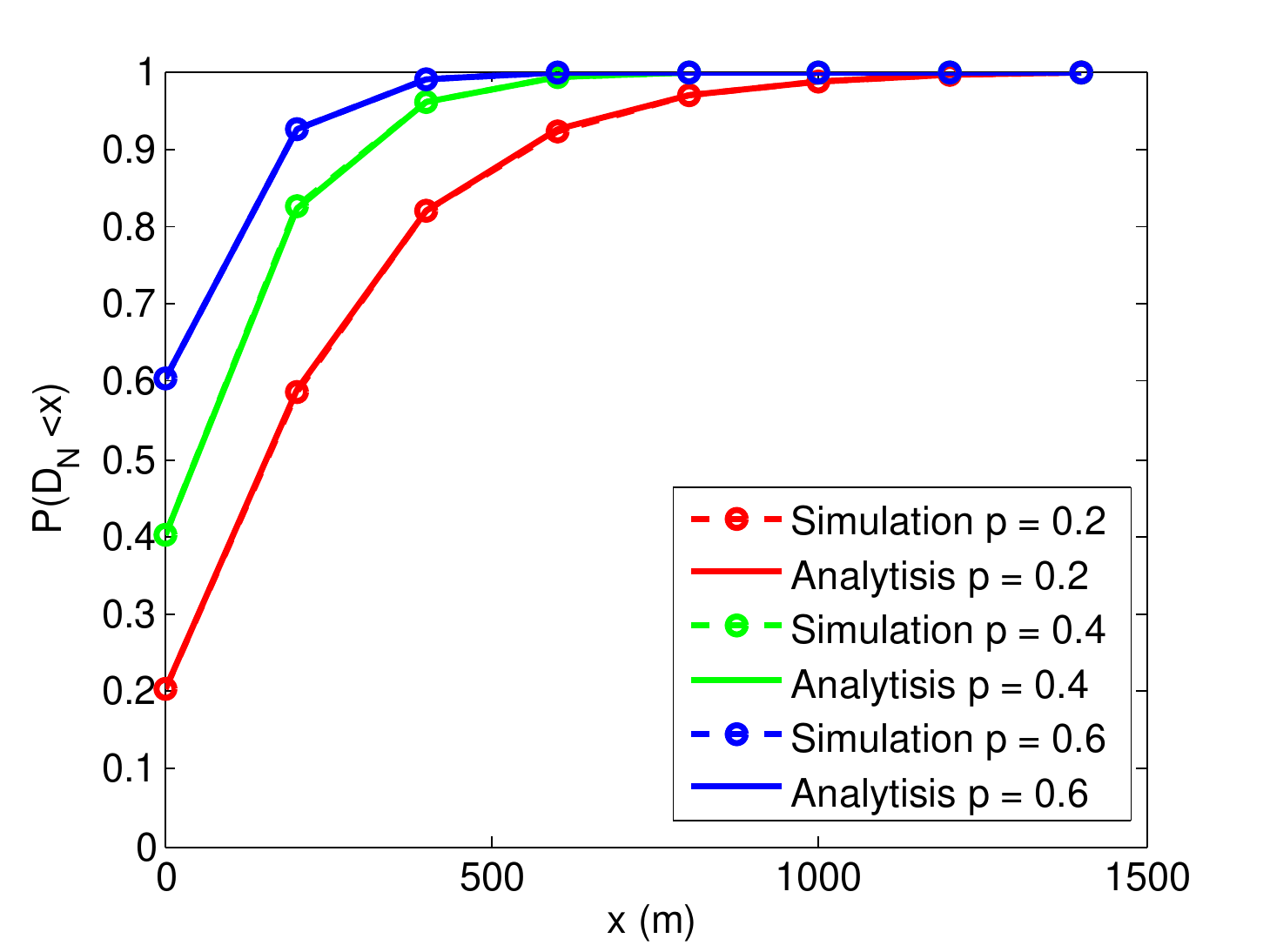}%
}\hfill
\subfloat[Western Chicago\label{sfig:chi}]{%
  \includegraphics[width=0.8\linewidth,keepaspectratio]{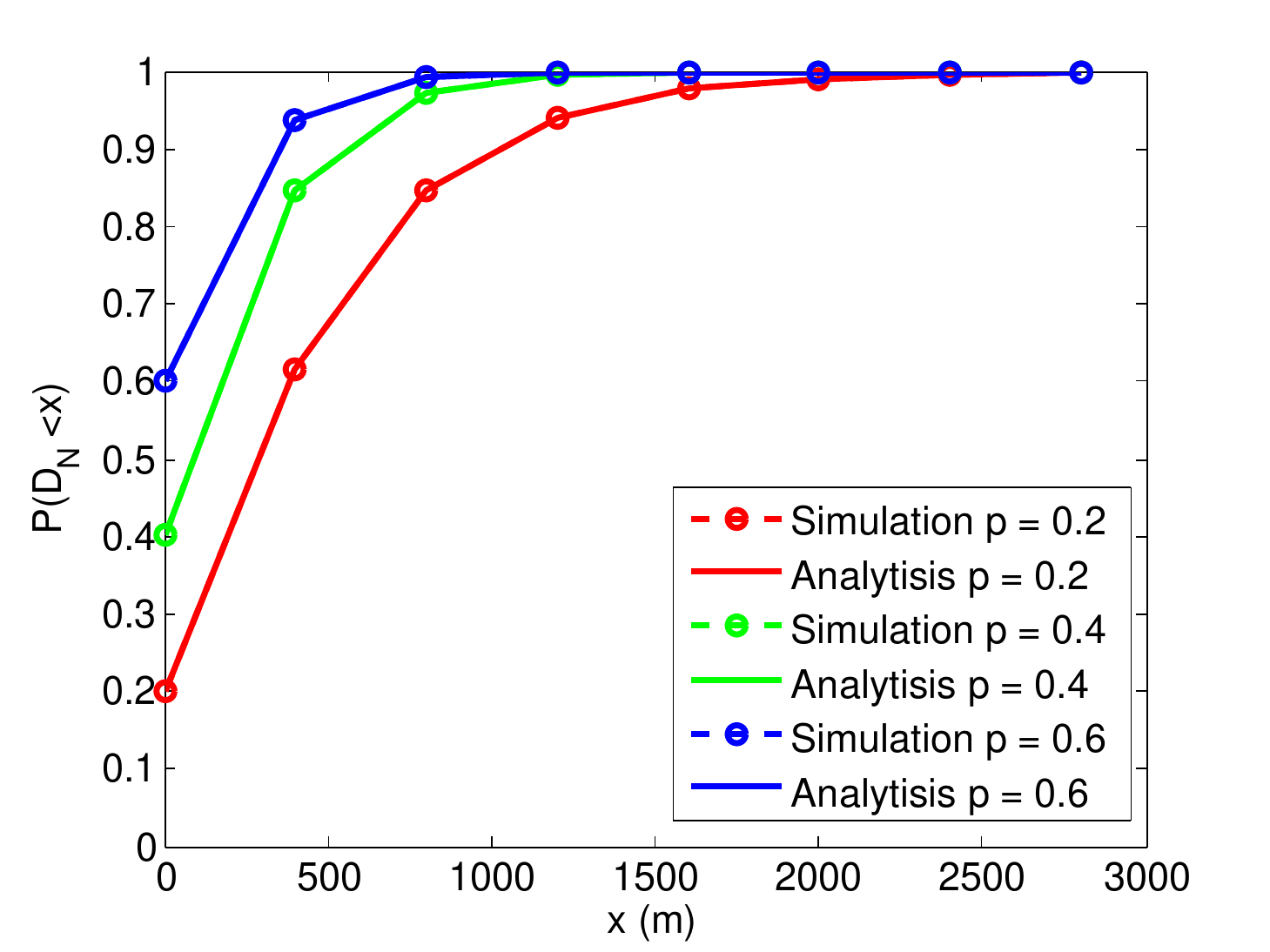}%
}
\caption{The probability $\mathbb{P}(D_n <x)$ in two urban cities.}
\label{TexEd}
\end{figure}

Since our analysis focuses on the deployment of charging roads in a metropolitan setting, we choose to perform simulations that portray two areas: Manhattan (New York) and western Chicago. Based on the road network density studied in~\cite{spatialtemporalcha}, we estimate the density parameter $\lambda$ to be 0.016 (road/meter) in Manhattan and 0.006 (road/meter) in western Chicago. We also select $d_h = 2$, $d_v = 3$ (km) in the simulation of Manhattan, and $d_h = 4$, $d_v = 5$ (km) in the simulation of western Chicago to represent typical trips in these two areas. The distributions of the distance to the nearest charging road, i.e., $\mathbb{P}(D_n <x)$, for these parameter sets are shown in Figs.~\ref{sfig:man}-\ref{sfig:chi}, respectively. We observe the overall trend that as the density of charging road increases, i.e., higher values of $p$, the quicker $\mathbb{P}(D_n <x)$ goes to one, or in other words, the closer the nearest charging road is from the source. The probability also goes to one faster in Manhattan as it does in western Chicago, as Manhattan has a higher density of roads. In addition, for all the curves of $\mathbb{P}(D_n <x)$ with different values of $p$, for a small value of $x$, $\mathbb{P}(D_n <x)$ is $p$, which is intuitive since $p$ is exactly the probability that the source road is a charging road. Another interesting observation is that in Manhattan and western Chicago, when 20\% of the roads are charging roads, after about only 500m and 1km, respectively, a driver will have 80\% chance of coming across a charging road on his or her trip. These insightful findings may benefit urban planers and policy makers to design how densely it needs to deploy charging roads. Car manufacturers can also refer to this metric to customize the battery size for electric vehicles in a specific city.

\subsection{Probability that any given trip passes through at least one charging road}



\begin{figure}[!t]
\centering
\captionsetup[subfigure]{font=scriptsize,labelfont=normalsize}
\subfloat[Manhattan\label{sfig:tcman}]{%
  \includegraphics[width=0.8\linewidth,keepaspectratio]{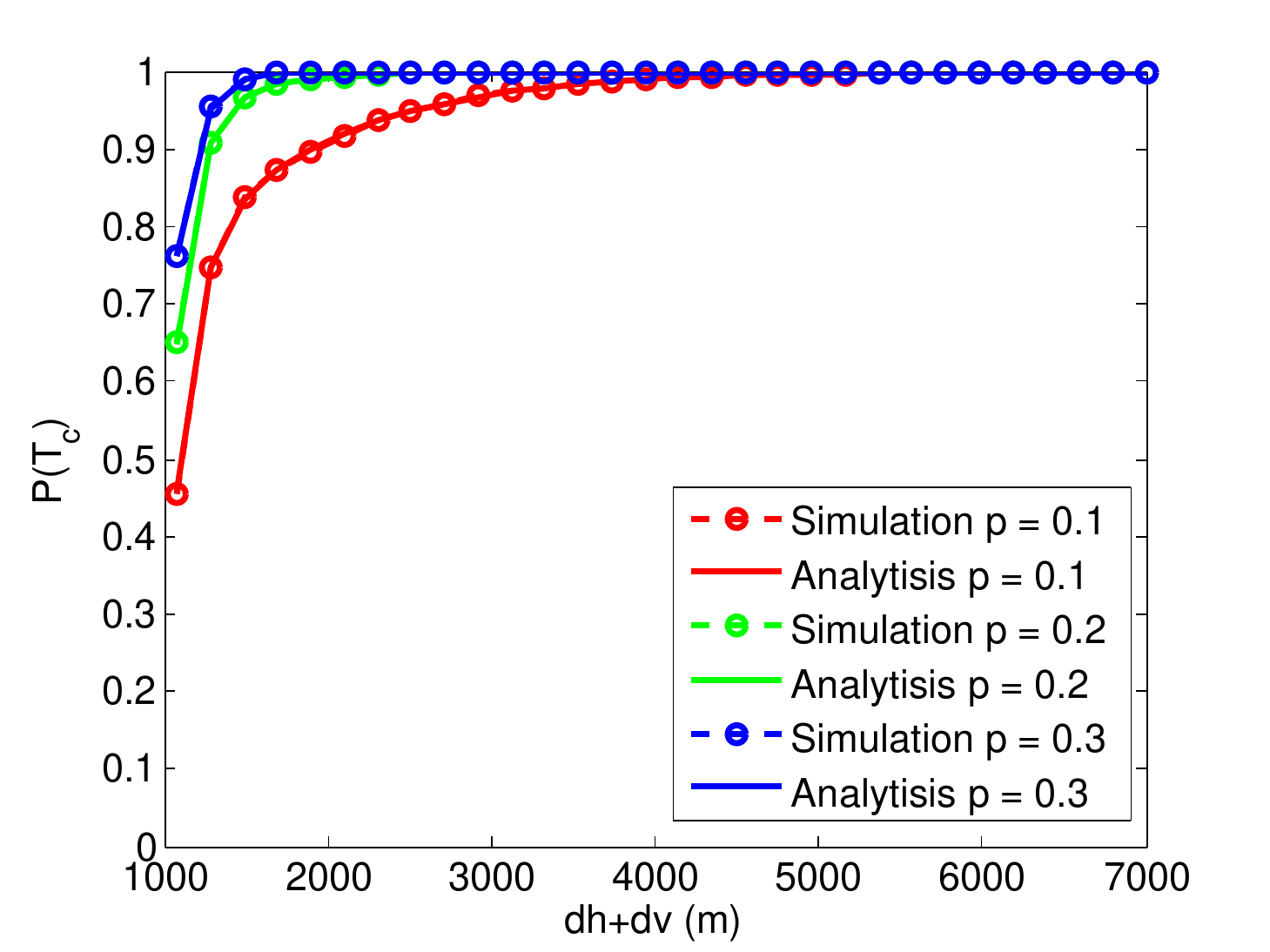}%
}\hfill
\subfloat[Western Chicago\label{sfig:tcchi}]{%
  \includegraphics[width=0.8\linewidth,keepaspectratio]{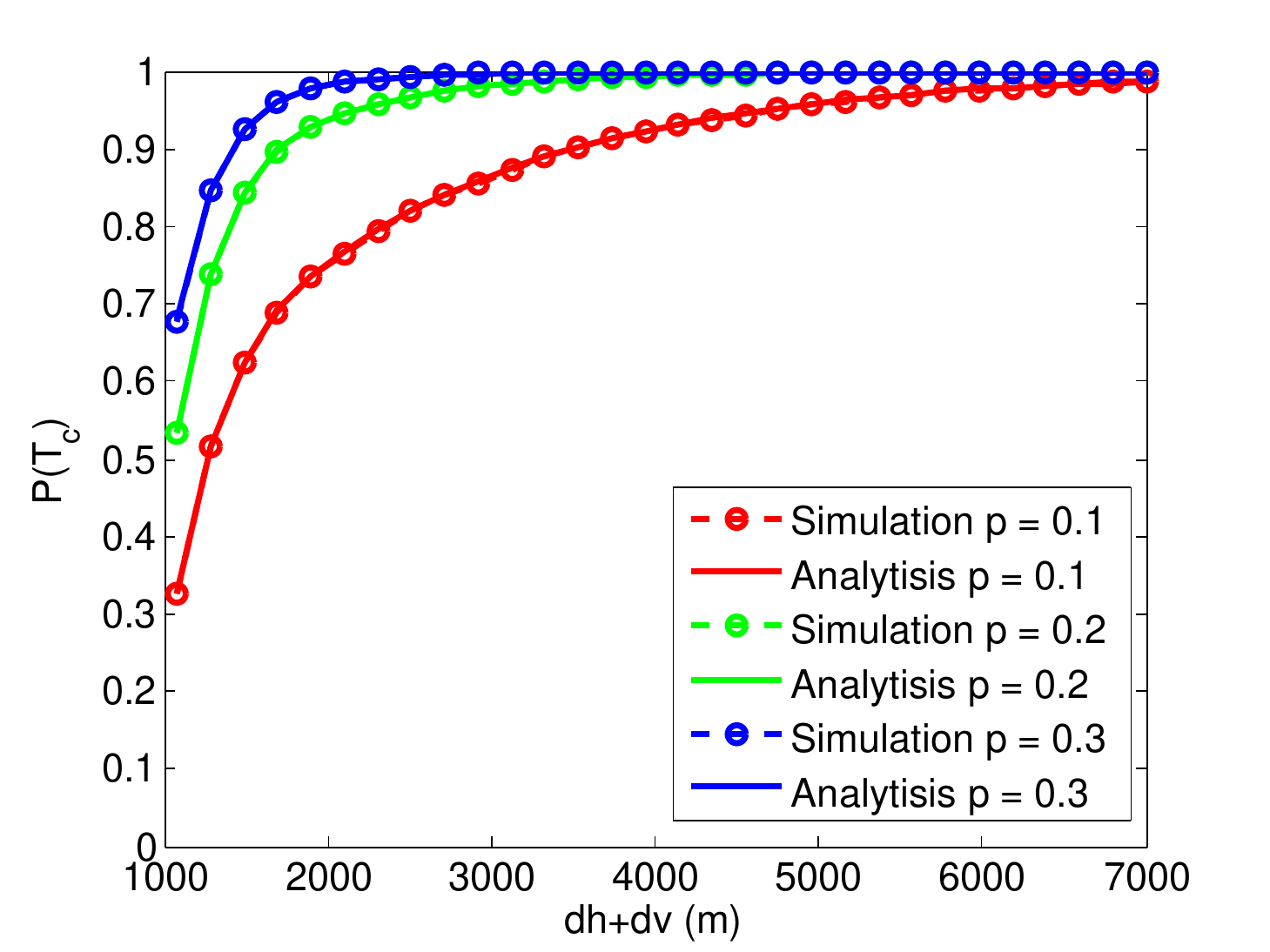}%
}
\caption{The probability $\mathbb{P}(T_c)$ as a function of the Manhattan distance of the trip.}
\label{TexEd}
\end{figure}

In this subsection we demonstrate the result for the probability that any given trip passes through at least one charging road, i.e., $\mathbb{P}(T_c)$. To maintain the metropolitan city setting, we keep the same road density $\lambda = 0.011$ for Manhattan and $\lambda = 0.006$ for western Chicago. Next, we simulate the probability $\mathbb{P}(T_c)$ for a trip distance from 1km to 7km. The result of our simulation is presented in Figs.~\ref{sfig:tcman}-\ref{sfig:tcchi}. We plot $\mathbb{P}(T_c)$ as a function of the Manhattan distance between the source and the destination, i.e., $d_h + d_v$. The trend is that as the distance between source and destination increases, it is more certain that the trip passes through at least one charging road. Furthermore, $\mathbb{P}(T_c)$ significantly increases with $p$.

\section{Conclusion \& Future Work}
\label{conclusion}

In this paper, we introduced a framework using stochastic geometry to assess the deployment of charging roads in a metropolitan setting. We provided a routing policy for drivers such that the shortest route is always selected and the time spent on charging roads throughout the trip is maximized. Then, we proposed analytical solutions to the two performance metrics: (i) the distribution of the distance from the source to the nearest charging road, and (ii) the probability that any given trip passes through at least one charging road. This analytical framework takes an important step towards a better understanding of the charging roads deployment in metropolitan cities and provides insights for various groups such as city planners, policy makers, car manufacturers, and drivers.

Further extension to this paper may include a more general system setup, in which important factors such as human mobility~\cite{limitsofpredictability} are considered for the placement of charging roads. In addition, spatial and temporal information about the traffic flow, congestion, and charging price can also be taken into account to formulate a more accurate routing policy and update the two performance metrics.   

\appendices
\section{proof of propositions~\ref{distx1} and~\ref{distx2}}
\label{appendxx1x2}
In this appendix, we outline the proof for the distribution of $X_2$ as given in Proposition \ref{distx2}. The proof for the distribution of $X_1$ is similar to that of $X_2$. 
\begin{align*}
    &\mathbb{P}(X_2 < x) = \\& \mathbb{P}(D_\mathrm{N-HC}-D_\mathrm{N-HNC}<x | D_\mathrm{N-HNC} < D_\mathrm{N-HC} < d_v) \\& = 
    \mathbb{E_{D_\mathrm{N-HC}}}[\mathbb{P}(D_\mathrm{N-HNC}>t-x | D_\mathrm{N-HC} = t, D_\mathrm{N-HNC} < t)\\&\times\mathbbm{1}\{t>x\}] \\& = 
    \mathbb{E_{D_\mathrm{N-HC}}}[1-\mathbb{P}(D_\mathrm{N-HNC}\leq t-x | D_\mathrm{N-HC} = t,\\& D_\mathrm{N-HNC} < t)\mathbbm{1}\{t>x\}] \\& =
     1- \int_{x}^{d_v} \frac{1-e^{-\lambda (1-p) (t-x)}}{1-e^{-\lambda (1-p) t}} \times\frac{\lambda p e^{-\lambda p t}}{1-e^{-\lambda p d_v}} {\rm d} t, 
\end{align*}
where $f_{D_\mathrm{N-HC}}(t|0<D_\mathrm{N-HC}<d_v) = \frac{\lambda p e^{-\lambda p t}}{1-e^{-\lambda p d_v}}$, and $F_{D_\mathrm{N-HNC}}(t-x|0<D_\mathrm{N-HC}<t) =\frac{1-e^{-\lambda (1-p) (t-x)}}{1-e^{-\lambda (1-p) t}}$.
\begin{align*}
    &f_{X_2}(x) = \frac{{\rm d}}{{\rm d}x}\mathbb{P}(X_2 < x) \\&
    = \int_{x}^{d_v} \frac{\lambda^2 (1-p) p e^{-\lambda p t - \lambda (1-p) (t-x)}}{(1-e^{-\lambda p d_v})(1-e^{-\lambda (1-p) t})}  {\rm d} t.
\end{align*}

\bgroup
\def\arraystretch{1.5}
\begin{table*}
\centering
    \caption{Frequently-used functions}
    \begin{tabular}{p{5cm}| p{10cm}}
        \hline
        \textbf{Function Name}       & \textbf{Definition}    \\
        \hline
        $f_{1}(a_1,a_2,a_3,a_4,a_5,a_6)$ & $\int_{a_1}^{a_2} \int_{a_3}^{a_4} (F_{D_\mathrm{N-HC}} (a_5) -F_{D_\mathrm{N-HC}} (a_6)) f_{D_\mathrm{N-HNC}}(y) f_{d_L}(t) {\rm d}y {\rm d}t$ \\ \hline
        $f_{2}(a_1,a_2,a_3,a_4,a_5,a_6)$     & $\int_{a_1}^{a_2} \int_{a_3}^{a_4} (F_{D_\mathrm{N-HC}} (a_5) -F_{D_\mathrm{N-HC}} (a_6)) f_{d_L}(y) f_{D_\mathrm{N-HNC}}(t) {\rm d}y {\rm d}t$       \\ \hline
        $f_{3}(a_1,a_2,a_3,a_4,a_5,a_6)$        & $\int_{a_1}^{a_2} \int_{a_3}^{a_4} (F_{D_\mathrm{N-HNC}} (a_5) -F_{D_\mathrm{N-HNC}} (a_6)) f_{D_\mathrm{N-HC}}(y)  f_{d_L}(t) {\rm d}y {\rm d}t$ \\ \hline
        $f_{4}(a_1,a_2,a_3,a_4,a_5,a_6)$         & $\int_{a_1}^{a_2} \int_{a_3}^{a_4} (F_{D_\mathrm{N-HNC}} (a_5) -F_{D_\mathrm{N-HNC}} (a_6)) f_{d_L}(y) f_{D_\mathrm{N-HC}}(t) {\rm d}y {\rm d}t$ \\ \hline
        $f_{5}(a_1,a_2,a_3,a_4,a_5)$    & $\int_{a_1}^{a_2} \int_{a_3}^{a_4} (1-F_{D_\mathrm{N-HC}} (a_5)) f_{D_\mathrm{N-HNC}}(y)  f_{d_L}(t) {\rm d}y {\rm d}t$      \\ \hline
        $f_{6}(a_1,a_2,a_3,a_4)$	& $\int_{a_1}^{a_2} (F_{D_\mathrm{N-HC}} (a_3) -F_{D_\mathrm{N-HC}} (a_4)) f_{D_\mathrm{N-HNC}}(y) {\rm d}y$ \\ \hline
        $f_{7}(a_1,a_2,a_3)$	& $\int_{a_1}^{a_2} F_{D_\mathrm{N-HC}} (a_3) f_{D_\mathrm{N-HNC}}(y) {\rm d}y$ \\ \hline
        $f_{8}(a_1,a_2,a_3,a_4,a_5)$	& $\int_{a_1}^{a_2} \int_{a_3}^{a_4} F_{D_\mathrm{N-VC}} (a_5) f_{D_\mathrm{N-HC}}(y) f_{D_\mathrm{N-HNC}}(t) {\rm d}y {\rm d}t$ \\ \hline
       $f_{9}(a_1,a_2,a_3,a_4,a_5,a_6)$		& $\int_{a_1}^{a_2} \int_{a_3}^{a_4} (F_{D_\mathrm{N-HC}} (a_5) -F_{D_\mathrm{N-HC}} (a_6)) f_{D_\mathrm{N-HNC}}(y) f_{D_\mathrm{N-VC}}(t) {\rm d}y {\rm d}t$ \\ \hline
       $f_{10}(a_1,a_2,a_3)$	& $\int_{a_1}^{a_2} F_{D_\mathrm{N-HNC}} (a_3) f_{D_\mathrm{N-HC}}(y) {\rm d}y$ \\ \hline
		$f_{11}(a_1,a_2,a_3)$	& $\int_{a_1}^{a_2} F_{D_\mathrm{N-HC}} (a_3) f_{D_\mathrm{N-VNC}}(y) {\rm d}y$ \\ \hline
		$f_{12}(a_1,a_2,a_3,a_4,a_5)$ &  $\int_{a_1}^{a_2} \int_{a_3}^{a_4} F_{D_\mathrm{N-HC}} (a_5) f_{D_\mathrm{N-VC}}(y) f_{D_\mathrm{N-VNC}}(t) {\rm d}y {\rm d}t$ \\ \hline
		$f_{13}(a_1,a_2,a_3,a_4,a_5,a_6)$   &   $\int_{a_1}^{a_2} \int_{a_3}^{a_4} (F_{D_\mathrm{N-VC}} (a_5) - F_{D_\mathrm{N-VC}} (a_6)) f_{D_\mathrm{N-VNC}}(y) f_{D_\mathrm{N-HC}}(t) {\rm d}y {\rm d}t$ \\ \hline
		$g_1$	& $f_1(x-d_v,\infty,x,\infty,x,d_v)$ \\ \hline
		$g_2$	& $f_1(x-d_v,\infty,d_v,{\rm min}(x,t+d_v),y,d_v)$ \\ \hline
		$g_3$	& $f_1(0,x-d_v,t+d_v,\infty,t+d_v,d_v)$ \\ \hline
		$g_4$	& $f_3(x-d_h-d_v,\infty,x- d_h, \infty, x-d_h,d_v)$ \\ \hline
		$g_5$	& $f_3(0,\infty,d_v,{\rm min}(x-d_h,t+d_v),y,d_v)$ \\ \hline
		$g_6$	& $f_3(0,x-d_h-d_v,t+d_v,\infty,t+d_v,d_v)$ \\ 
        \hline
    \end{tabular}
    \label{table: functions}
\end{table*}
\egroup

\section{proof of Lemma \ref{DnE1}}
\label{appendxe1}
\begin{figure}[!h]
\centering
\includegraphics[width=0.5\columnwidth]{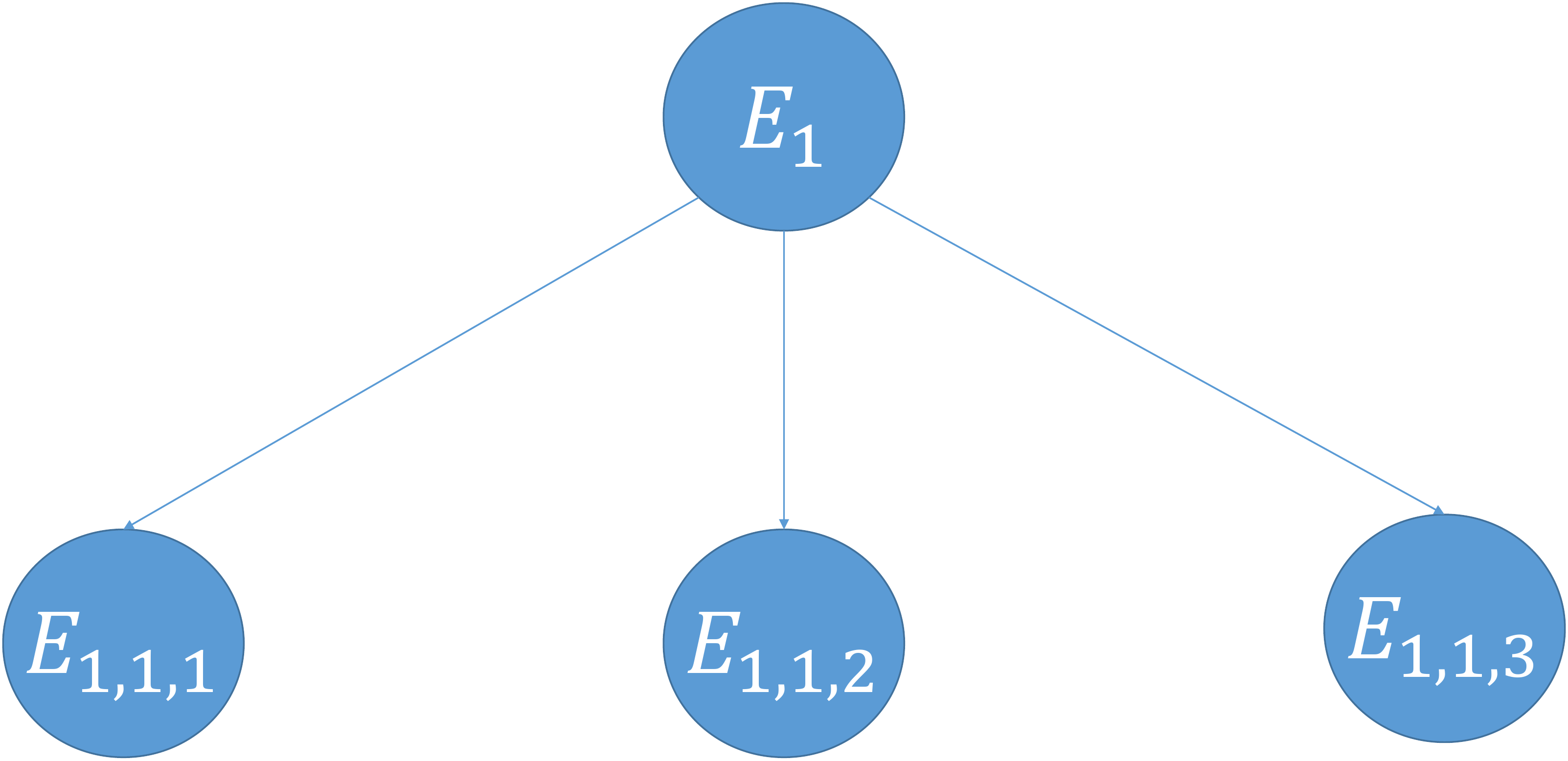}
\caption{Tree $E_1$: both source and destination roads are on two parallel roads and are charging.}
\label{E1}
\end{figure}

\begin{figure}[h]
\centering
\captionsetup[subfigure]{font=scriptsize,labelfont=normalsize}
\subfloat[Event $E_{1,1,1}$\label{E111}]{%
  \includegraphics[width=0.33\columnwidth,keepaspectratio]{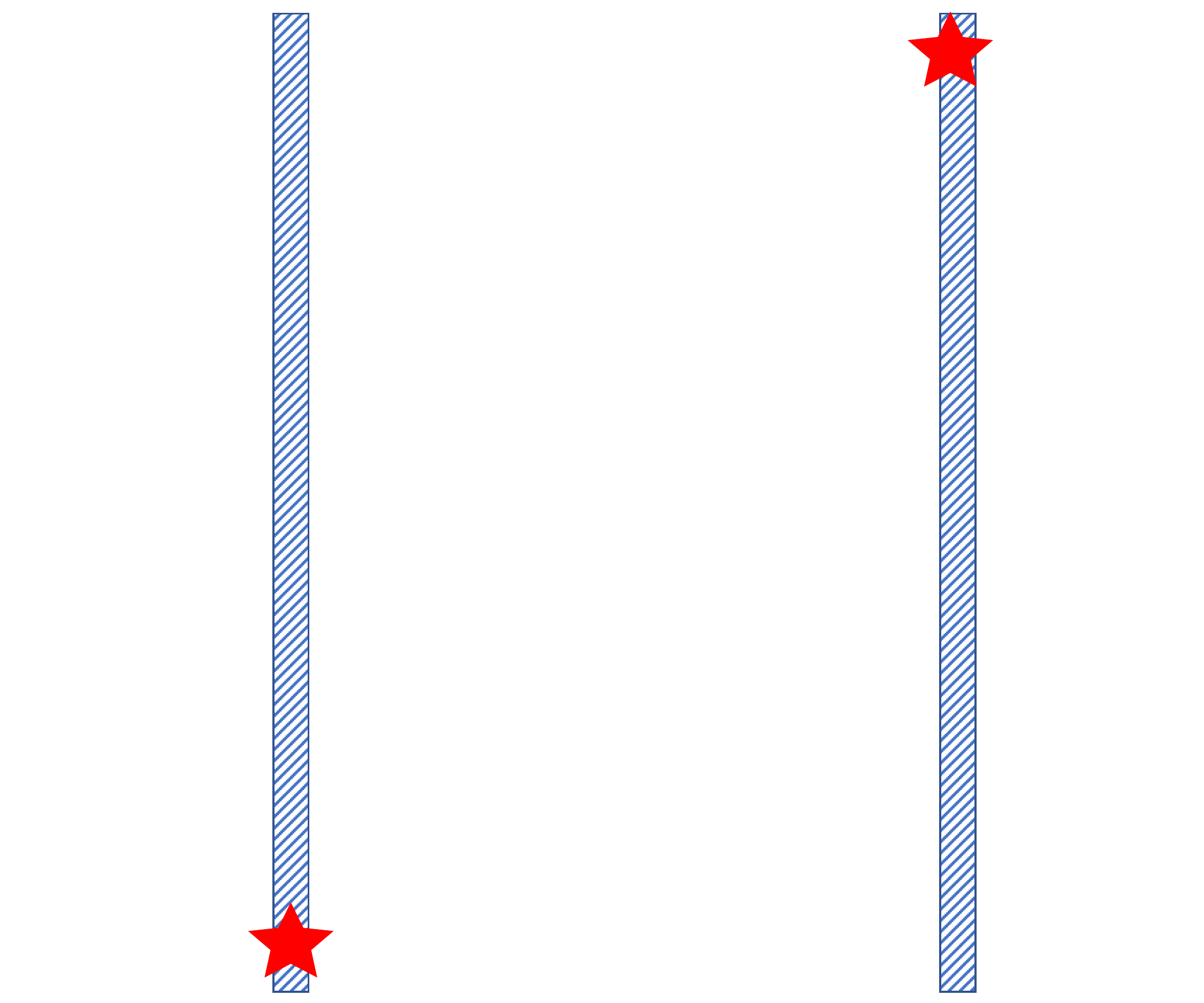}%
}\hfill
\subfloat[Event $E_{1,1,2}$\label{E112}]{%
  \includegraphics[width=0.33\columnwidth,keepaspectratio]{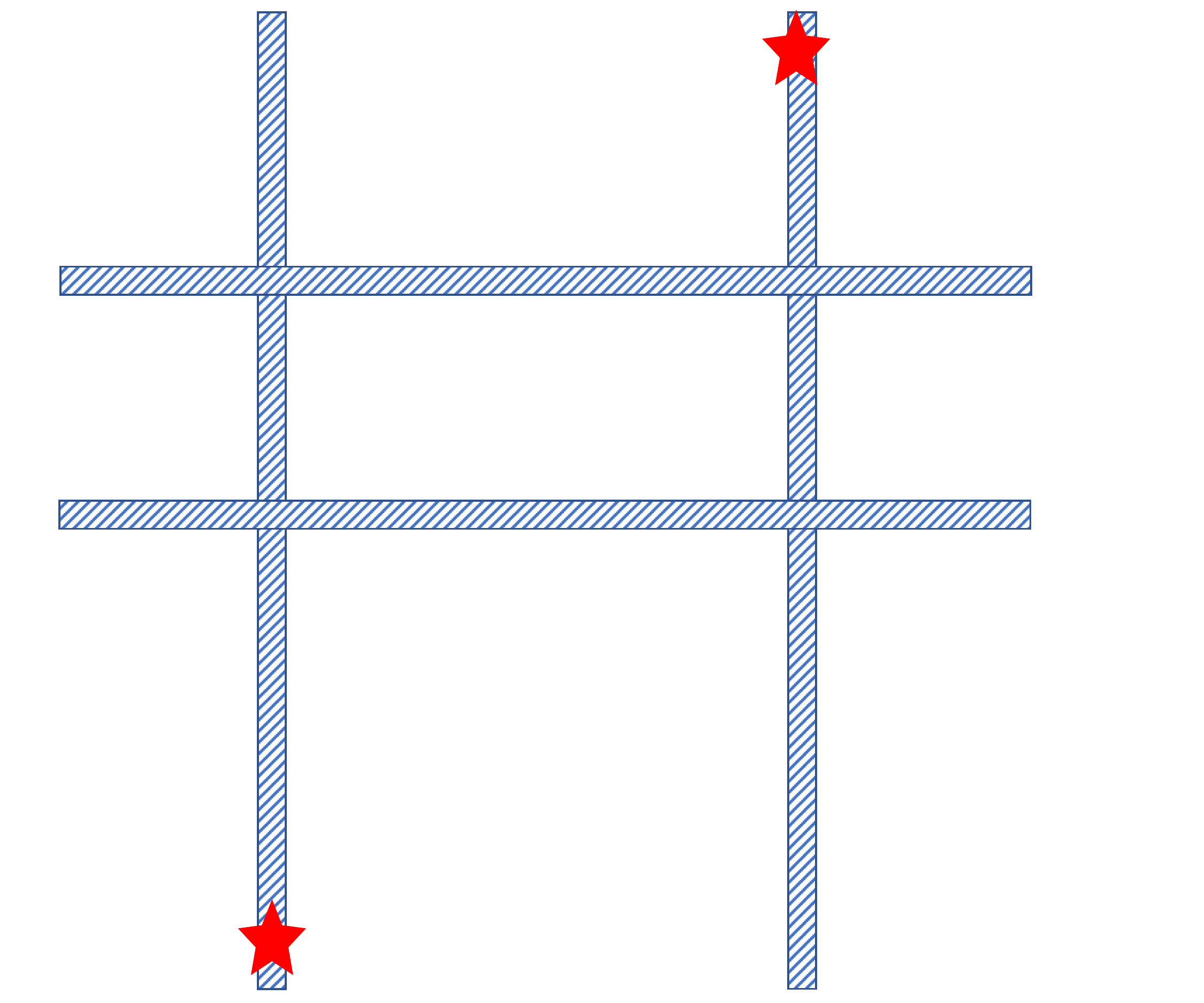}%
}
\hfill
\subfloat[Event $E_{1,1,3}$\label{E113}]{%
  \includegraphics[width=0.33\columnwidth,keepaspectratio]{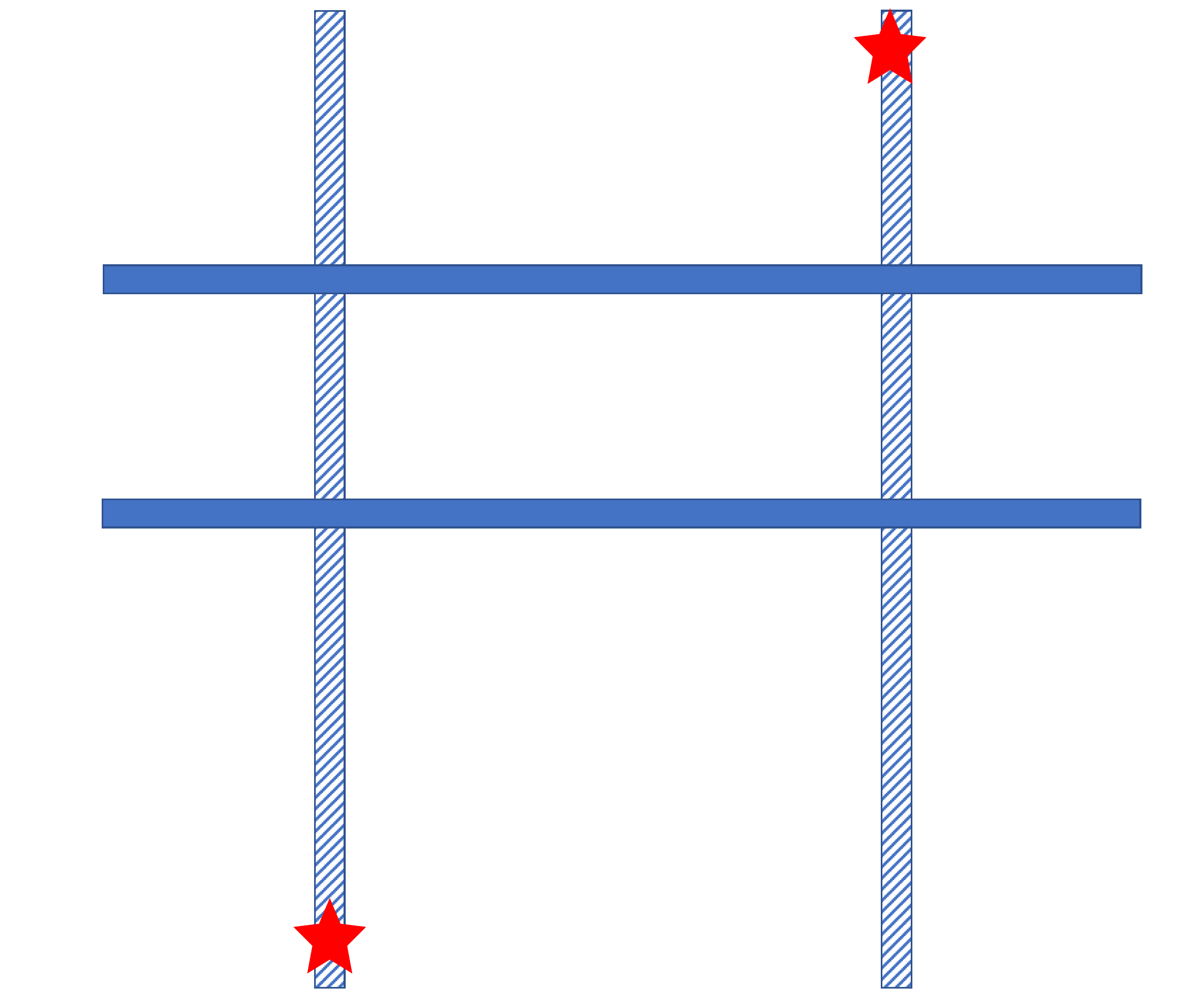}%
}

\subfloat{%
  \includegraphics[width=0.8\columnwidth,keepaspectratio]{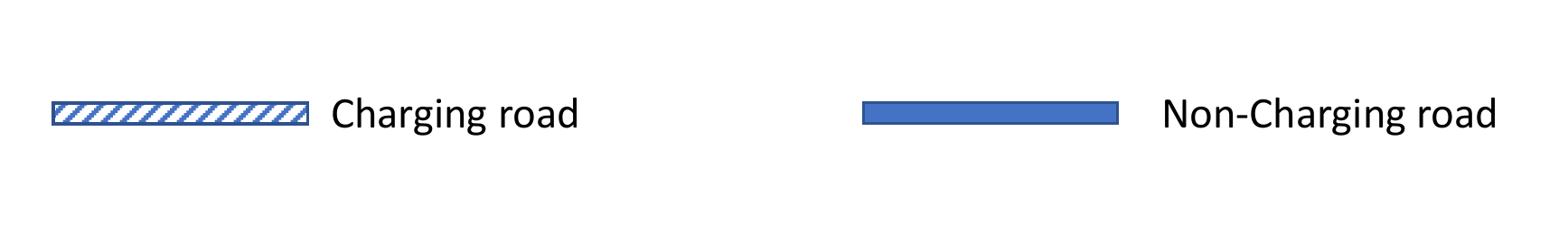}%
}

\caption{Subcases of tree $E_1$.}
\label{e1sub}
\end{figure}

In this appendix, we outline the proof for the probability that the distance to the nearest charging road is less than a positive real number $x$ given the event $E_1$, i.e., $\mathbb{P}(D_n < x|E_1)$. As shown in Fig.~\ref{E1}, we hereby denote subevents as $E_{1,i,j}$, in which $i$ is the level of depth of the event in the probability tree and $j$ is the index of the event at that level. Representative figures for $E_1$ are shown in Fig.~\ref{e1sub}. The distribution of $D_n$ given $E_1$ can be derived as follows:

\begin{align*}
\mathbb{P}(D_n < x | E_1)\mathbb{P}(E_1) = \sum_{i=1}^{N_1} \mathbb{P}(D_n < x | L_{1,i})\mathbb{P}(L_{1,i}),
\end{align*} 
where $N_1$ denotes the number of leaves of tree $E_1$, i.e., $N_1=3$, and $L_{1,i}$'s are successive events ending at the leaves of tree $E_1$ as shown in Fig. \ref{E1}. The definition for each event $L_{1,i}$ will be given in more details as we visit each leaf of the tree. 


\begin{itemize}[wide, labelwidth=!, labelindent=0pt]
    \item Event $E_{1,1,1}$: \\\textit{Description}: If there are no horizontal roads between S and D, as shown in Fig. \ref{E111}; 
    \\Event $L_{1,1} = E_{1,1,1} \cap E_1$;
    \\\textit{Probability}: $\mathbb{P}(E_{1,1,1}|E_1)=e^{-\lambda d_v}$, $\mathbb{P}(L_{1,1})=\mathbb{P}(E_{1,1,1}|E_1)\mathbb{P}(E_1)$; 
    \\\textit{Action}: we simply use the nearest horizontal road, whether it is below the source or above the destination. 
    
    \item Event $E_{1,1,2}$: \\\textit{Description}: If there is at least one horizontal charging road between S and D, as shown in Fig. \ref{E112}; 
    \\Event $L_{1,2} = E_{1,1,2} \cap E_1$;
    \\\textit{Probability}: $\mathbb{P}(E_{1,1,2}|E_1)=1-e^{-\lambda p d_v}$, $\mathbb{P}(L_{1,2})=\mathbb{P}(E_{1,1,2}|E_1)\mathbb{P}(E_1)$; \\\textit{Action}: we can take any horizontal charging road between S and D.
    \item Event $E_{1,1,3}$: \\\textit{Description}: If there are no horizontal charging roads but at least one horizontal non-charging road between S and D, as shown in Fig. \ref{E113}; 
    \\Event $L_{1,3} = E_{1,1,3} \cap E_1$;
    \\\textit{Probability}: $\mathbb{P}(E_{1,1,3}|E_1)=e^{-\lambda p d_v}(1-e^{-\lambda (1-p) d_v})$, $\mathbb{P}(L_{1,3})=\mathbb{P}(E_{1,1,3}|E_1)\mathbb{P}(E_1)$; \\\textit{Action}: we take any horizontal non-charging road between S and D.  
\end{itemize}

Since the source road is already a charging road, $\mathbb{P}(D_n < x | L_{1,i}) = 1$ for all $i$.

\section{proof of Lemma \ref{DnE2}}
\label{appendxe2}
\begin{figure}[!h]
\centering
\includegraphics[width=0.8\columnwidth]{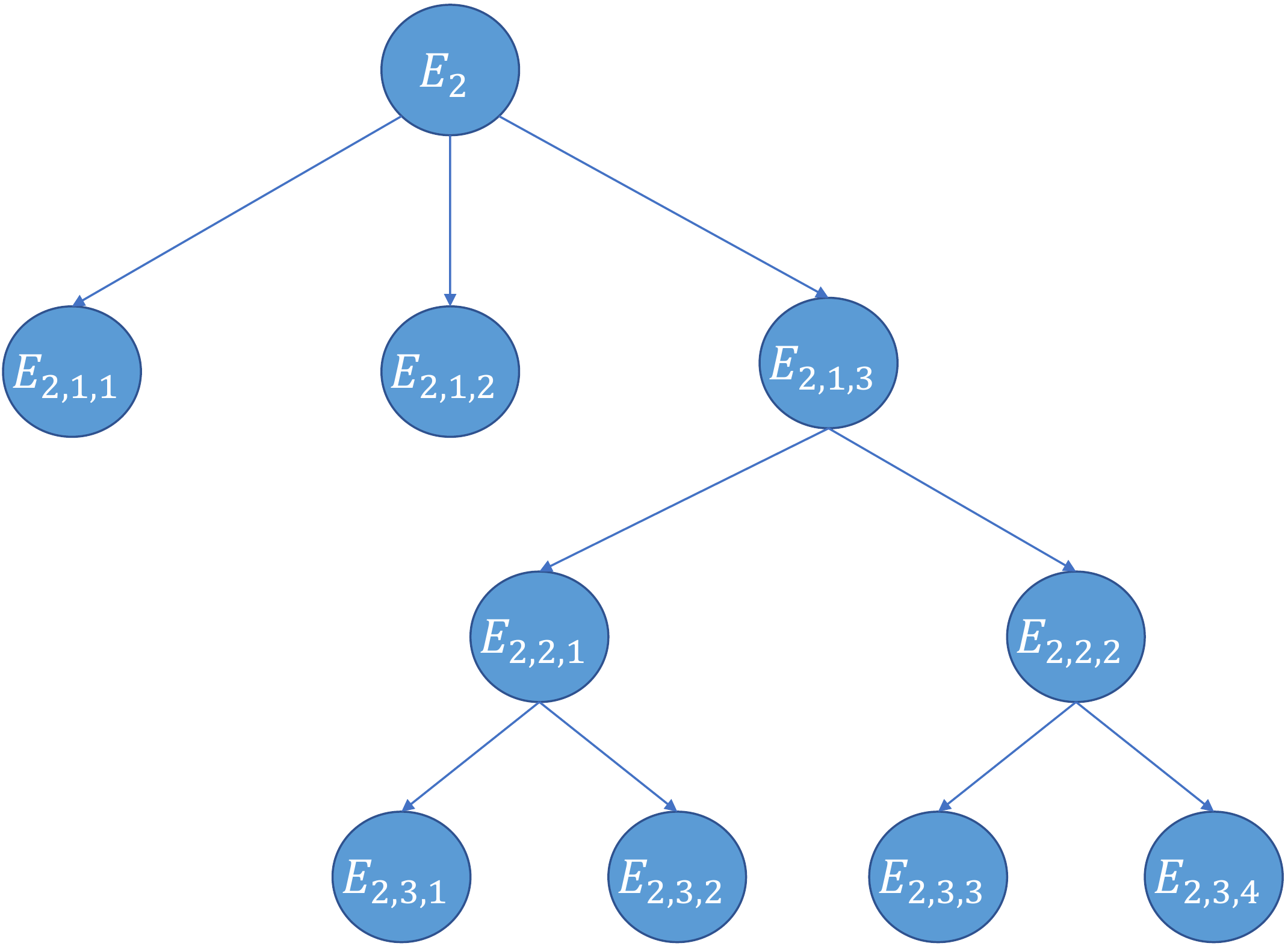}
\caption{Tree $E_2$: both source and destination roads are on two parallel roads and only the source road is charging.}
\label{E2}
\end{figure}

\begin{figure}[h]
\centering
\captionsetup[subfigure]{font=scriptsize,labelfont=normalsize}
\subfloat[Event $E_{2,1,1}$\label{E211}]{%
  \includegraphics[width=0.33\columnwidth,keepaspectratio]{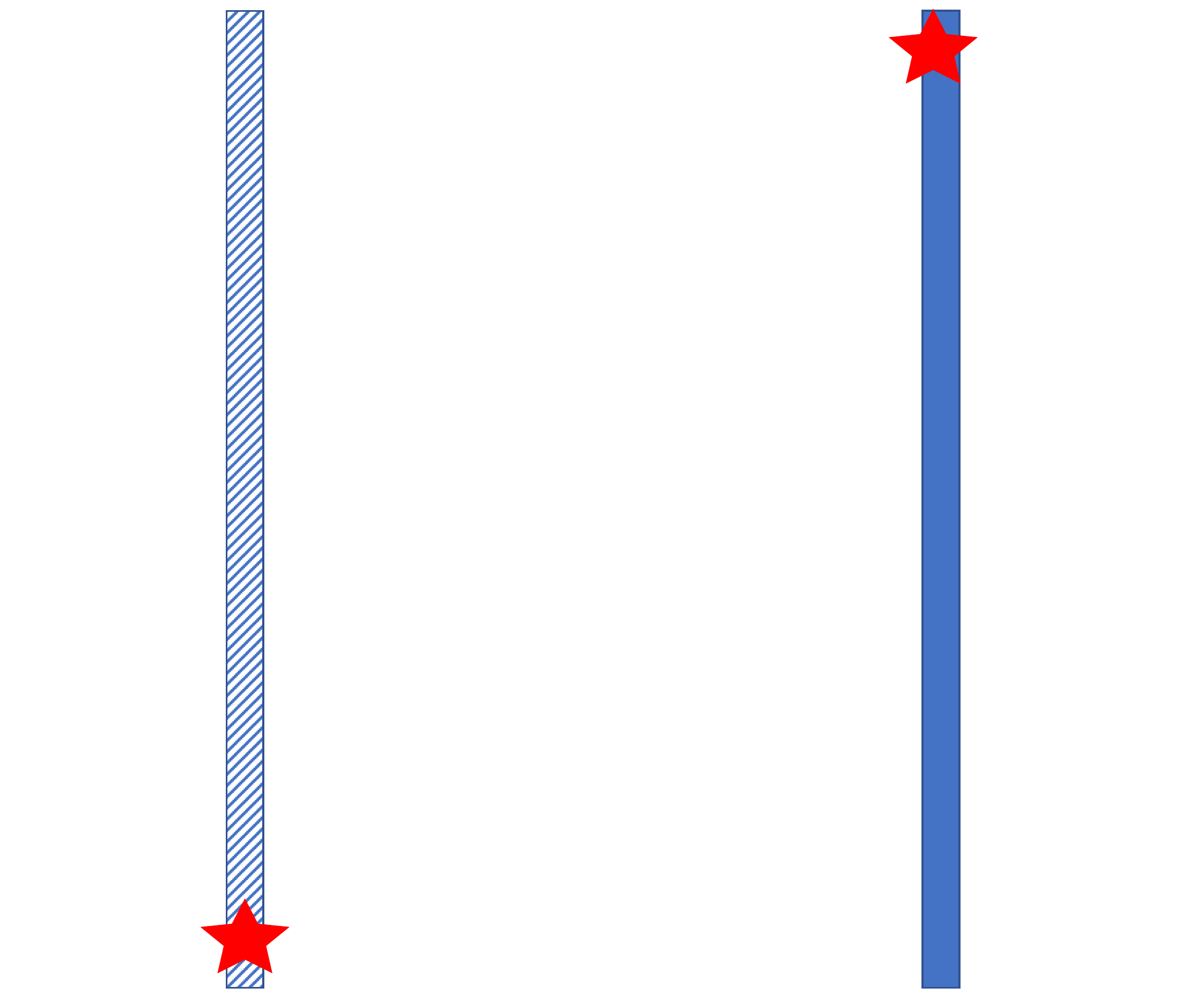}%
}\hfill
\subfloat[Event $E_{2,1,2}$\label{E212}]{%
  \includegraphics[width=0.33\columnwidth,keepaspectratio]{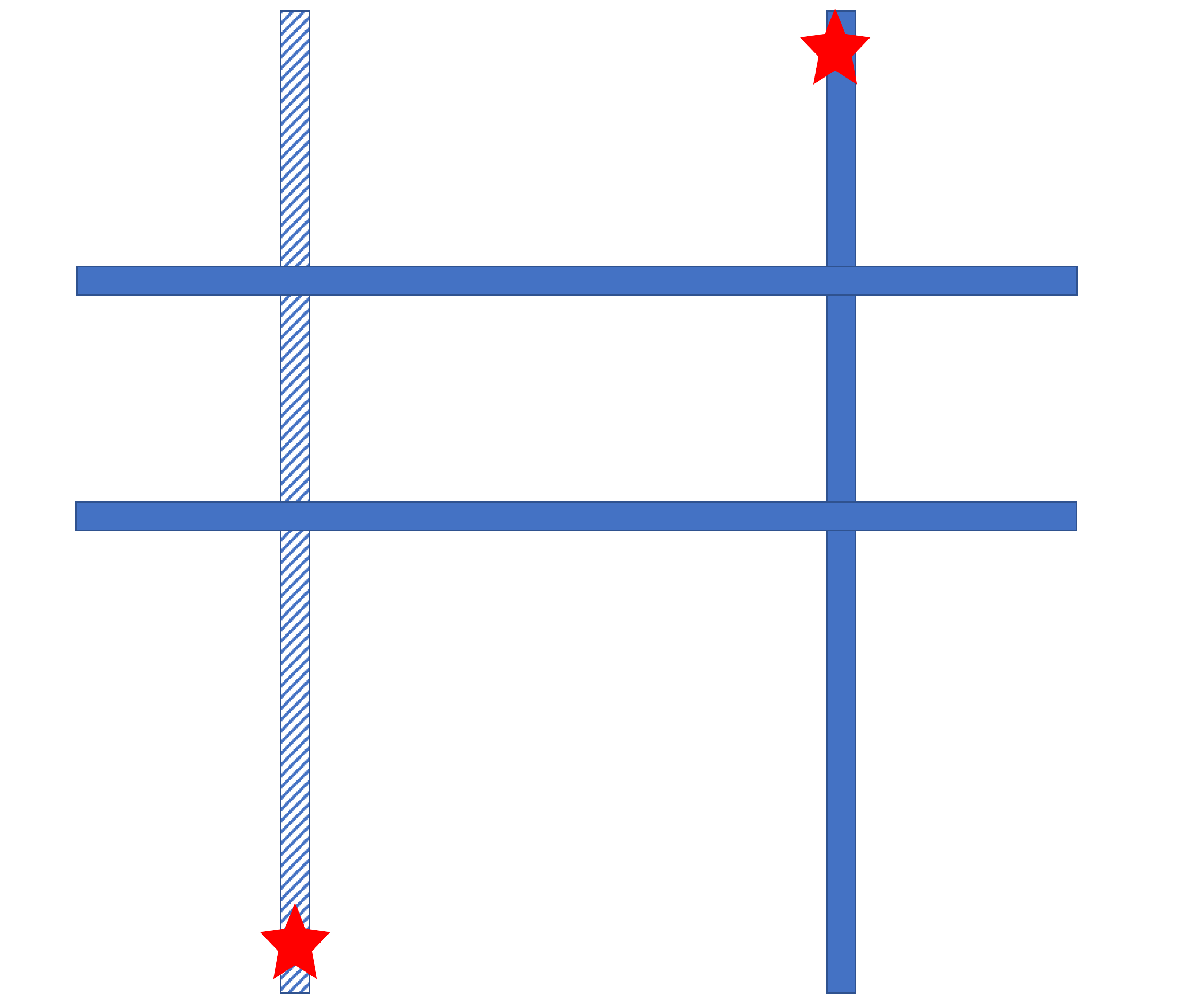}%
}
\hfill
\subfloat[Event $E_{2,3,1}$\label{E231}]{%
  \includegraphics[width=0.33\columnwidth,keepaspectratio]{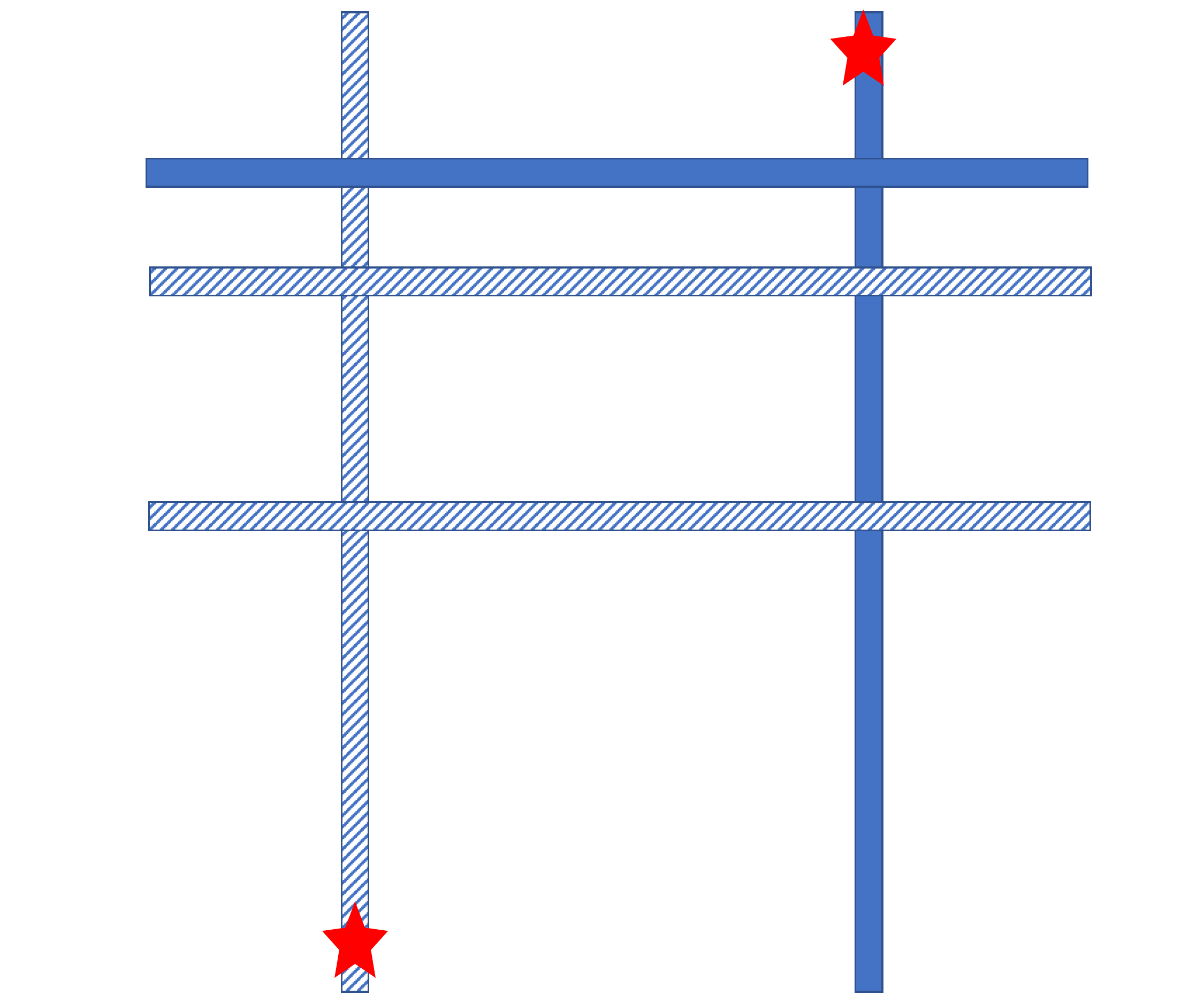}%
}

\subfloat[Event $E_{2,3,2}$\label{E232}]{%
  \includegraphics[width=0.33\columnwidth,keepaspectratio]{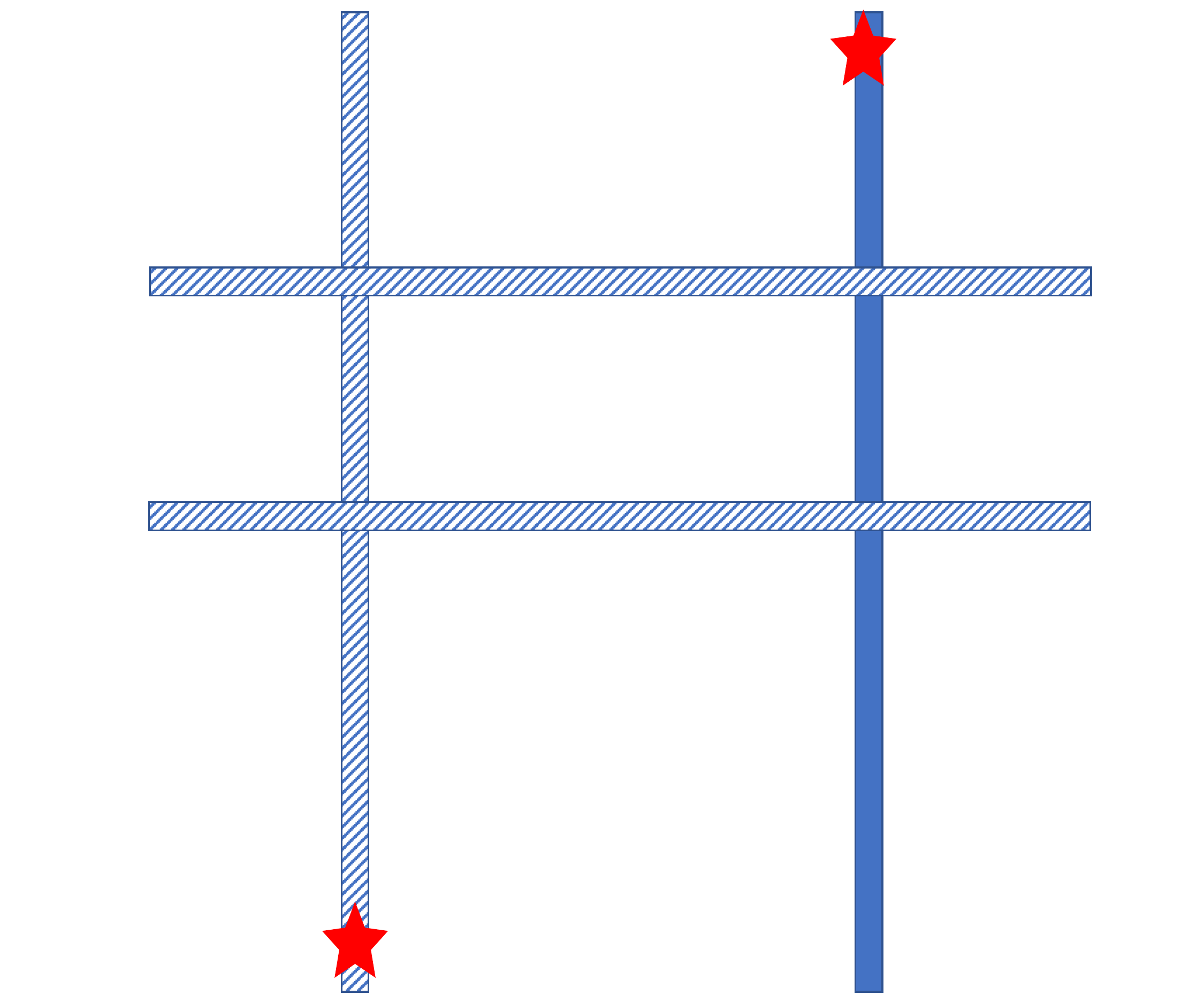}%
}\hfill
\subfloat[Event $E_{2,3,3}$\label{E233}]{%
  \includegraphics[width=0.33\columnwidth,keepaspectratio]{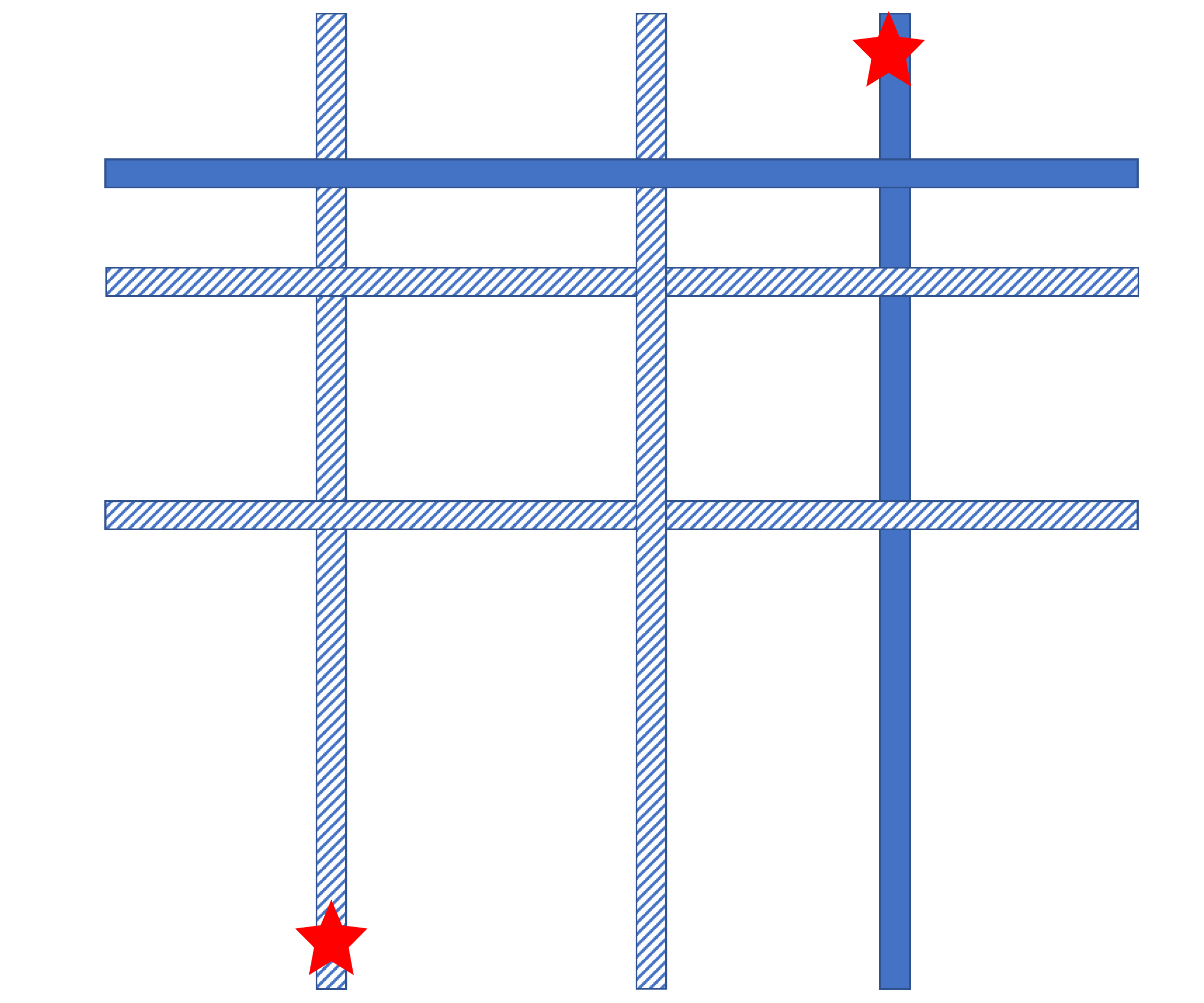}%
}
\hfill
\subfloat[Event $E_{2,3,4}$\label{E234}]{%
  \includegraphics[width=0.33\columnwidth,keepaspectratio]{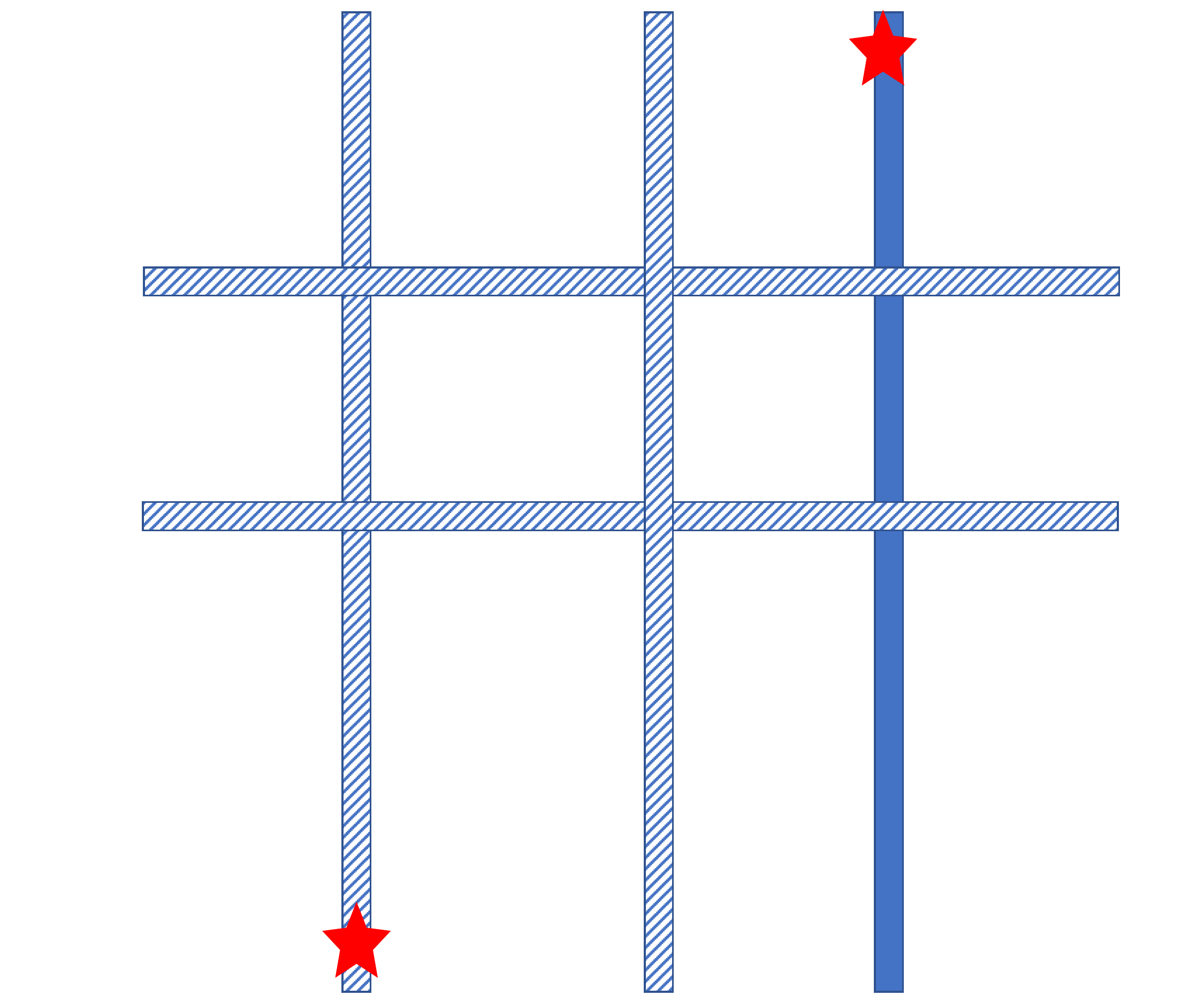}%
}

\subfloat{%
  \includegraphics[width=0.8\columnwidth,keepaspectratio]{dnx_legend_2.pdf}%
}

\caption{Subcases of tree $E_2$.}
\label{e2sub}
\end{figure}

In this appendix, we outline the proof for the probability that the distance to the nearest charging road is less than a positive real number $x$ given the event $E_2$, i.e., $\mathbb{P}(D_n < x|E_2)$. As shown in Fig.~\ref{E2}, we hereby denote subevents as $E_{2,i,j}$, in which $i$ is the level of depth of the event in the probability tree and $j$ is the index of the event at that level. Representative figures for $E_2$ are shown in Fig.~\ref{e2sub}. The distribution of $D_n$ given $E_2$ can be derived as follows:

\begin{align*}
\mathbb{P}(D_n < x | E_2)\mathbb{P}(E_2) = \sum_{i=1}^{N_2} \mathbb{P}(D_n < x | L_{2,i})\mathbb{P}(L_{2,i}),
\end{align*} 
where $N_2$ denotes the number of leaves of tree $E_2$, i.e., $N_2=6$, and $L_{2,i}$'s are successive events ending at the leaves of tree $E_2$ as shown in Fig. \ref{E2}. The definition for each event $L_{2,i}$ will be given in more details as we visit each leaf of the tree. 

%

\begin{itemize}[wide, labelwidth=!, labelindent=0pt]
    \item Event $E_{2,1,1}$: \\\textit{Description}: If there are no horizontal roads between S and D, as shown in Fig. \ref{E211}; \\Event $L_{2,1} = E_{2,1,1} \cap E_2$; \\\textit{Probability}: $\mathbb{P}(E_{2,1,1}|E_2)=e^{-\lambda d_v}$, $\mathbb{P}(L_{2,1})=\mathbb{P}(E_{2,1,1}|E_2)\mathbb{P}(E_2)$; \\\textit{Action}: we simply take the shortest path from S to D (either upper path or lower path).
    \item Event $E_{2,1,2}$: \\\textit{Description}: If there are no horizontal charging roads but at least one horizontal non-charging road between S and D, as shown in Fig. \ref{E212};\\Event $L_{2,2} = E_{2,1,2} \cap E_2$; \\\textit{Probability}: $\mathbb{P}(E_{2,1,2}|E_2)=e^{-\lambda p d_v}(1-e^{-\lambda (1-p) d_v})$, $\mathbb{P}(L_{2,2})=\mathbb{P}(E_{2,1,2}|E_2)\mathbb{P}(E_2)$; \\\textit{Action}: we take the furthest horizontal non-charging road from S.
    \item Event $E_{2,1,3}$: \\\textit{Description}: If there is at least one horizontal charging road between S and D; \\\textit{Probability}: $\mathbb{P}(E_{2,1,3}|E_2)=1-e^{-\lambda p d_v}$.
    \begin{itemize}[wide, labelwidth=!, labelindent=0pt]
        \item Event $E_{2,2,1}$: \\\textit{Description}: If there are no vertical charging roads between S and D; \\\textit{Probability}: $\mathbb{P}(E_{2,2,1}|E_{2,1,3},E_2)=e^{-\lambda p d_h}$.
        \begin{itemize}[wide, labelwidth=!, labelindent=0pt]
            \item Event $E_{2,3,1}$: \\\textit{Description}: If there exists at least one horizontal non-charging road above the furthest horizontal charging road from S, as shown in Fig. \ref{E231};
            \\Event $L_{2,3} = E_{2,3,1} \cap E_{2,2,1} \cap E_{2,1,3} \cap E_2$;
            \\\textit{Probability}: $\mathbb{P}(E_{2,3,1}|E_{2,2,1},E_{2,1,3},E_2)=1 - \frac{p - pe^{-\lambda d_v}}{1 -e^{-\lambda p d_v}}$, \\$\mathbb{P}(L_{2,3})=\mathbb{P}(E_{2,3,1}|E_{2,2,1},E_{2,1,3},E_2)\times
            \\\mathbb{P}(E_{2,2,1}|E_{2,1,3},E_2)\mathbb{P}(E_{2,1,3}|E_2)\mathbb{P}(E_2)$;
             \\\textit{Action}: we compare (i) the vertical distance between the furthest horizontal charging road and the furthest horizontal non charging road, and (ii) $d_h$, to take the longer one. 
            \item Event $E_{2,3,2}$: \\\textit{Description}: If there does not exist horizontal non-charging roads above the furthest horizontal charging road from S, as shown in Fig. \ref{E232}; 
            \\Event $L_{2,4} = E_{2,3,2} \cap E_{2,2,1} \cap E_{2,1,3} \cap E_2$;
            \\\textit{Probability}: $\mathbb{P}(E_{2,3,2}|E_{2,2,1},E_{2,1,3},E_2)=\frac{p - pe^{-\lambda d_v}}{1 -e^{-\lambda p d_v}}$,
            \\$\mathbb{P}(L_{2,4})=\mathbb{P}(E_{2,3,2}|E_{2,2,1},E_{2,1,3},E_2)\times
            \\\mathbb{P}(E_{2,2,1}|E_{2,1,3},E_2)\mathbb{P}(E_{2,1,3}|E_2)\mathbb{P}(E_2)$;
            \\\textit{Action}: we simply take the furthest horizontal charging road.
        \end{itemize}
        \item Event $E_{2,2,2}$: \\\textit{Description}: If there is at least one vertical charging road between S and D; \\\textit{Probability}: $\mathbb{P}(E_{2,2,2}|E_{2,1,3},E_2)=1- e^{-\lambda p d_h}$.
        \begin{itemize}[wide, labelwidth=!, labelindent=0pt]
            \item Event $E_{2,3,3}$: \\\textit{Description}: If there exists at least one horizontal non-charging road above the furthest horizontal charging road from S, as shown in Fig. \ref{E233};
            \\Event $L_{2,5} = E_{2,3,3} \cap E_{2,2,2} \cap E_{2,1,3} \cap E_2$;
            \\\textit{Probability}: $\mathbb{P}(E_{2,3,3}|E_{2,2,2},E_{2,1,3},E_2)=1 - \frac{p - pe^{-\lambda d_v}}{1 -e^{-\lambda p d_v}}$,
            \\$\mathbb{P}(L_{2,5})=\mathbb{P}(E_{2,3,3}|E_{2,2,2},E_{2,1,3},E_2)\times
            \\\mathbb{P}(E_{2,2,2}|E_{2,1,3},E_2)\mathbb{P}(E_{2,1,3}|E_2)\mathbb{P}(E_2)$; 
            \\\textit{Action}: we compare (i) the distance between the furthest horizontal charging road and the furthest horizontal non-charging road and (ii) the horizontal distance between the furthest vertical charging road and destination, to take the longer one. 
            \item Event $E_{2,3,4}$: \\\textit{Description}: If there does not exist horizontal non-charging roads above the furthest horizontal charging road from S, as shown in Fig. \ref{E234}; 
            \\Event $L_{2,6} = E_{2,3,4} \cap E_{2,2,2} \cap E_{2,1,3} \cap E_2$;
            \\\textit{Probability}: $\mathbb{P}(E_{2,3,4}|E_{2,2,2},E_{2,1,3},E_2)=\frac{p - pe^{-\lambda d_v}}{1 -e^{-\lambda p d_v}}$,
            \\$\mathbb{P}(L_{2,6})=\mathbb{P}(E_{2,3,4}|E_{2,2,2},E_{2,1,3},E_2)\times
            \\\mathbb{P}(E_{2,2,2}|E_{2,1,3},E_2)\mathbb{P}(E_{2,1,3}|E_2)\mathbb{P}(E_2)$; 
            \\\textit{Action}: we simply go with the furthest horizontal charging road.
        \end{itemize}
    \end{itemize}
\end{itemize}

Since the source road is already a charging road, $\mathbb{P}(D_n < x | L_{2,i}) = 1$ for all $i$.

\section{proof of Lemma \ref{DnE3}}
\label{appendxb}

In this appendix, we first provide a table of functions that are frequently used in later proofs in Table \ref{table: functions}. Next, the complete form of Lemma~\ref{DnE3}, i.e., 
the distribution of $D_n$ given $E_3$, is given as follows:
\begin{align}
\label{lemma3eqn}
&\mathbb{P}(D_n < x | E_3)\mathbb{P}(E_3) = \Psi_1 (p, \lambda, d_h, d_v, x) = \sum_{i=1}^{8} C_i,
\end{align}
where
\begin{align*}
&C_1 = 
\nonumber\\&\biggl( {\textstyle \frac{p(g_1 + g_2 + g_3)\mathbbm{1}\{x> d_v\}}{f_1(0,\infty,d_v,t+d_v,y,d_v)+f_1(0,\infty,t+d_v,\infty,t+d_v,d_v)}}
      	\nonumber\\& + 
      	 {\textstyle \frac{(1-p)(g_4+g_5+g_6)\mathbbm{1}\{x-d_h-d_v> 0\}}{f_3(0,\infty,d_v,t+d_v,y,d_v)+f_3(0,\infty,t+d_v,\infty,t+d_v,d_v)}}
      	\nonumber\\& + p \frac{f_2(d_v,\infty,0,{\rm min}(x,t-d_v),t,y+d_v)}{f_2(d_v,\infty,0,t-d_v,t,y+d_v)}
      	\nonumber\\& + p\frac{f_4(d_v,\infty,0,{\rm min}(x,t-d_v),t,y+d_v)}{f_4(d_v,\infty,0,t-d_v,t,y+d_v)}
      	\nonumber\\& + (1-p)\frac{f_2(d_v,\infty,0,{\rm min}(x-d_h,t-d_v),t,y+d_v)}{f_2(d_v,\infty,0,t-d_v,t,y+d_v)}\times \nonumber\\& \mathbbm{1}\{x-d_h> 0\}
      	\nonumber\\& + (1-p)\times \nonumber\\& \frac{f_5(0,x-d_h,t+d_v,\infty,y) + f_5(0,x-d_h,d_v,t+d_v,t+d_v)}{f_5(0,\infty,t+d_v,\infty,y) + f_5(0,\infty,d_v,t+d_v,t+d_v)} \nonumber\\& \times\mathbbm{1}\{x-d_h> 0\}\biggr)e^{-\lambda d_v}\frac{p(1-p)}{2},
\\&C2 = 
\nonumber\\& \biggl(\frac{1-e^{-\lambda (1-p) (x-d_h)}}{1-e^{-\lambda (1-p) d_v}}\mathbbm{1}\{d_h<x<d_h+ d_v \} \nonumber\\& + \mathbbm{1}\{d_h+ d_v<x \}\biggr)e^{-\lambda p d_v}\biggl(\lambda(1-p)d_v e^{-\lambda (1-p) d_v}
\\&+ e^{-\lambda p d_h}(1-e^{-\lambda (1-p) d_v}-\lambda(1-p)d_v e^{-\lambda (1-p) d_v})\biggr)\times\\&\frac{p(1-p)}{2},
\\&C_3 = 
\nonumber\\& \biggl(\biggl(\frac{\int_{{\rm max}(x-d_v,0)}^{{\rm min}(d_h,x)} F_{D_\mathrm{N-HNC}} (x-y) f_{D_\mathrm{N-VC}}(y) {\rm d}y}{F_{D_\mathrm{N-HNC}} (d_v)F_{D_\mathrm{N-VC}} (d_h)}
\nonumber\\& + \frac{F_{D_\mathrm{N-VC}}({\rm min}(d_h,x-d_v)) \mathbbm{1}\{x>d_v\}}{F_{D_\mathrm{N-VC}} (d_h)}\biggr) \times\nonumber\\& \mathbbm{1}\{x < d_h + d_v\}+ \mathbbm{1}\{x> d_h + d_v \}\biggr)(1-e^{-\lambda p d_h})e^{-\lambda p d_v}\\&\times(1-e^{-\lambda (1-p) d_v}-\lambda(1-p)d_v e^{-\lambda (1-p) d_v})\frac{p(1-p)}{2},
\\&C_4 = 
\nonumber\\&\biggl({\textstyle \frac{f_6(0,{\rm max}(x-d_h,0),d_h+y,y)+f_6({\rm max}(x-d_h,0),x,x,y)}{f_6(0,{\rm max}(d_v-d_h,0),d_h+y,y)+f_6({\rm max}(d_v-d_h,0),d_v,d_v,y)}}\times
\nonumber\\&\mathbbm{1}\{x <d_v\} + \mathbbm{1}\{x> d_v \}\biggr)F_{X_2}(d_h)\biggl(1 - \frac{p - pe^{-\lambda d_v}}{1 -e^{-\lambda p d_v}}\biggr)\times\\&e^{-\lambda p d_h}(1-e^{-\lambda p d_v})\frac{p(1-p)}{2},
\\&C_5 = 
\nonumber\\& \biggl(\frac{f_{10}(x,d_v,x-d_h) + f_{10}(d_h,{\rm min}(x,d_v),y-d_h)}{f_6(0,d_v-d_h,d_v,d_h + y)}\times \nonumber\\& \mathbbm{1}\{d_h< x < d_v\} + \mathbbm{1}\{x> d_v \}\biggr)(1-F_{X_2}(d_h))\times\\&\biggl(1 - \frac{p - pe^{-\lambda d_v}}{1 -e^{-\lambda p d_v}}\biggr)e^{-\lambda p d_h}(1-e^{-\lambda p d_v})\frac{p(1-p)}{2},
\\&C_6 = 
\nonumber\\& \biggl(\frac{f_7(x,\infty,x) + f_7(0,x,y)}{f_7(d_v,\infty,d_v) + f_7(0,d_v,y)}\mathbbm{1}\{x < d_v\} + \mathbbm{1}\{x>d_v \}\biggr)\\&\times\frac{p - pe^{-\lambda d_v}}{1 -e^{-\lambda p d_v}}(1-e^{-\lambda p d_v})\frac{p(1-p)}{2},
\\&C_7 = 
\nonumber\\& \biggl(\frac{f_8(0,x,x,d_v,x-t)+f_8(0,x,t,x,y-t)}{f_9(0,d_v,0,d_v-t,d_v,t+y)}\mathbbm{1}\{x < d_v\} \nonumber\\&+\mathbbm{1}\{x>d_v \}\biggr)\biggl(1- \int_{0}^{d_h} F_{X_2}(x) f_{D_\mathrm{N-VC}}(x) {\rm d}x\biggr)\times\\&\biggl(1 - \frac{p - pe^{-\lambda d_v}}{1 -e^{-\lambda p d_v}}\biggr)(1- e^{-\lambda p d_h})(1- e^{-\lambda p d_v})\frac{p(1-p)}{2},
\\&C_8 = 
\nonumber\\& \biggl({\textstyle \frac{f_9 (0,x,0,{\rm max}(x-t,0),y+t,y)+f_9 (0,\infty, {\rm max}(x-t,0),x,x,y)}{f_9 (0,\infty,0,{\rm max}(d_v-t,0),y+t,y)+f_9 (0,\infty,{\rm max}(d_v-t,0),d_v,x,y)}}\times
\nonumber\\& \mathbbm{1}\{x < d_v\} + \mathbbm{1}\{x>d_v \}\biggr)\int_{0}^{d_h} F_{X_2}(x) f_{D_\mathrm{N-VC}}(x) {\rm d}x\\&\times\biggl(1 - \frac{p - pe^{-\lambda d_v}}{1 -e^{-\lambda p d_v}}\biggr)(1- e^{-\lambda p d_h})(1- e^{-\lambda p d_v})\frac{p(1-p)}{2}.
\end{align*}

\begin{IEEEproof}
We prove (\ref{lemma3eqn}) by further dividing it into subevents, as shown in Fig.~\ref{E3}. We hereby denote subevents as $E_{3,i,j}$, in which $i$ is the level of depth of the event in the probability tree and $j$ is the index of the event at that level. Representative figures for $E_3$ are shown in Fig.~\ref{e3sub}. The distribution of $D_n$ given $E_3$ can be derived as follows:

\begin{align*}
\mathbb{P}(D_n < x | E_3)\mathbb{P}(E_3) = \sum_{i=1}^{N_3} \mathbb{P}(D_n < x | L_{3,i})\mathbb{P}(L_{3,i}),
\end{align*} 
where $N_3$ denotes the number of leaves of tree $E_3$, i.e., $N_3=10$, and $L_{3,i}$'s are successive events ending at the leaves of tree $E_3$ as shown in Fig. \ref{E3}. The definition for each event $L_{3,i}$ will be given in more details as we visit each leaf of the tree.


\begin{figure}[!h]
\centering
\includegraphics[width=1\columnwidth]{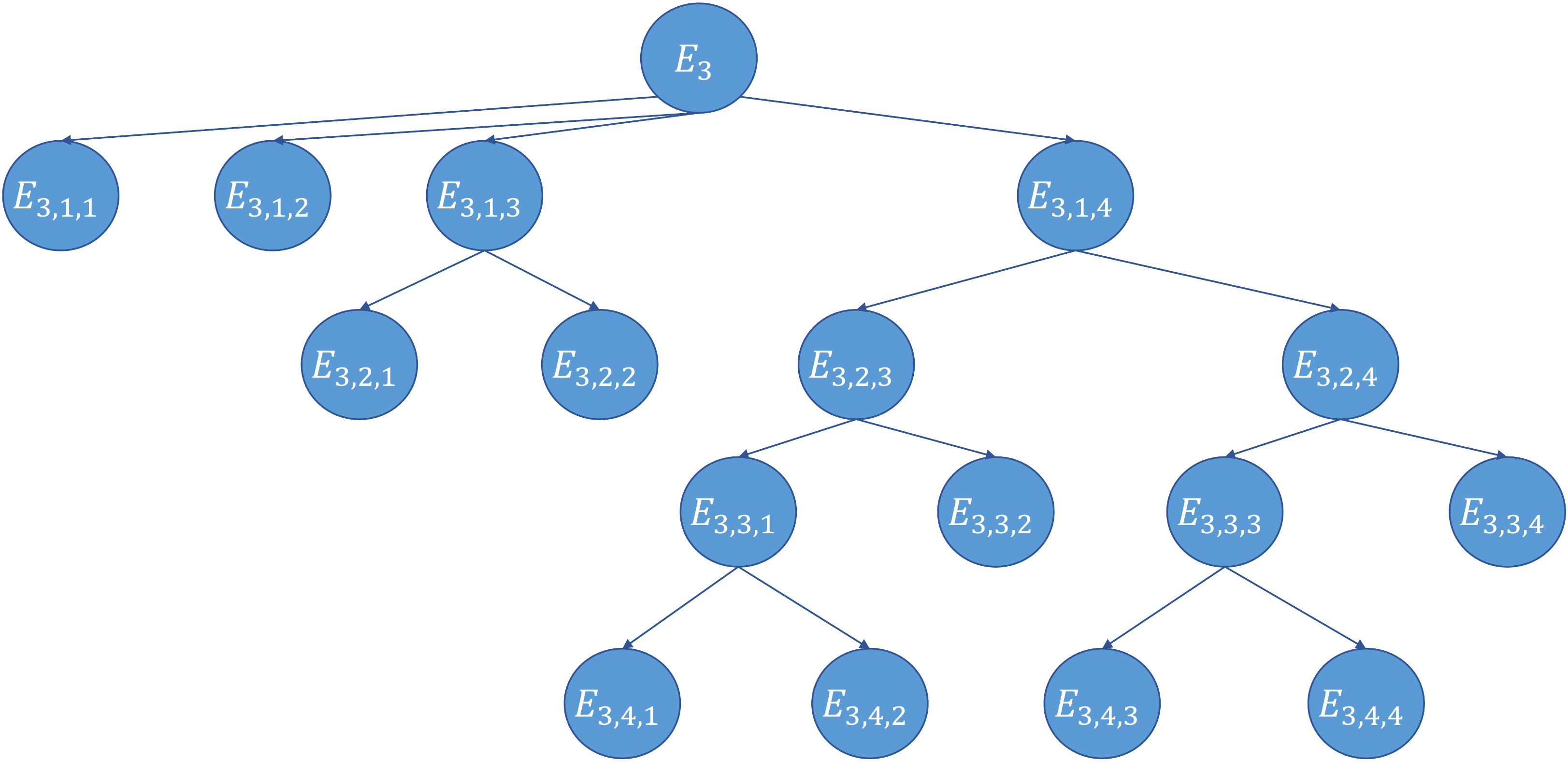}
\caption{Tree $E_3$: both source and destination roads are on two parallel roads and only the destination road is charging.}
\label{E3}
\end{figure}

\begin{figure}[h]
\centering
\captionsetup[subfigure]{font=scriptsize,labelfont=normalsize}
\subfloat[Event $E_{3,1,1}$\label{E311}]{%
  \includegraphics[width=0.33\columnwidth,keepaspectratio]{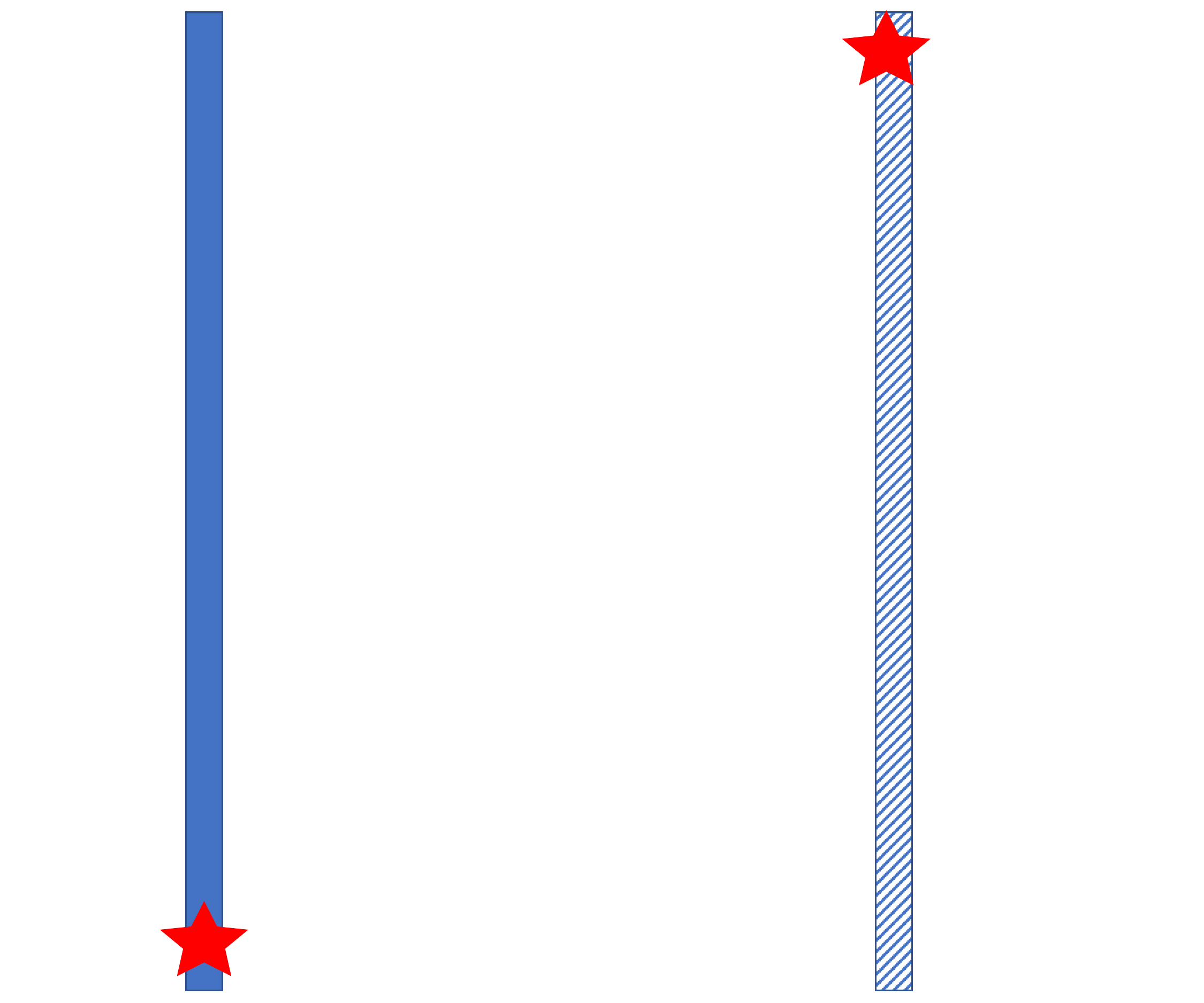}%
}\hfill
\subfloat[Event $E_{3,1,2}$\label{E312}]{%
  \includegraphics[width=0.33\columnwidth,keepaspectratio]{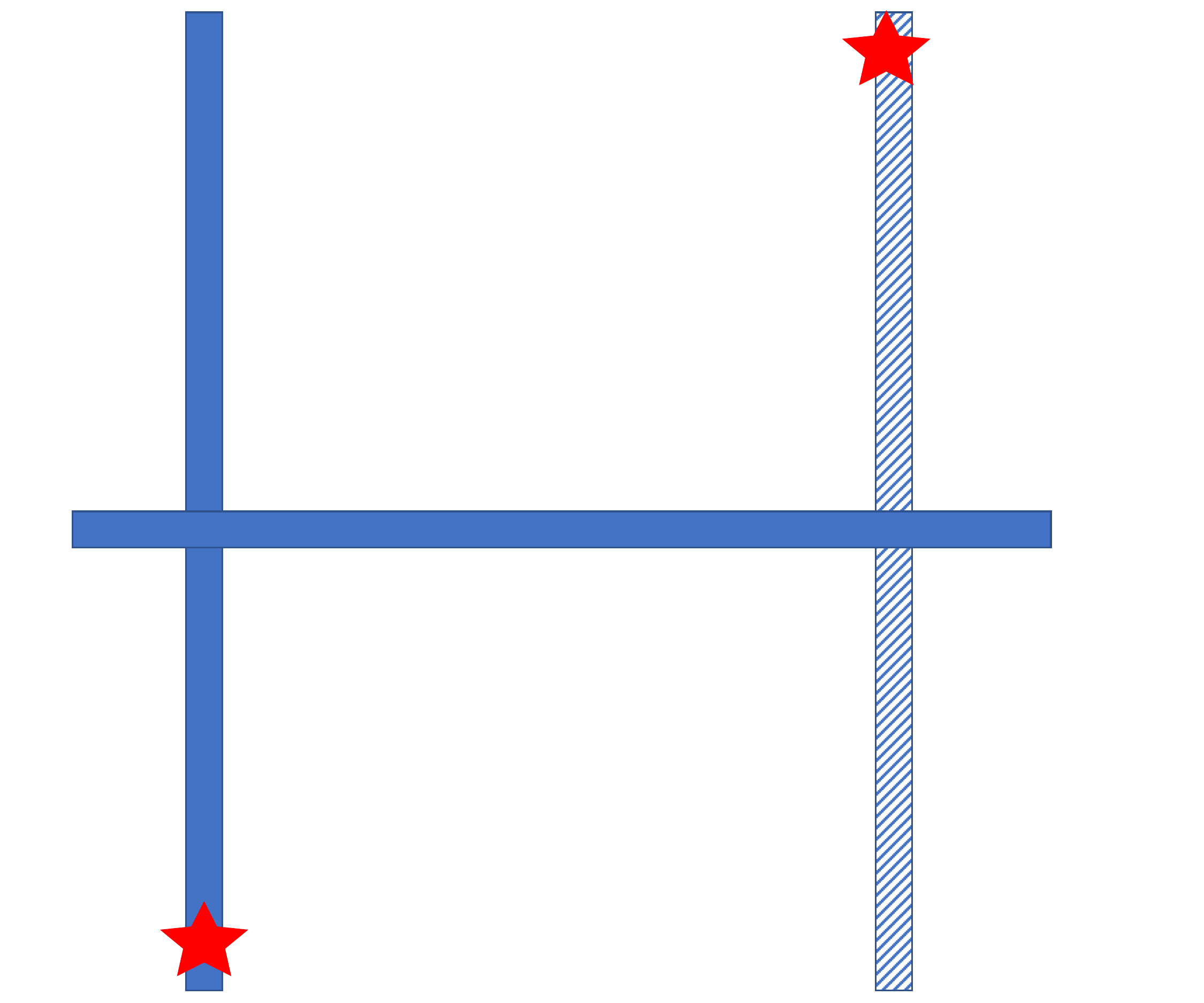}%
}
\hfill
\subfloat[Event $E_{3,2,1}$\label{E321}]{%
  \includegraphics[width=0.33\columnwidth,keepaspectratio]{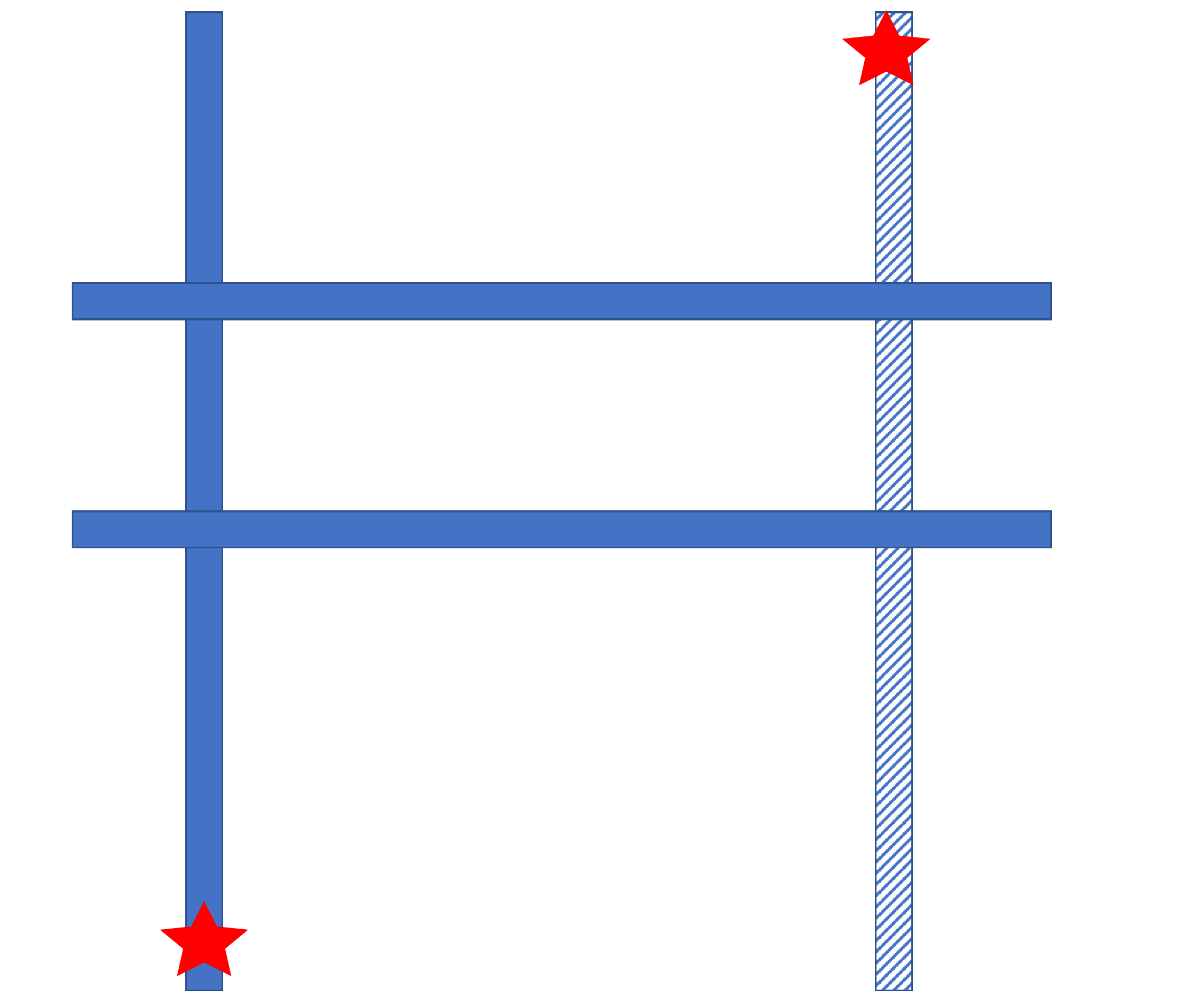}%
}

\subfloat[Event $E_{3,2,2}$\label{E322}]{%
  \includegraphics[width=0.33\columnwidth,keepaspectratio]{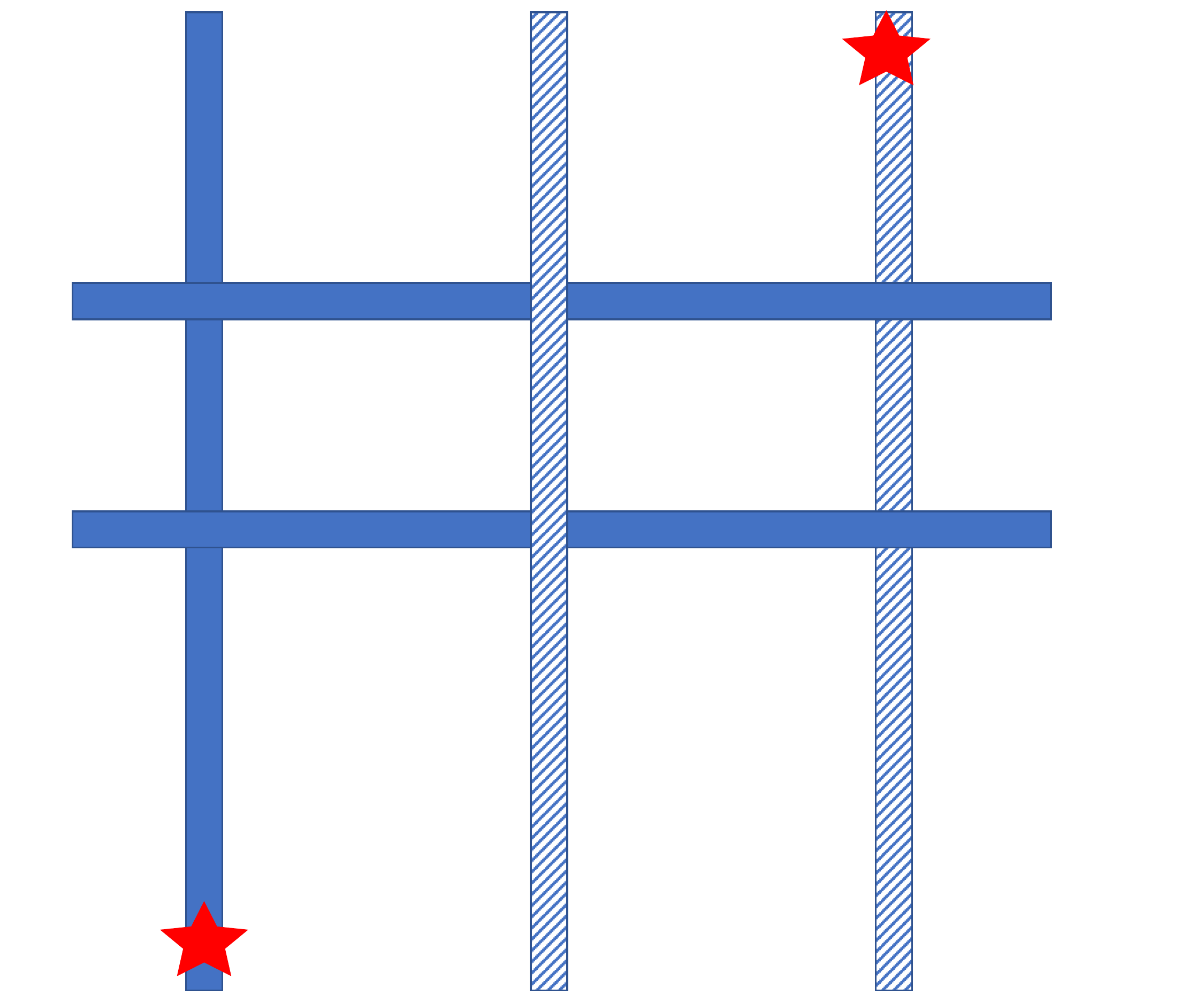}%
}\hfill
\subfloat[Event $E_{3,3,1}$\label{E331}]{%
  \includegraphics[width=0.33\columnwidth,keepaspectratio]{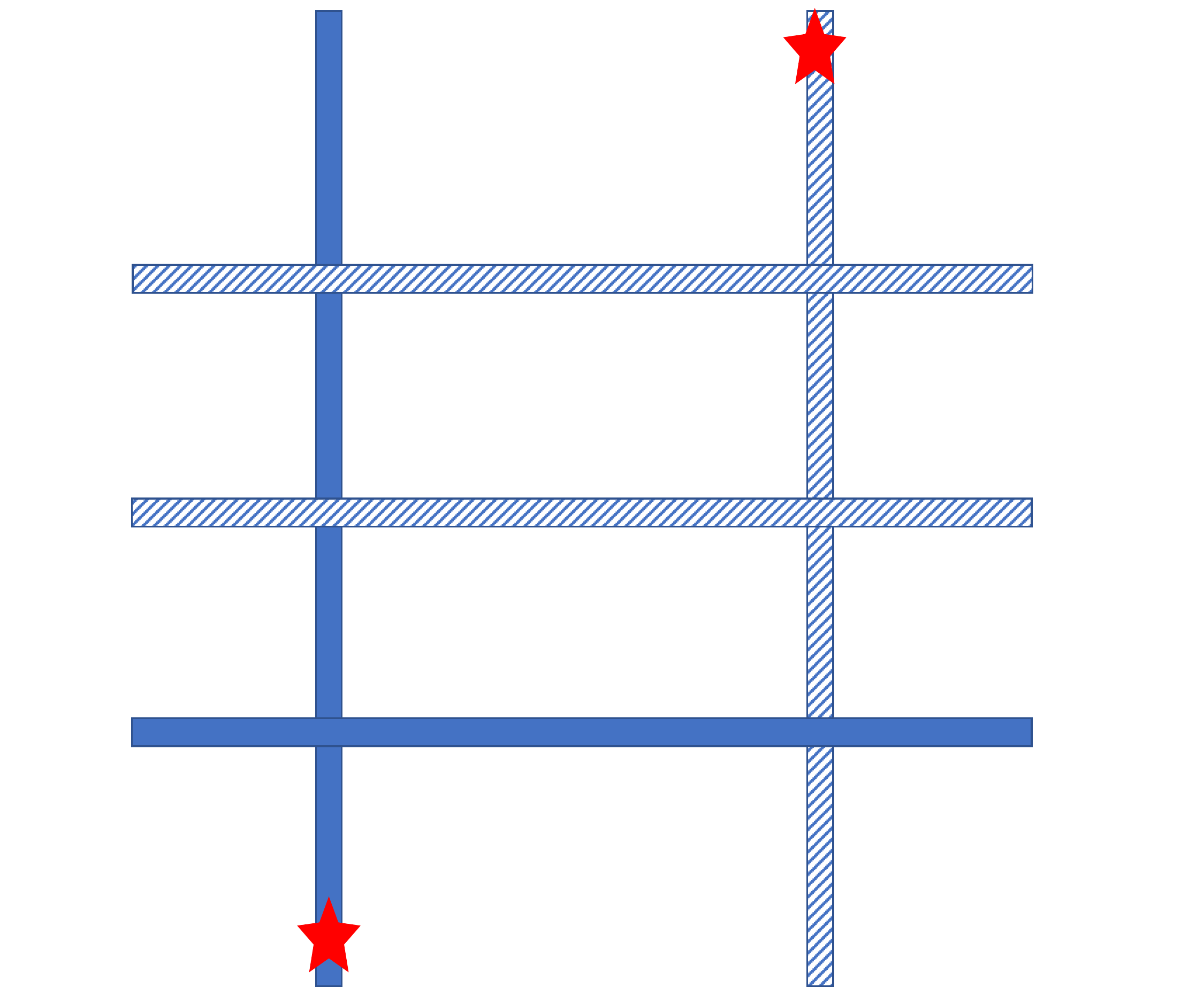}%
}
\hfill
\subfloat[Event $E_{3,3,2}$\label{E332}]{%
  \includegraphics[width=0.33\columnwidth,keepaspectratio]{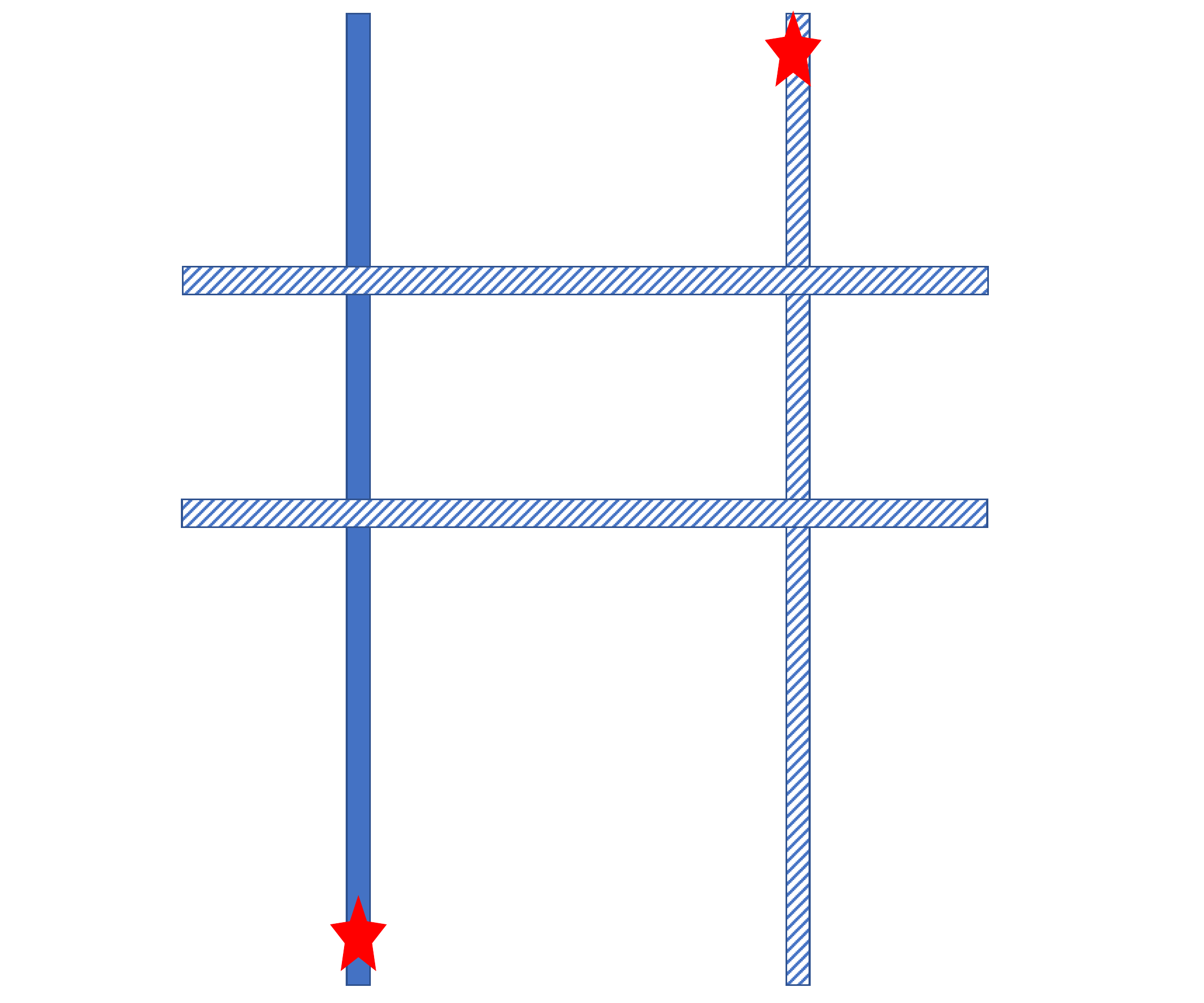}%
}

\subfloat[Event $E_{3,3,3}$\label{E333}]{%
  \includegraphics[width=0.33\columnwidth,keepaspectratio]{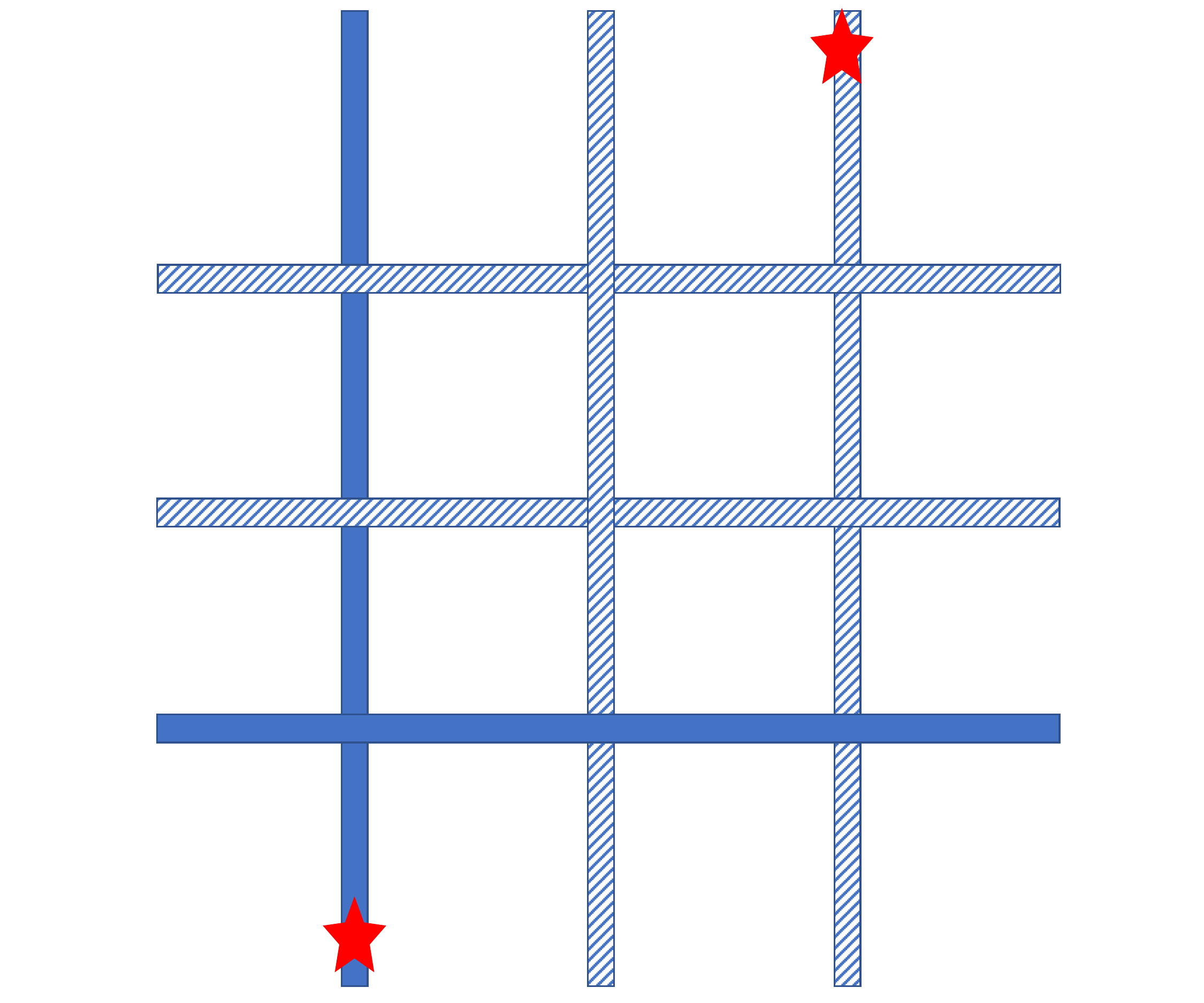}%
}\hfill
\subfloat[Event $E_{3,3,4}$\label{E334}]{%
  \includegraphics[width=0.33\columnwidth,keepaspectratio]{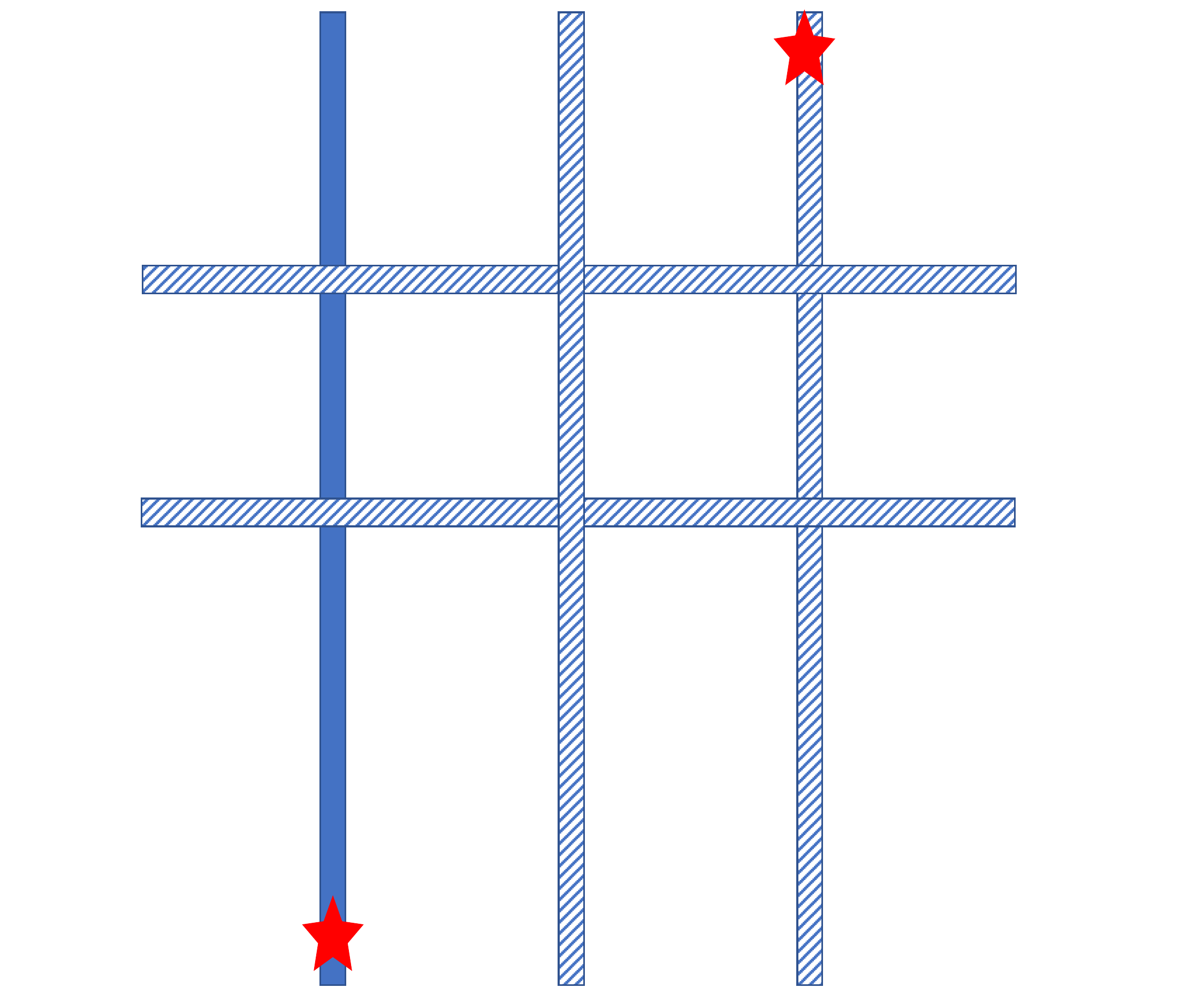}%
}\hfill
\subfloat{%
  \includegraphics[width=0.33\columnwidth,keepaspectratio]{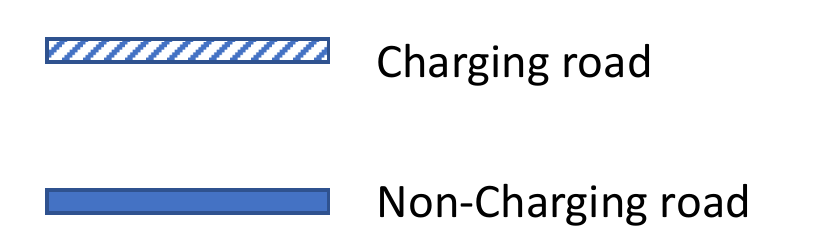}%
}

\caption{Subcases of tree $E_3$.}
\label{e3sub}
\end{figure}

\begin{itemize}[wide, labelwidth=!, labelindent=0pt]
    \item Event $E_{3,1,1}$: \\\textit{Description}: If there are no horizontal roads between S and D, as shown in Fig. \ref{E311}; 
    \\Event $L_{3,1} = E_{3,1,1} \cap E_3$;
    \\\textit{Probability}: $\mathbb{P}(E_{3,1,1}|E_3)=e^{-\lambda d_v}$,
    $\mathbb{P}(L_{3,1})=\mathbb{P}(E_{3,1,1}|E_3)\mathbb{P}(E_3)$; \\\textit{Action}: we simply take the shortest path from S to D.
    \\$\mathbb{P}(D_n < x|L_{3,1}) = \mathbb{P}(D_n < x|E_{3,1,1},E_3)$ \begin{align}
     \label{311}
        &= \mathbb{P}(D_n < x |D_\mathrm{N-HC} > d_v, D_\mathrm{N-HNC} > d_v) \nonumber
        \\&= p\mathbb{P}(D_\mathrm{N-HC} < x| D_\mathrm{N-HC} < D_\mathrm{N-HNC}, D_\mathrm{N-HC} > d_v, \nonumber\\& D_\mathrm{N-HNC} > d_v, D_\mathrm{N-HC} < d_L + d_v) \nonumber
      	\\& + (1-p) \mathbb{P}(D_\mathrm{N-HNC} + d_h < x| D_\mathrm{N-HC} > D_\mathrm{N-HNC}, \nonumber\\& D_\mathrm{N-HC} > d_v, D_\mathrm{N-HNC} > d_v, D_\mathrm{N-HNC} < d_L + d_v)
      	\nonumber\\&+ p\mathbb{P}(d_L < x| D_\mathrm{N-HC} > d_v, D_\mathrm{N-HNC} > d_v, \nonumber\\& D_\mathrm{N-HC} < D_\mathrm{N-HNC}, d_L < D_\mathrm{N-HC} - d_v)
      	\nonumber\\&+ p\mathbb{P}(d_L < x| D_\mathrm{N-HC} > d_v, D_\mathrm{N-HNC} > d_v, \nonumber\\& D_\mathrm{N-HC} > D_\mathrm{N-HNC}, d_L < D_\mathrm{N-HNC} - d_v)
      	\nonumber\\&+ (1-p)\mathbb{P}(d_L +d_h < x| D_\mathrm{N-HC} > d_v, D_\mathrm{N-HNC} > d_v, \nonumber\\& D_\mathrm{N-HC} < D_\mathrm{N-HNC}, d_L < D_\mathrm{N-HC} - d_v)
      	\nonumber\\&+ (1-p)\mathbb{P}(d_L +d_h < x| D_\mathrm{N-HC} > d_v, D_\mathrm{N-HNC} > d_v, \nonumber\\& D_\mathrm{N-HC} > D_\mathrm{N-HNC}, d_L < D_\mathrm{N-HNC} - d_v)
      	\nonumber\\& = {\textstyle \frac{p(g_1 + g_2 + g_3)}{f_1(0,\infty,d_v,t+d_v,y,d_v)+f_1(0,\infty,t+d_v,\infty,t+d_v,d_v)}} \times\nonumber\\&\mathbbm{1}\{x> d_v\}
      	\nonumber\\& + 
      	 {\textstyle \frac{(1-p)(g_4+g_5+g_6)}{f_3(0,\infty,d_v,t+d_v,y,d_v)+f_3(0,\infty,t+d_v,\infty,t+d_v,d_v)}} \times \nonumber\\&\mathbbm{1}\{x-d_h-d_v> 0\}
      	\nonumber\\& + p \frac{f_2(d_v,\infty,0,{\rm min}(x,t-d_v),t,y+d_v)}{f_2(d_v,\infty,0,t-d_v,t,y+d_v)}
      	\nonumber\\& + p\frac{f_4(d_v,\infty,0,{\rm min}(x,t-d_v),t,y+d_v)}{f_4(d_v,\infty,0,t-d_v,t,y+d_v)}
      	\nonumber\\& + (1-p)\frac{f_2(d_v,\infty,0,{\rm min}(x-d_h,t-d_v),t,y+d_v)}{f_2(d_v,\infty,0,t-d_v,t,y+d_v)}\times \nonumber\\& \mathbbm{1}\{x-d_h> 0\}
      	\nonumber\\& + (1-p)\times \nonumber\\& \frac{f_5(0,x-d_h,t+d_v,\infty,y) + f_5(0,x-d_h,d_v,t+d_v,t+d_v)}{f_5(0,\infty,t+d_v,\infty,y) + f_5(0,\infty,d_v,t+d_v,t+d_v)} \nonumber\\& \times\mathbbm{1}\{x-d_h> 0\}
    \end{align}
 
    \item Event $E_{3,1,2}$: \\\textit{Description}: If there are no horizontal charging roads but only one horizontal non-charging road between S and D, as shown in Fig. \ref{E312}; 
    \\Event $L_{3,2} = E_{3,1,2} \cap E_3$;
    \\\textit{Probability}: $\mathbb{P}(E_{3,1,2}|E_3)=e^{-\lambda p d_v} \times \\ \lambda(1-p)d_v e^{-\lambda (1-p) d_v}$, $\mathbb{P}(L_{3,2})=\mathbb{P}(E_{3,1,2}|E_3)\mathbb{P}(E_3)$; \\\textit{Action}: we take the nearest horizontal non-charging road from S. 
    \\$\mathbb{P}(D_n < x|L_{3,2}) = \mathbb{P}(D_n < x|E_{3,1,2},E_3)$ \begin{align}
    \label{312}
        &=\mathbb{P}(D_\mathrm{N-HNC} + d_h < x| D_\mathrm{N-HNC} < d_v, D_\mathrm{N-HC} > d_v) \times \nonumber\\& \mathbbm{1}\{d_h < x < d_h + d_v\} + \mathbbm{1}\{x> d_h + d_v \}
        \nonumber\\& = \frac{\mathbb{P}(D_\mathrm{N-HNC} < {\rm min}(x-d_h,d_v), D_\mathrm{N-HC} > d_v)}{\mathbb{P}(D_\mathrm{N-HNC} < d_v, D_\mathrm{N-HC}>d_v)} \times \nonumber\\& \mathbbm{1}\{d_h < x < d_h + d_v\} + \mathbbm{1}\{x> d_h + d_v \} 
        \nonumber\\& = \frac{\mathbb{P}(D_\mathrm{N-HNC} < x - d_h)\mathbbm{1}\{d_h<x< d_h + d_v \}}{\mathbb{P}(D_\mathrm{N-HNC} < d_v)\mathbb{P}(D_\mathrm{N-HC} > d_v)}
        \nonumber\\& + \frac{\mathbb{P}(D_\mathrm{N-HNC} < d_v)\mathbbm{1}\{x> d_h +  d_v \}\mathbb{P}(D_\mathrm{N-HC} > d_v)}{\mathbb{P}(D_\mathrm{N-HNC} < d_v)\mathbb{P}(D_\mathrm{N-HC} > d_v)} 
        \nonumber\\&= \frac{F_{D_\mathrm{N-HNC}} (x-d_h)\mathbbm{1}\{d_h<x<d_h + d_v \}}{F_{D_\mathrm{N-HNC}} (d_v)}
        \nonumber\\& + \frac{F_{D_\mathrm{N-HNC}} (d_v)\mathbbm{1}\{d_h + d_v<x \}}{F_{D_\mathrm{N-HNC}} (d_v)}
        \nonumber\\& = \frac{1-e^{-\lambda (1-p) (x-d_h)}}{1-e^{-\lambda (1-p) d_v}}\mathbbm{1}\{d_h<x<d_h+ d_v \} \nonumber\\& + \mathbbm{1}\{d_h+ d_v<x \}
    .
    \end{align}
    
    \item Event $E_{3,1,3}$: \\\textit{Description}: If there are no horizontal charging roads but at least two horizontal non-charging road between S and D; \\\textit{Probability}: $\mathbb{P}(E_{3,1,3}|E_3)=e^{-\lambda p d_v}(1-e^{-\lambda (1-p) d_v}-\lambda(1-p)d_v e^{-\lambda (1-p) d_v})$;
    \begin{itemize}[wide, labelwidth=!, labelindent=0pt]
    \item Event $E_{3,2,1}$: \\\textit{Description}: If there are no vertical charging roads between S and D, as shown in Fig. \ref{E321}; 
    \\Event $L_{3,3} = E_{3,2,1} \cap E_{3,1,3} \cap E_3$;
    \\\textit{Probability}: $\mathbb{P}(E_{3,2,1}|E_{3,1,3},E_3)=e^{-\lambda p d_h}$, \\$\mathbb{P}(L_{3,3})=\mathbb{P}(E_{3,2,1}|E_{3,1,3},E_3)\mathbb{P}(E_{3,1,3}|E_3)\mathbb{P}(E_3)$; \\\textit{Action}: we take the nearest horizontal non-charging road from S. 
        \begin{align}
    \label{321}
    	&\mathbb{P}(D_n < x|L_{3,3}) = \mathbb{P}(D_n < x|E_{3,2,1},E_{3,1,3},E_3) \nonumber\\&= \frac{1-e^{-\lambda (1-p) (x-d_h)}}{1-e^{-\lambda (1-p) d_v}}\times \mathbbm{1}\{d_h<x<d_h+ d_v \} \nonumber\\&+ \mathbbm{1}\{d_h+ d_v<x \}
    .
    \end{align}
    The proof of $\mathbb{P}(D_n < x|E_{3,2,1},E_{3,1,3},E_{3})$ is similar to that of $\mathbb{P}(D_n < x|E_{3,1,2},E_{3})$ given in (\ref{312}).  
    
    \item Event $E_{3,2,2}$: \\\textit{Description}: If there is at least one vertical charging road between S and D, as shown in Fig. \ref{E322}; 
    \\Event $L_{3,4} = E_{3,2,2} \cap E_{3,1,3} \cap E_3$;
    \\\textit{Probability}: $\mathbb{P}(E_{3,2,2}|E_{3,1,3},E_3)=1- e^{-\lambda p d_h}$, \\$\mathbb{P}(L_{3,4})=\mathbb{P}(E_{3,2,2}|E_{3,1,3},E_3)\mathbb{P}(E_{3,1,3}|E_3)\mathbb{P}(E_3)$; \\\textit{Action}: we take the nearest horizontal non-charging road from S, then switch to the nearest vertical charging road. 
    \\$\mathbb{P}(D_n < x|L_{3,4}) = \mathbb{P}(D_n < x|E_{3,2,2},E_{3,1,3},E_3) =$ 
    \begin{align}
    \label{322}
        &\mathbb{P}(D_\mathrm{N-HNC} + D_\mathrm{N-VC} < x| D_\mathrm{N-HNC} < d_v, D_\mathrm{N-VC}<d_h, \nonumber\\& D_\mathrm{N-HC} > d_v)\mathbbm{1}\{x < d_h + d_v\}  + \mathbbm{1}\{x> d_h + d_v \}
        \nonumber\\& = {\textstyle\frac{\mathbb{P}(D_\mathrm{N-HNC} + D_\mathrm{N-VC} < x, D_\mathrm{N-HNC} < d_v, D_\mathrm{N-VC}<d_h, D_\mathrm{N-HC} > d_v)}{\mathbb{P}(D_\mathrm{N-HNC} < d_v, D_\mathrm{N-VC}<d_h, D_\mathrm{N-HC} > d_v)}}\times
        \nonumber\\&\mathbbm{1}\{x < d_h + d_v\} + \mathbbm{1}\{x> d_h + d_v \} 
        \nonumber\\& = \biggl(\frac{\int_{{\rm max}(x-d_v,0)}^{{\rm min}(d_h,x)} F_{D_\mathrm{N-HNC}} (x-y) f_{D_\mathrm{N-VC}}(y) {\rm d}y}{F_{D_\mathrm{N-HNC}} (d_v)F_{D_\mathrm{N-VC}} (d_h)}
        \nonumber\\&+ \frac{\int_{0}^{{\rm min}(d_h,x-d_v)} F_{D_\mathrm{N-HNC}} (d_v)\mathbbm{1}\{x>d_v\} f_{D_\mathrm{N-VC}}(r) {\rm d}r}{F_{D_\mathrm{N-HNC}} (d_v)F_{D_\mathrm{N-VC}} (d_h)}\biggr)
        \times\nonumber\\& \mathbbm{1}\{x < d_h + d_v\}+ \mathbbm{1}\{x> d_h + d_v \}
        \nonumber\\& = \biggl(\frac{\int_{{\rm max}(x-d_v,0)}^{{\rm min}(d_h,x)} F_{D_\mathrm{N-HNC}} (x-y) f_{D_\mathrm{N-VC}}(y) {\rm d}y}{F_{D_\mathrm{N-HNC}} (d_v)F_{D_\mathrm{N-VC}} (d_h)}
       \nonumber\\& + \frac{F_{D_\mathrm{N-VC}}({\rm min}(d_h,x-d_v)) \mathbbm{1}\{x>d_v\}}{F_{D_\mathrm{N-VC}} (d_h)}\biggr) \times\nonumber\\& \mathbbm{1}\{x < d_h + d_v\}+ \mathbbm{1}\{x> d_h + d_v \}
    .
    \end{align}
    
	\end{itemize}     
    
    \item Event $E_{3,1,4}$: \\\textit{Description}: If there is at least one horizontal charging road between S and D; \\\textit{Probability}: $\mathbb{P}(E_{3,1,4}|E_3)=1-e^{-\lambda p d_v}$.
    \begin{itemize}[wide, labelwidth=!, labelindent=0pt]
        \item Event $E_{3,2,3}$: \\\textit{Description}: If there are no vertical charging roads between S and D; \\\textit{Probability}: $\mathbb{P}(E_{3,2,3}|E_{3,1,4},E_3)=e^{-\lambda p d_h}$.
        \begin{itemize}[wide, labelwidth=!, labelindent=0pt]
            \item Event $E_{3,3,1}$: \\\textit{Description}: If there exists at least one horizontal non-charging road below the nearest horizontal charging road from S, as shown in Fig. \ref{E331}; \\\textit{Probability}: $\mathbb{P}(E_{3,3,1}|E_{3,2,3},E_{3,1,4},E_3)=1 - \frac{p - pe^{-\lambda d_v}}{1 -e^{-\lambda p d_v}}$; \\\textit{Action}: we compare (i) the vertical distance between the nearest horizontal charging road and the nearest horizontal non charging road, and (ii) $d_h$, to take the longer one. 
            \begin{itemize}[wide, labelwidth=!, labelindent=0pt]
                \item Event $E_{3,4,1}$: \\\textit{Description}: If we decide to take the nearest horizontal charging road; 
                \\Event $L_{3,5} = E_{3,4,1} \cap E_{3,3,1} \cap E_{3,2,3} \cap E_{3,1,4} \cap E_3$;
                \\\textit{Probability}: $\mathbb{P}(E_{3,4,1}|E_{3,3,1},E_{3,2,3},E_{3,1,4},E_3)=F_{X_2}(d_h)$;
                \\$\mathbb{P}(L_{3,5})=\mathbb{P}(E_{3,4,1}|E_{3,3,1},E_{3,2,3},E_{3,1,4},E_3)\times
                \\\mathbb{P}(E_{3,3,1}|E_{3,2,3},E_{3,1,4},E_3)\times
                \\\mathbb{P}(E_{3,2,3}|E_{3,1,4},E_3)\mathbb{P}(E_{3,1,4}|E_3)\mathbb{P}(E_3)$;
                \\$\mathbb{P}(D_n < x|L_{3,5}) = \mathbb{P}(D_n < x|E_{3,4,1},E_{3,3,1},E_{3,2,3},E_{3,1,4},E_3) = $
                \begin{align}
                \label{341}
                    &  {\scriptscriptstyle \mathbb{P}(D_\mathrm{N-HC} < x| D_\mathrm{N-HC} < d_v, D_\mathrm{N-HNC} < D_\mathrm{N-HC}, D_\mathrm{N-HNC} + d_h>D_\mathrm{N-HC})} \times
                    \nonumber\\&\mathbbm{1}\{x <d_v\} + \mathbbm{1}\{x> d_v \}
                    \nonumber\\&={\textstyle \frac{f_6(0,{\rm max}(x-d_h,0),d_h+y,y)+f_6({\rm max}(x-d_h,0),x,x,y)}{f_6(0,{\rm max}(d_v-d_h,0),d_h+y,y)+f_6({\rm max}(d_v-d_h,0),d_v,d_v,y)}}\times
                    \nonumber\\&\mathbbm{1}\{x <d_v\} + \mathbbm{1}\{x> d_v \}.
                \end{align}
                
                \item Event $E_{3,4,2}$: \\\textit{Description}: If we take the nearest horizontal non-charging road; 
                \\Event $L_{3,6} = E_{3,4,2} \cap E_{3,3,1} \cap E_{3,2,3} \cap E_{3,1,4} \cap E_3$;
                \\\textit{Probability}: $\mathbb{P}(E_{3,4,2}|E_{3,3,1},E_{3,2,3},E_{3,1,4},E_3)=1-F_{X_2}(d_h)$;
                \\$\mathbb{P}(L_{3,6})=\mathbb{P}(E_{3,4,2}|E_{3,3,1},E_{3,2,3},E_{3,1,4},E_3)\times
                \\\mathbb{P}(E_{3,3,1}|E_{3,2,3},E_{3,1,4},E_3)\times\\\mathbb{P}(E_{3,2,3}|E_{3,1,4},E_3)\mathbb{P}(E_{3,1,4}|E_3)\mathbb{P}(E_3)$;
                \\$\mathbb{P}(D_n < x|L_{3,6}) = \mathbb{P}(D_n < x|E_{3,4,2},E_{3,3,1},E_{3,2,3},E_{3,1,4},E_3)$ 
                \newline If $d_h > d_v$, $\mathbb{P}(D_n < x|E_{3,4,2},E_{3,3,1},E_{3,2,3},E_{3,1,4},E_{3}) = 0$.
      \\If $d_h < d_v$,
    \begin{align}
    \label{342}
                    &\mathbb{P}(D_n < x|E_{3,4,2},E_{3,3,1},E_{3,2,3},E_{3,1,4},E_{3}) 
                    \nonumber\\&= \mathbb{P}(D_\mathrm{N-HNC} + d_h < x| D_\mathrm{N-HC} < d_v, D_\mathrm{N-HNC} < D_\mathrm{N-HC}, \nonumber\\& d_h<D_\mathrm{N-HC}-D_\mathrm{N-HNC}) \mathbbm{1}\{d_h< x <d_v\} + \mathbbm{1}\{x> d_v \}=
                    \nonumber\\&{\scriptstyle \frac{\mathbb{P}(D_\mathrm{N-HNC} + d_h < x, D_\mathrm{N-HC} < d_v, D_\mathrm{N-HNC} < D_\mathrm{N-HC}, d_h<D_\mathrm{N-HC}-D_\mathrm{N-HNC})}{\mathbb{P}( D_\mathrm{N-HC} < d_v, D_\mathrm{N-HNC} < D_\mathrm{N-HC}, d_h<D_\mathrm{N-HC}-D_\mathrm{N-HNC})}}\nonumber\\& \times\mathbbm{1}\{d_h< x < d_v\} + \mathbbm{1}\{x>d_v \}
                    \nonumber\\&= \frac{\mathbb{P}(D_\mathrm{N-HNC} + d_h < {\rm min}(x, D_\mathrm{N-HC}), D_\mathrm{N-HC} < d_v)}{\mathbb{P}( D_\mathrm{N-HNC} +d_h < D_\mathrm{N-HC}<d_v)} \times \nonumber\\& \mathbbm{1}\{d_h< x < d_v\} + \mathbbm{1}\{x> d_v \}
                    \nonumber\\& = \frac{f_{10}(x,d_v,x-d_h) + f_{10}(d_h,{\rm min}(x,d_v),y-d_h)}{f_6(0,d_v-d_h,d_v,d_h + y)}\times \nonumber\\& \mathbbm{1}\{d_h< x < d_v\} + \mathbbm{1}\{x> d_v \}
                .
                \end{align}
                
            \end{itemize}
            \item Event $E_{3,3,2}$: \\\textit{Description}: If there does not exist horizontal non-charging roads below the nearest horizontal charging road from S, as shown in Fig. \ref{E332}; 
            \\Event $L_{3,7} = E_{3,3,2} \cap E_{3,2,3} \cap E_{3,1,4} \cap E_3$;
            \\\textit{Probability}: $\mathbb{P}(E_{3,3,2}|E_{3,2,3},E_{3,1,4},E_3)=\frac{p - pe^{-\lambda d_v}}{1 -e^{-\lambda p d_v}}$,
            \\$\mathbb{P}(L_{3,7})=\mathbb{P}(E_{3,3,2}|E_{3,2,3},E_{3,1,4},E_3)\times
            \\\mathbb{P}(E_{3,2,3}|E_{3,1,4},E_3)\mathbb{P}(E_{3,1,4}|E_3)\mathbb{P}(E_3)$;
            \\\textit{Action}: we simply take the nearest horizontal charging road.
            \begin{align}
            \label{332}
                &\mathbb{P}(D_n < x|L_{3,7}) = \mathbb{P}(D_n < x|E_{3,3,2},E_{3,2,3},E_{3,1,4},E_3)
                \nonumber\\&= \mathbb{P}(D_\mathrm{N-HC} < x|D_\mathrm{N-HC}<D_\mathrm{N-HNC}, D_\mathrm{N-HC} < d_v)
                \nonumber\\&= \frac{\mathbb{P}(D_\mathrm{N-HC} < {\rm min}(x,D_\mathrm{N-HNC},d_v))}{\mathbb{P}(D_\mathrm{N-HC} < {\rm min}(D_\mathrm{N-HNC},d_v))}\mathbbm{1}\{x < d_v\} \nonumber\\&+ \mathbbm{1}\{x>d_v \}
				\nonumber\\& = \frac{f_7(x,\infty,x) + f_7(0,x,y)}{f_7(d_v,\infty,d_v) + f_7(0,d_v,y)}\mathbbm{1}\{x < d_v\} + \mathbbm{1}\{x>d_v \}
            .
            \end{align}
            
        \end{itemize}
        \item Event $E_{3,2,4}$: \\\textit{Description}: If there is at least one vertical charging road between S and D; \\\textit{Probability}: $\mathbb{P}(E_{3,2,4}|E_{3,1,4},E_3)=1- e^{-\lambda p d_h}$.
        \begin{itemize}[wide, labelwidth=!, labelindent=0pt]
            \item Event $E_{3,3,3}$: \\\textit{Description}: If there exists at least one horizontal non-charging road below the nearest horizontal charging road from S, as shown in Fig. \ref{E333}; \\\textit{Probability}: $\mathbb{P}(E_{3,3,3}|E_{3,2,4},E_{3,1,4},E_3)=1 - \frac{p - pe^{-\lambda d_v}}{1 -e^{-\lambda p d_v}}$; \\\textit{Action}: we compare (i) the distance between the nearest horizontal charging road and the nearest horizontal non-charging road, and (ii) the horizontal distance between the nearest vertical charging road and source, to take the longer one. 
            \begin{itemize}[wide, labelwidth=!, labelindent=0pt]
                \item Event $E_{3,4,3}$: \\\textit{Description}: If we take the nearest vertical charging road; 
                \\Event $L_{3,8} = E_{3,4,3} \cap E_{3,3,3} \cap E_{3,2,4} \cap E_{3,1,4} \cap E_3$;
                \\\textit{Probability}: $\mathbb{P}(E_{3,4,3}|E_{3,3,3},E_{3,2,4},E_{3,1,4},E_3)=1- \int_{0}^{d_h} F_{X_2}(x) f_{D_\mathrm{N-VC}}(x) {\rm d}x$;
                \\$\mathbb{P}(L_{3,8})=\mathbb{P}(E_{3,4,3}|E_{3,3,3},E_{3,2,4},E_{3,1,4},E_3)\times \\\mathbb{P}(E_{3,3,3}|E_{3,2,4},E_{3,1,4},E_3)\times\\\mathbb{P}(E_{3,2,4}|E_{3,1,4},E_3)\mathbb{P}(E_{3,1,4}|E_3)\mathbb{P}(E_3)$;
                \begin{align}
    \label{343}
                &\mathbb{P}(D_n < x|L_{3,8}) \nonumber\\&= \mathbb{P}(D_n < x|E_{3,4,3},E_{3,3,3},E_{3,2,4},E_{3,1,4},E_3)
                \nonumber\\&= \mathbb{P}(D_\mathrm{N-HNC} + D_\mathrm{N-VC} < x|D_\mathrm{N-HC} < d_v, \nonumber\\& D_\mathrm{N-HNC} < D_\mathrm{N-HC}, D_\mathrm{N-HC} > D_\mathrm{N-VC} + D_\mathrm{N-HNC})\times\nonumber\\&\mathbbm{1}\{x < d_v\} + \mathbbm{1}\{x >  d_v\}
                \nonumber\\&=\frac{\mathbb{P}(D_\mathrm{N-HNC} + D_\mathrm{N-VC} < {\rm min}(x,D_\mathrm{N-HC}),D_\mathrm{N-HC} < d_v)}{\mathbb{P}(D_\mathrm{N-HNC} + D_\mathrm{N-VC} < D_\mathrm{N-HC} < d_v)} \nonumber\\& \times\mathbbm{1}\{x < d_v\} + \mathbbm{1}\{x>d_v \}
				\nonumber\\& = \frac{f_8(0,x,x,d_v,x-t)+f_8(0,x,t,x,y-t)}{f_9(0,d_v,0,d_v-t,d_v,t+y)}\mathbbm{1}\{x < d_v\} + \nonumber\\& \mathbbm{1}\{x>d_v \}
                .
                \end{align}
                
                \item Event $E_{3,4,4}$: \\\textit{Description}: If we take the nearest horizontal charging road; \\Event $L_{3,9} = E_{3,4,4} \cap E_{3,3,3} \cap E_{3,2,4} \cap E_{3,1,4} \cap E_3$;
                \\\textit{Probability}: $\mathbb{P}(E_{3,4,4}|E_{3,3,3},E_{3,2,4},E_{3,1,4},E_3)=\int_{0}^{d_h} F_{X_2}(x) f_{D_\mathrm{N-VC}}(x) {\rm d}x$;
                \\$\mathbb{P}(L_{3,9})=\mathbb{P}(E_{3,4,4}|E_{3,3,3},E_{3,2,4},E_{3,1,4},E_3)\times
                \\\mathbb{P}(E_{3,3,3}|E_{3,2,4},E_{3,1,4},E_3)\times\\\mathbb{P}(E_{3,2,4}|E_{3,1,4},E_3)\mathbb{P}(E_{3,1,4}|E_3)\mathbb{P}(E_3)$;
                \begin{align}
                \label{344}
                    &\mathbb{P}(D_n < x|L_{3,9}) = \mathbb{P}(D_n < x|E_{3,4,4},E_{3,3,3},E_{3,2,4},E_{3,1,4},E_3) 
                    \nonumber\\&= \mathbb{P}(D_\mathrm{N-HC} < x|D_\mathrm{N-HC} < d_v, D_\mathrm{N-HNC} < D_\mathrm{N-HC}, \nonumber\\& D_\mathrm{N-HC} - D_\mathrm{N-HNC} < D_\mathrm{N-VC}) 
                    \nonumber\\&= \frac{\mathbb{P}(D_\mathrm{N-HNC} < D_\mathrm{N-HC} < {\rm min}(x,d_v, D_\mathrm{N-HNC} + D_\mathrm{N-VC}))}{\mathbb{P}(D_\mathrm{N-HNC} < D_\mathrm{N-HC} < {\rm min}(d_v, D_\mathrm{N-HNC} + D_\mathrm{N-VC}))} \nonumber\\&\times\mathbbm{1}\{x < d_v\} + \mathbbm{1}\{x>d_v \}
					\nonumber\\& ={\textstyle \frac{f_9 (0,x,0,{\rm max}(x-t,0),y+t,y)+f_9 (0,\infty, {\rm max}(x-t,0),x,x,y)}{f_9 (0,\infty,0,{\rm max}(d_v-t,0),y+t,y)+f_9 (0,\infty,{\rm max}(d_v-t,0),d_v,x,y)}}\times
					\nonumber\\& \mathbbm{1}\{x < d_v\} + \mathbbm{1}\{x>d_v \}
                .
                \end{align}
                
            \end{itemize}
            \item Event $E_{3,3,4}$: \\\textit{Description}: If there does not exist horizontal non-charging roads below the nearest horizontal charging road from S, as shown in Fig. \ref{E334}; 
            \\Event $L_{3,10} = E_{3,3,4} \cap E_{3,2,4} \cap E_{3,1,4} \cap E_3$;
            \\\textit{Probability}: $\mathbb{P}(E_{3,3,4}|E_{3,2,4},E_{3,1,4},E_3)=\frac{p - pe^{-\lambda d_v}}{1 -e^{-\lambda p d_v}}$,
            \\$\mathbb{P}(L_{3,10})=\mathbb{P}(E_{3,3,4}|E_{3,2,4},E_{3,1,4},E_3)\times
            \\\mathbb{P}(E_{3,2,4}|E_{3,1,4},E_3)\mathbb{P}(E_{3,1,4}|E_3)\mathbb{P}(E_3)$;
            \\\textit{Action}: we simply take the nearest horizontal charging road. 
            \begin{align}
            \label{334}
                    &\mathbb{P}(D_n < x|L_{3,10}) = \mathbb{P}(D_n < x|E_{3,3,4},E_{3,2,4},E_{3,1,4},E_3) 
                    \nonumber\\&= \mathbb{P}(D_\mathrm{N-HC} < x|D_\mathrm{N-HC} < D_\mathrm{N-HNC}, D_\mathrm{N-HC} < d_v) 
                    \nonumber\\&= \frac{\mathbb{P}(D_\mathrm{N-HC} <{\rm min}(x, D_\mathrm{N-HNC})}{\mathbb{P}(D_\mathrm{N-HC}<{\rm min}(D_\mathrm{N-HNC}, d_v))} \mathbbm{1}\{x < d_v\} + \mathbbm{1}\{x>d_v \}
					\nonumber\\& = \frac{f_7(0,x,y) + f_7(x,\infty,x)}{f_7(0,d_v,y)+f_7(d_v,\infty,d_v)}\mathbbm{1}\{x < d_v\} + \mathbbm{1}\{x>d_v \}
                .
            \end{align}
        \end{itemize}
    \end{itemize}
\end{itemize}
\end{IEEEproof}
           
\section{proof of Lemma \ref{DnE4}}
\label{appendxc}
In this appendix, we first provide the complete form of Lemma~\ref{DnE4}, i.e., 
the distribution of $D_n$ given $E_4$, as follows:
\begin{align}
\label{lemma4eqn}
&\mathbb{P}(D_n < x | E_4)\mathbb{P}(E_4) = \sum_{i=1}^{6} C_i,
\end{align}
where
\begin{align*}
&C_1 = 
\nonumber\\& \biggl({\textstyle \frac{p(g_1 + g_2 + g_3)\mathbbm{1}\{x> d_v\}}{f_1(0,\infty,d_v,t+d_v,y,d_v)+f_1(0,\infty,t+d_v,\infty,t+d_v,d_v)}}
\nonumber\\& + p \frac{f_2(d_v,\infty,0,{\rm min}(x,t-d_v),t,y+d_v)}{f_2(d_v,\infty,0,t-d_v,t,y+d_v)}
\nonumber\\&+ p\frac{f_4(d_v,\infty,0,{\rm min}(x,t-d_v),t,y+d_v)}{f_4(d_v,\infty,0,t-d_v,t,y+d_v)}\biggr)e^{-\lambda d_v}\frac{(1-p)^2}{2},
\\&C_2 = 
\nonumber\\&\biggl(\frac{F_{D_\mathrm{N-HC}} (x)}{F_{D_\mathrm{N-HC}} (d_v)}\mathbbm{1}\{x < d_v\} + \mathbbm{1}\{x > d_v\}\biggr)\biggl(\lambda p d_v e^{-\lambda p d_v}\times\\&e^{-\lambda (1-p) d_v}+e^{-\lambda p d_h}\lambda p d_v e^{-\lambda p d_v}(1-e^{-\lambda (1-p) d_v})+\\&e^{-\lambda p d_h}(1-e^{-\lambda p d_v} - \lambda p d_v e^{-\lambda p d_v})\biggr)\frac{(1-p)^2}{2},
\\&C_3 = 
\nonumber\\& \biggl(\biggl(\frac{\int_{{\rm max}(x-d_v,0)}^{{\rm min}(d_h,x)} F_{D_\mathrm{N-HNC}} (x-y) f_{D_\mathrm{N-VC}}(y) {\rm d}y}{F_{D_\mathrm{N-HNC}} (d_v)F_{D_\mathrm{N-VC}} (d_h)}
\nonumber\\& + \frac{F_{D_\mathrm{N-VC}}({\rm min}(d_h,x-d_v)) \mathbbm{1}\{x>d_v\}}{F_{D_\mathrm{N-VC}} (d_h)}\biggr) \times 
\nonumber\\& \mathbbm{1}\{x < d_h + d_v\}+ \mathbbm{1}\{x> d_h + d_v \}\biggr)(1- e^{-\lambda p d_h})e^{-\lambda p d_v}\times\\&(1-e^{-\lambda (1-p) d_v} - \lambda (1-p) d_v e^{- \lambda (1-p) d_v})\frac{(1-p)^2}{2},
\\&C_4 = 
\nonumber\\& \biggl(\frac{f_8(0,x,x,d_v,x-t)+f_8(0,x,t,x,y-t)}{f_9(0,d_v,0,d_v-t,d_v,t+y)}\mathbbm{1}\{x < d_v\} \nonumber\\&+ \mathbbm{1}\{x>d_v \}\biggr)\biggl(1- \int_{0}^{d_h} F_{X_2}(x) f_{D_\mathrm{N-VC}}(x) {\rm d}x\biggr)\times\\&\biggl(1 - \frac{p - pe^{-\lambda d_v}}{1 -e^{-\lambda p d_v}}\biggr)(1-e^{-\lambda p d_h})\biggl(\lambda p d_v e^{-\lambda p d_v}\times\\&(1-e^{-\lambda (1-p) d_v})+(1-e^{-\lambda p d_v} - \lambda p d_v e^{-\lambda p d_v})\biggr)\frac{(1-p)^2}{2},
\\&C_5 = 
\nonumber\\& \biggl({\textstyle\frac{f_9 (0,x,0,{\rm max}(x-t,0),y+t,y)+f_9 (0,\infty, {\rm max}(x-t,0),x,x,y)}{f_9 (0,\infty,0,{\rm max}(d_v-t,0),y+t,y)+f_9 (0,\infty,{\rm max}(d_v-t,0),d_v,x,y)}}\times
\nonumber\\&\mathbbm{1}\{x < d_v\} + \mathbbm{1}\{x>d_v \}\biggr)\biggl(\int_{0}^{d_h} F_{X_2}(x) f_{D_\mathrm{N-VC}}(x) {\rm d}x\biggr)\\&\times\biggl(1 - \frac{p - pe^{-\lambda d_v}}{1 -e^{-\lambda p d_v}}\biggr)(1-e^{-\lambda p d_h})\biggl(\lambda p d_v e^{-\lambda p d_v}\times\\&(1-e^{-\lambda (1-p) d_v})+(1-e^{-\lambda p d_v} - \lambda p d_v e^{-\lambda p d_v})\biggr)\frac{(1-p)^2}{2},
\\&C_6 = 
\nonumber\\&\biggl(\frac{f_7(0,x,y) + f_7(x,\infty,x)}{f_7(0,d_v,y)+f_7(d_v,\infty,d_v)}\mathbbm{1}\{x < d_v\} + \mathbbm{1}\{x>d_v \} \biggr)\\&\times\frac{p - pe^{-\lambda d_v}}{1 -e^{-\lambda p d_v}}(1-e^{-\lambda p d_h})\biggl(\lambda p d_v e^{-\lambda p d_v}(1-e^{-\lambda (1-p) d_v})\\&+(1-e^{-\lambda p d_v} - \lambda p d_v e^{-\lambda p d_v})\biggr)\frac{(1-p)^2}{2}.
\end{align*}

\begin{IEEEproof}
We prove (\ref{lemma4eqn}) by further dividing it into subevents, as shown in Fig.~\ref{E4}, we hereby denote subevents as $E_{4,i,j}$, in which $i$ is the level of depth of the event in the probability tree and $j$ is the index of the event at that level. Representative figures for $E_4$ are shown in Fig.~\ref{e4sub}. The distribution of $D_n$ given $E_4$ can be derived as follows:

\begin{align*}
\mathbb{P}(D_n < x | E_4)\mathbb{P}(E_4) = \sum_{i=1}^{N_4} \mathbb{P}(D_n < x | L_{4,i})\mathbb{P}(L_{4,i}),
\end{align*} 
where $N_4$ denotes the number of leaves of tree $E_4$, i.e., $N_4=13$, and $L_{4,i}$'s are successive events ending at the leaves of tree $E_4$ as shown in Fig. \ref{E4}. The definition for each event $L_{4,i}$ will be given in more details as we visit each leaf of the tree. 


\begin{figure}[!h]
\centering
\includegraphics[width=1\columnwidth]{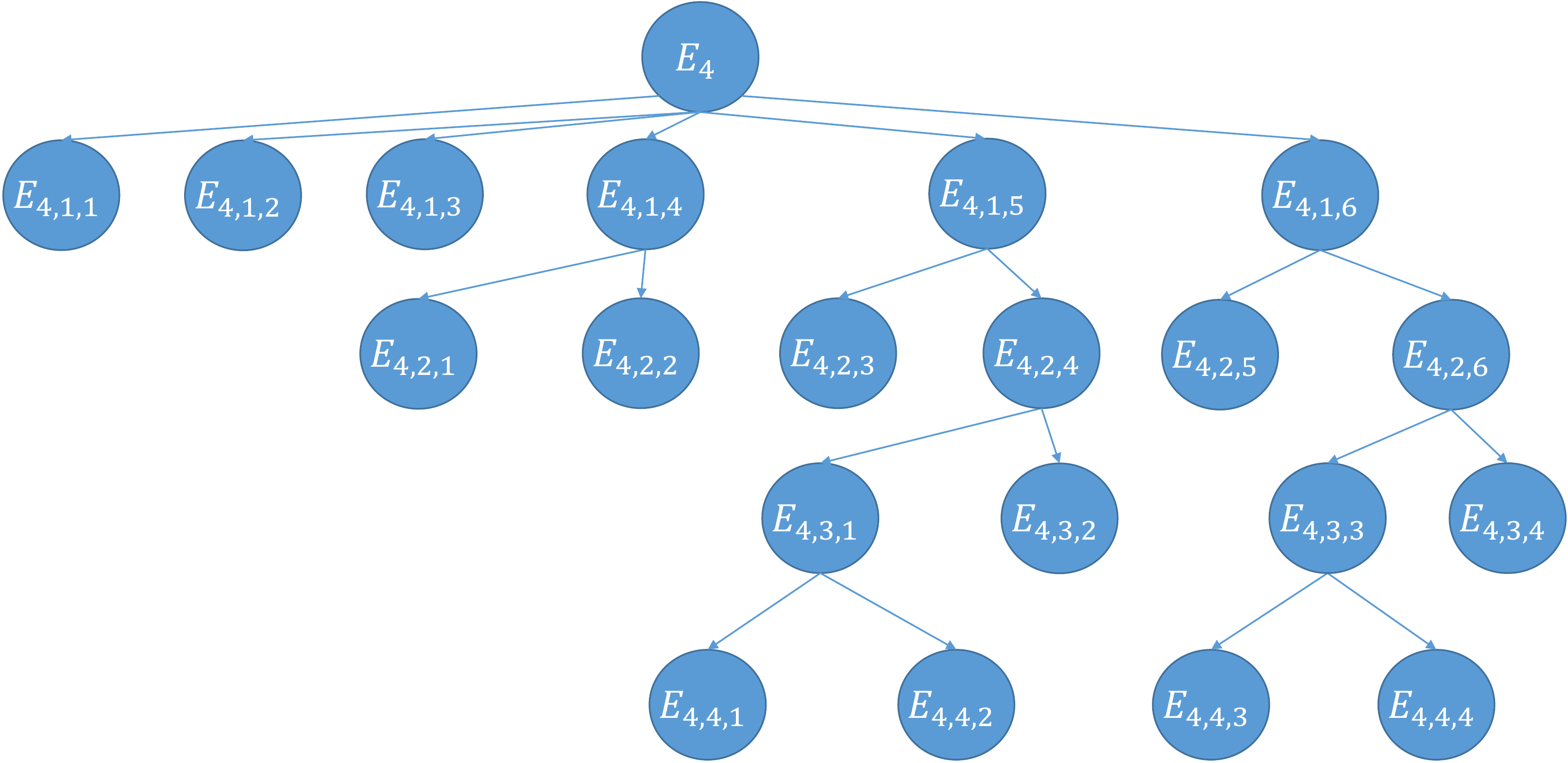}
\caption{Tree $E_4$: both source and destination roads are on two parallel roads and are not charging.}
\label{E4}
\end{figure}

\begin{figure}[t]
\centering
\captionsetup[subfigure]{font=scriptsize,labelfont=normalsize}
\subfloat[Event $E_{4,1,1}$\label{E411}]{%
  \includegraphics[width=0.33\columnwidth,keepaspectratio]{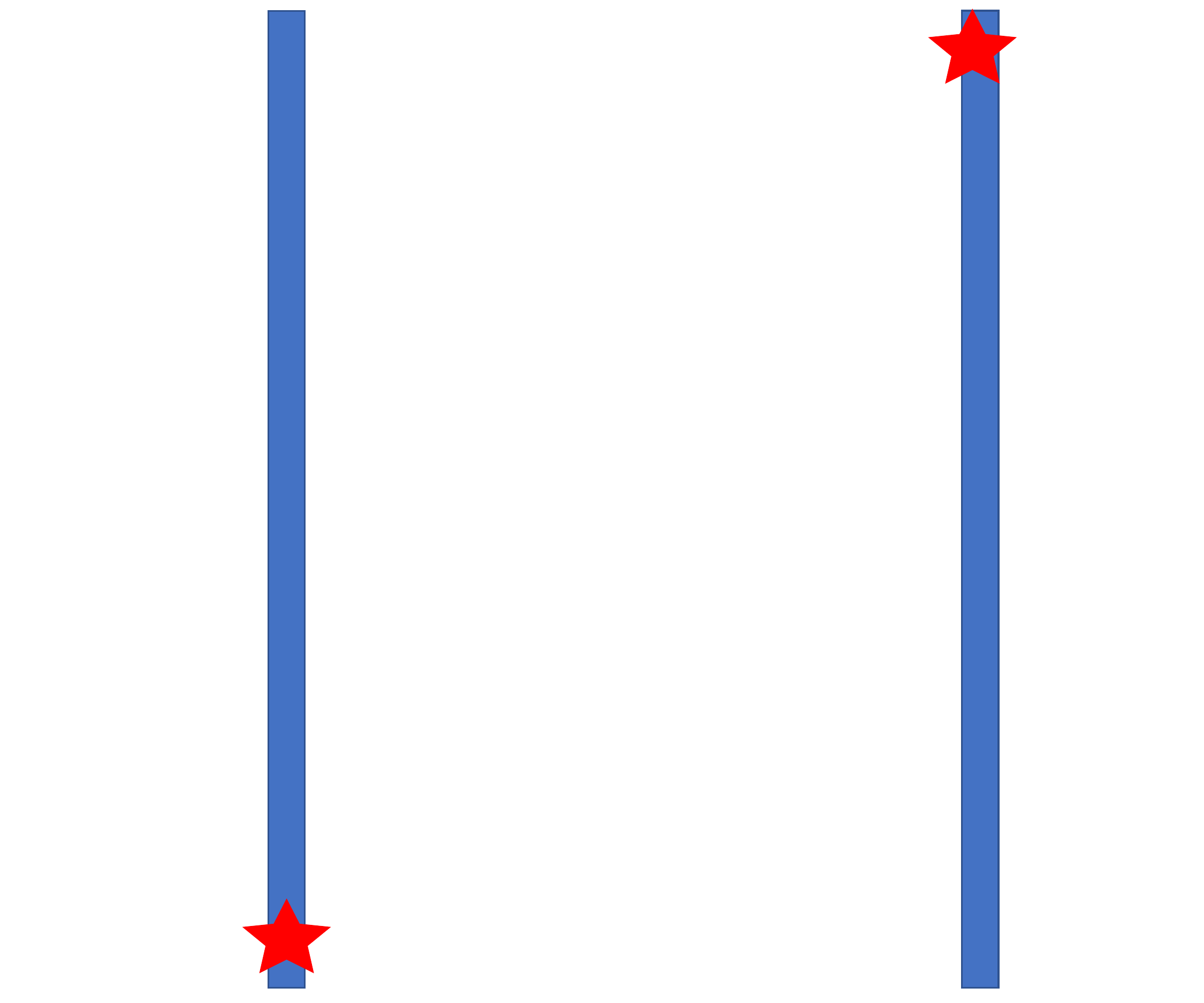}%
}\hfill
\subfloat[Event $E_{4,1,2}$\label{E412}]{%
  \includegraphics[width=0.33\columnwidth,keepaspectratio]{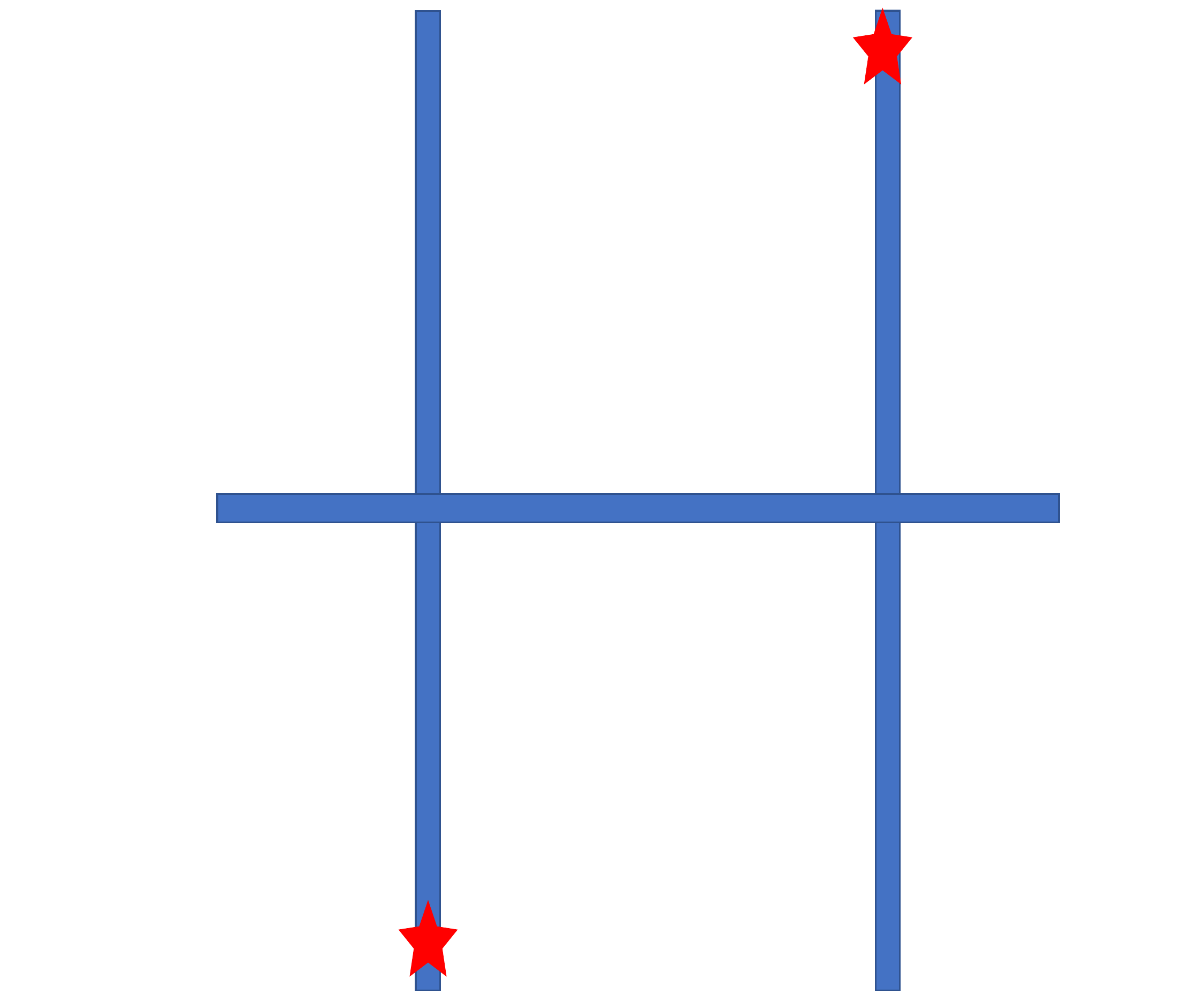}%
}
\hfill
\subfloat[Event $E_{4,1,3}$\label{E413}]{%
  \includegraphics[width=0.33\columnwidth,keepaspectratio]{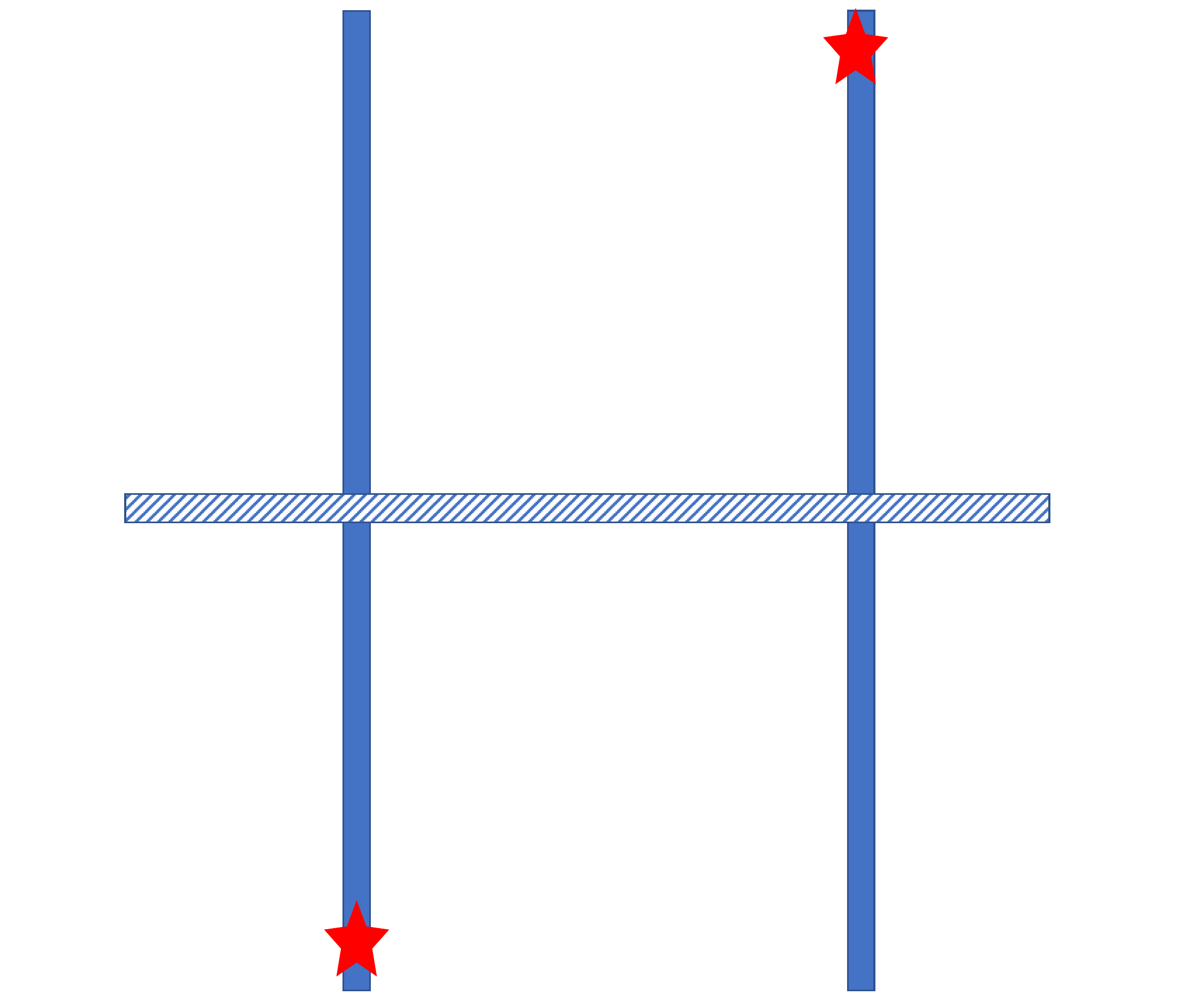}%
}

\subfloat[Event $E_{4,2,1}$\label{E421}]{%
  \includegraphics[width=0.33\columnwidth,keepaspectratio]{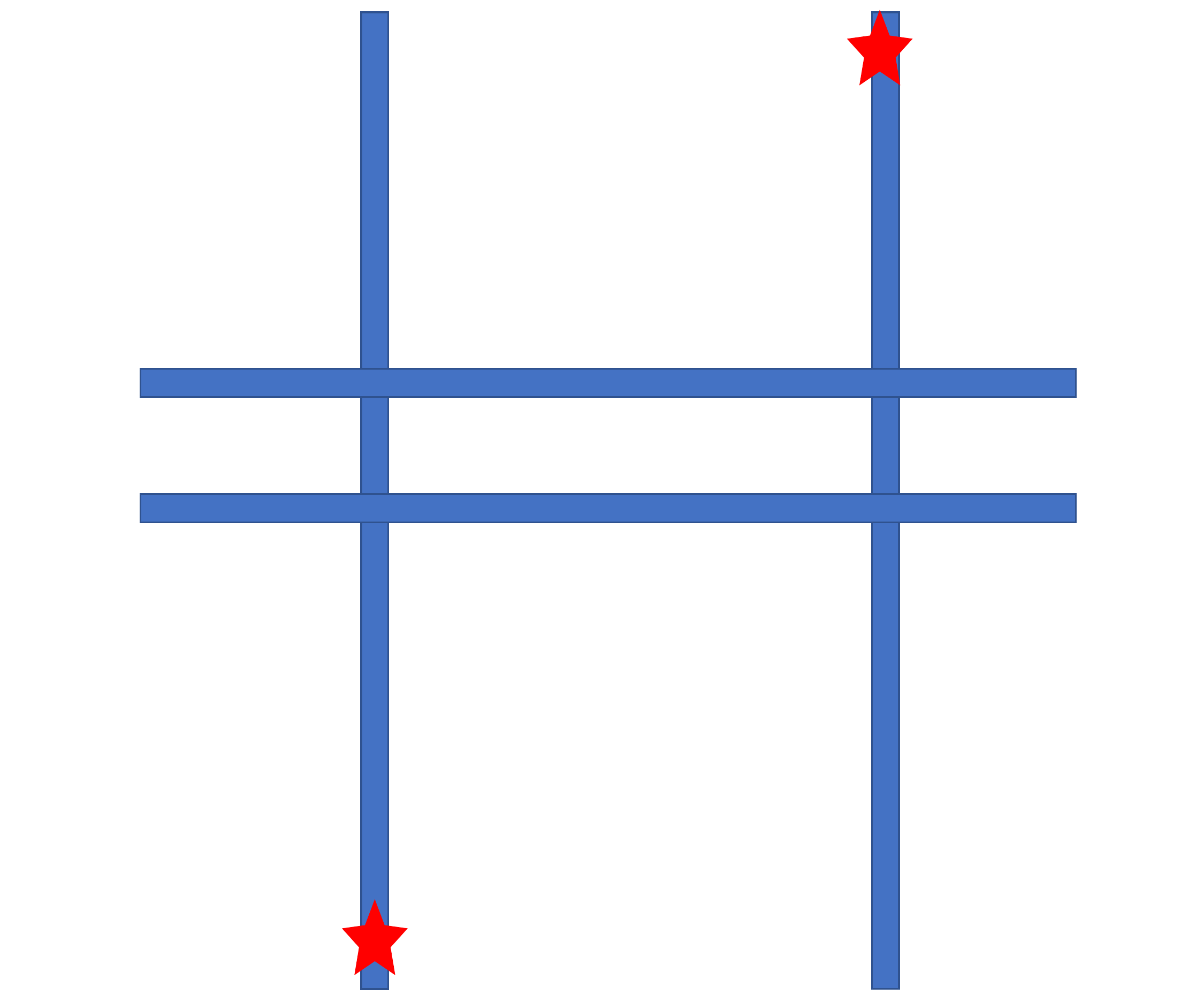}%
}\hfill
\subfloat[Event $E_{4,2,2}$\label{E422}]{%
  \includegraphics[width=0.33\columnwidth,keepaspectratio]{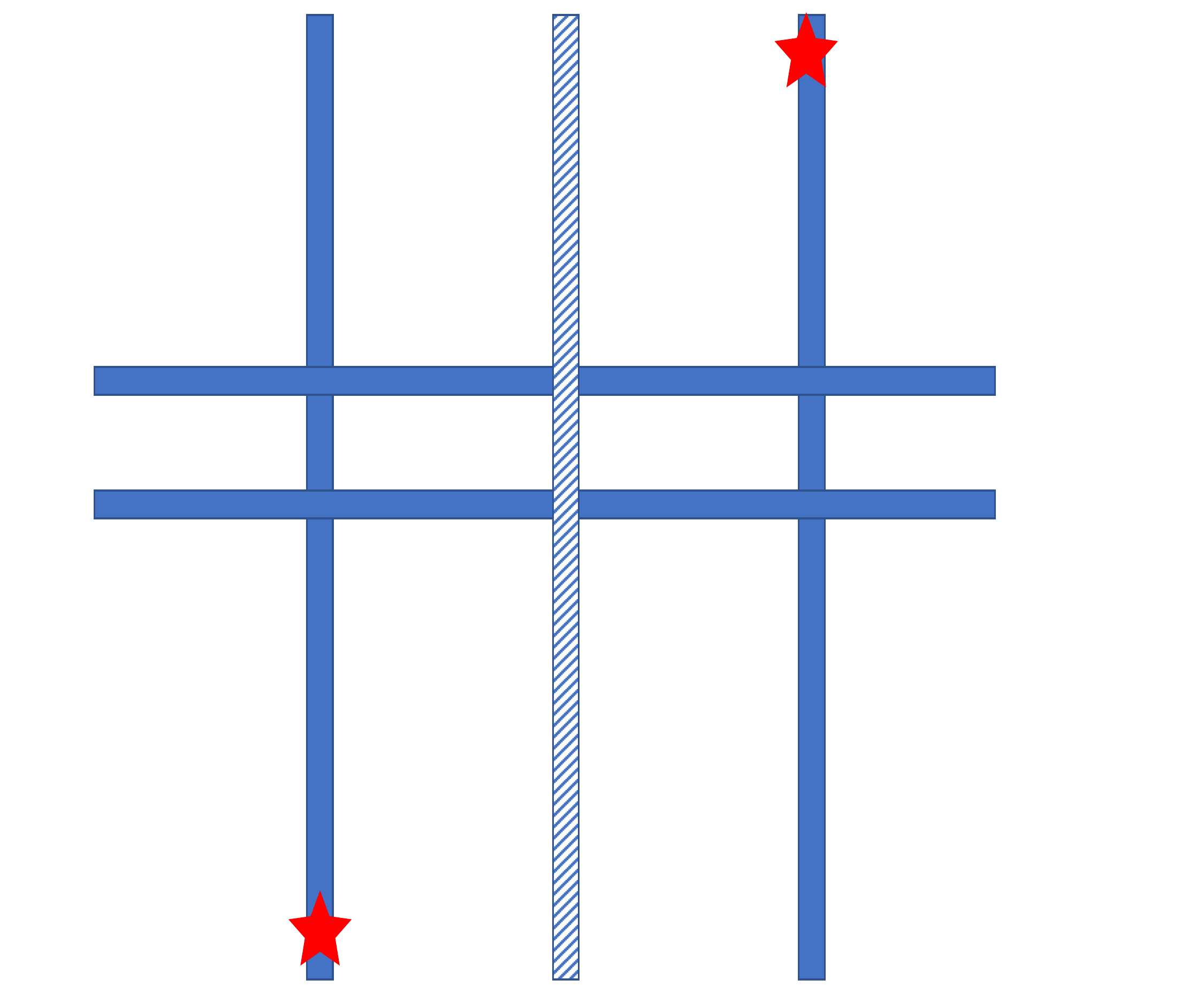}%
}
\hfill
\subfloat[Event $E_{4,2,3}$\label{E423}]{%
  \includegraphics[width=0.33\columnwidth,keepaspectratio]{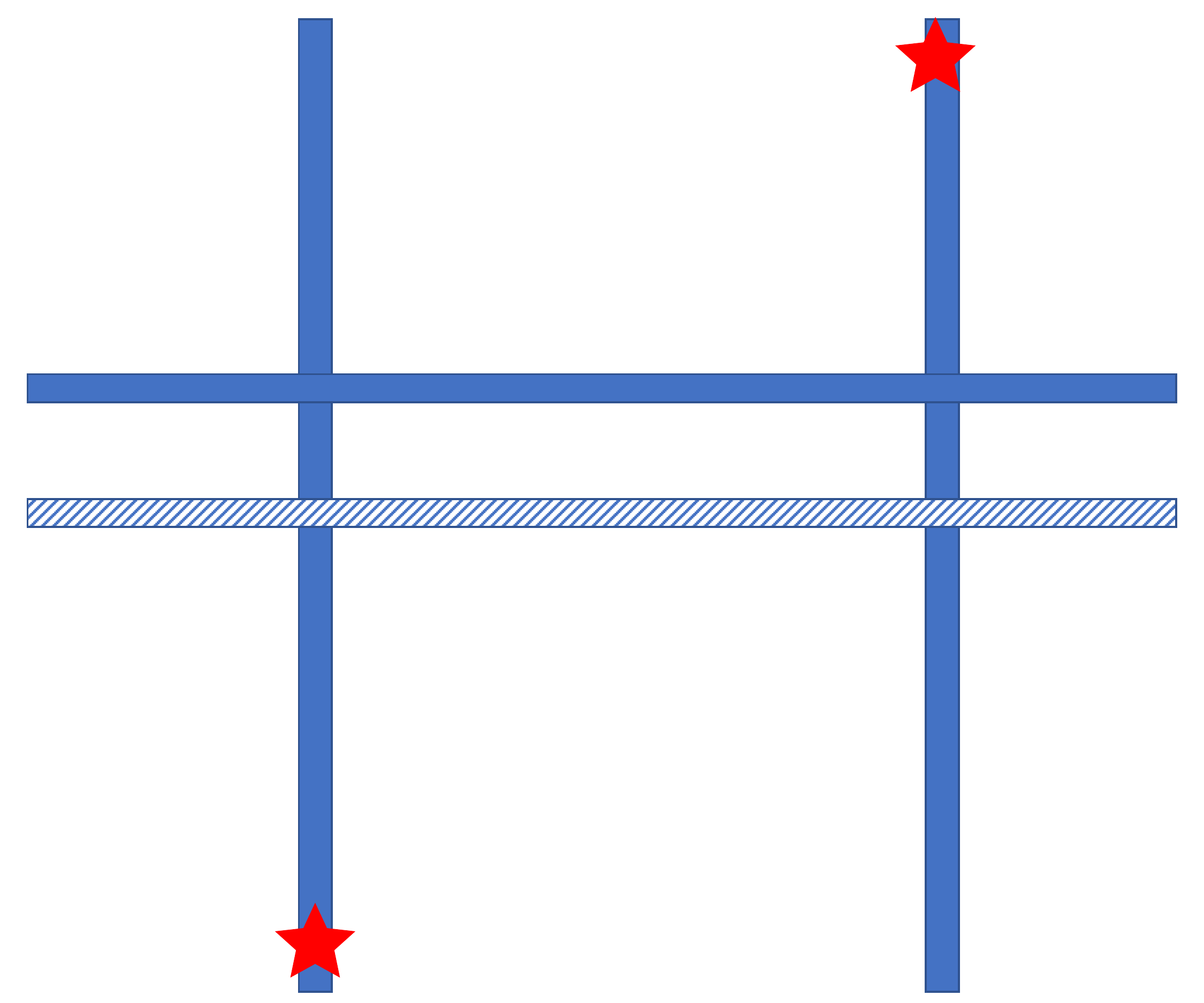}%
}

\subfloat[Event $E_{4,3,1}$\label{E431}]{%
  \includegraphics[width=0.33\columnwidth,keepaspectratio]{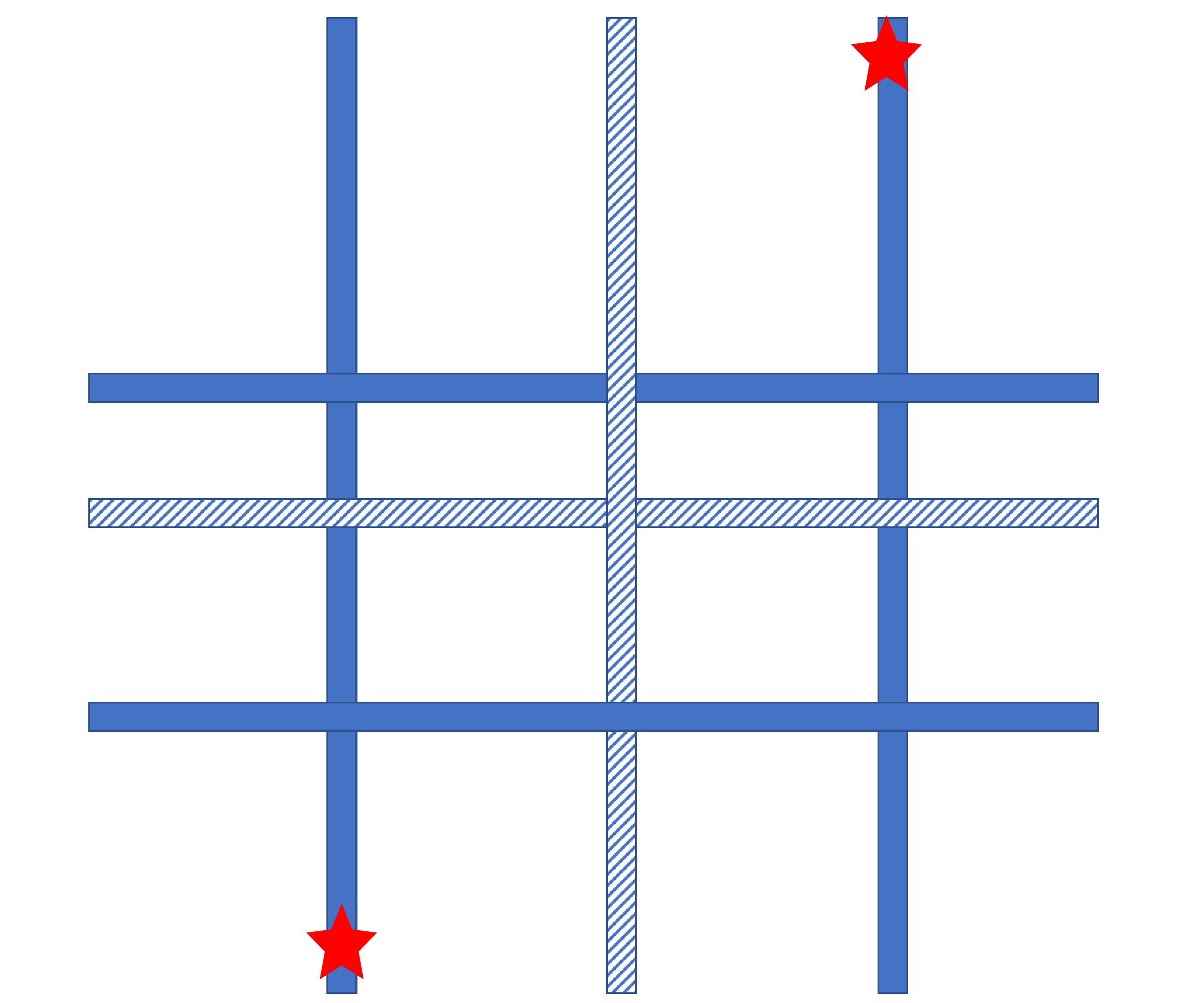}%
}\hfill
\subfloat[Event $E_{4,3,2}$\label{E432}]{%
  \includegraphics[width=0.33\columnwidth,keepaspectratio]{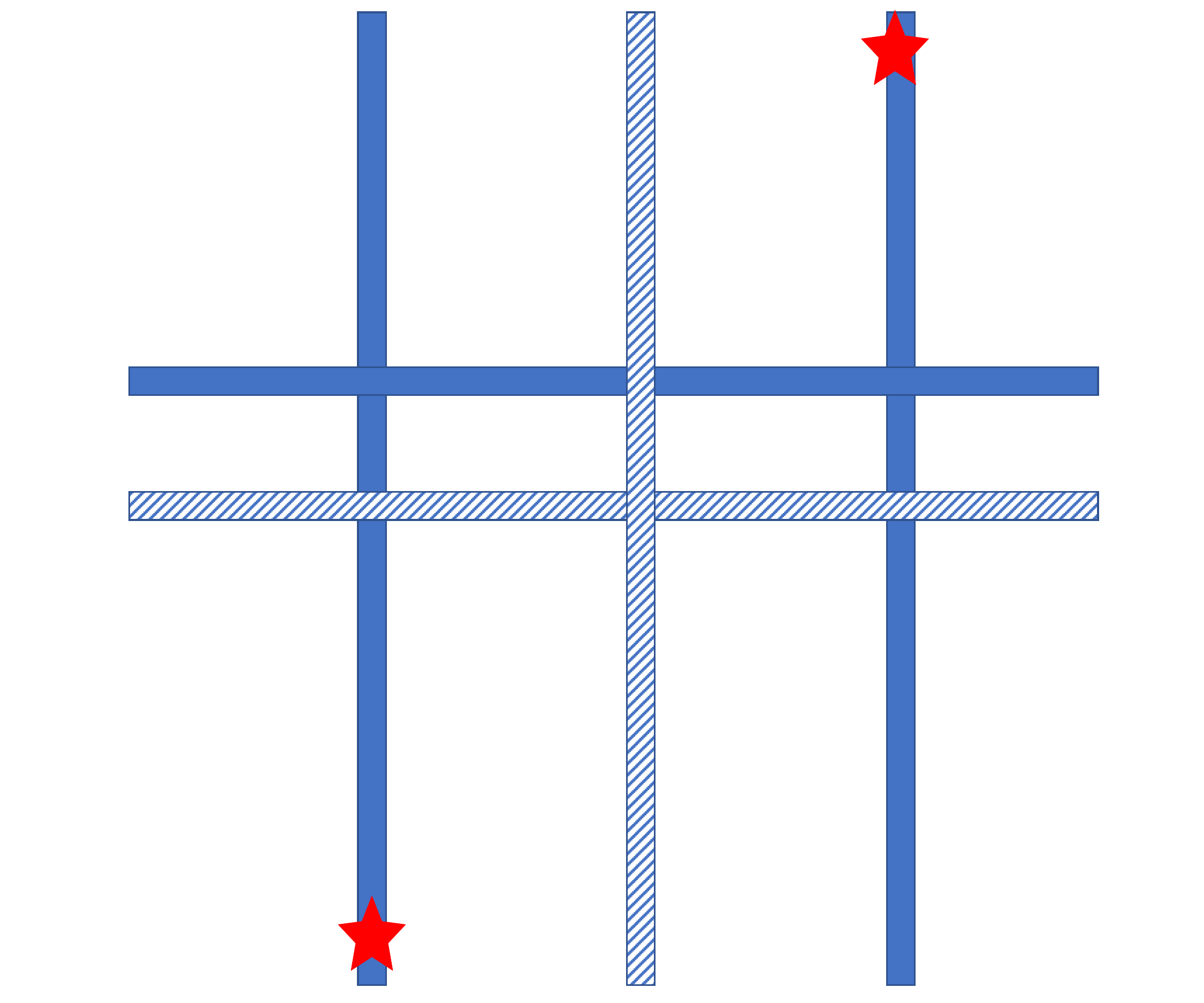}%
}\hfill
\subfloat[Event $E_{4,2,5}$\label{E425}]{%
  \includegraphics[width=0.33\columnwidth,keepaspectratio]{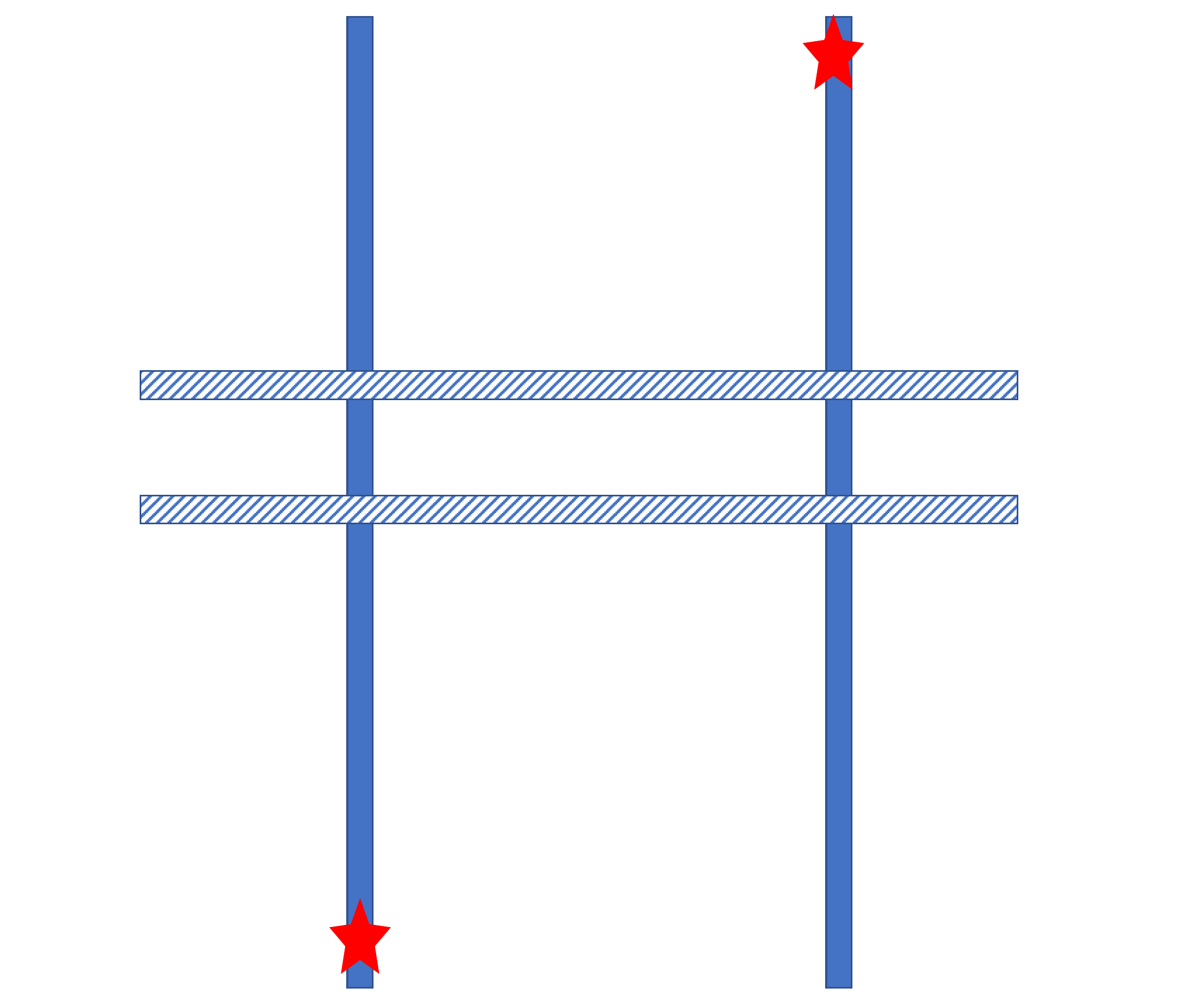}%
}

\subfloat[Event $E_{4,3,3}$\label{E433}]{%
  \includegraphics[width=0.33\columnwidth,keepaspectratio]{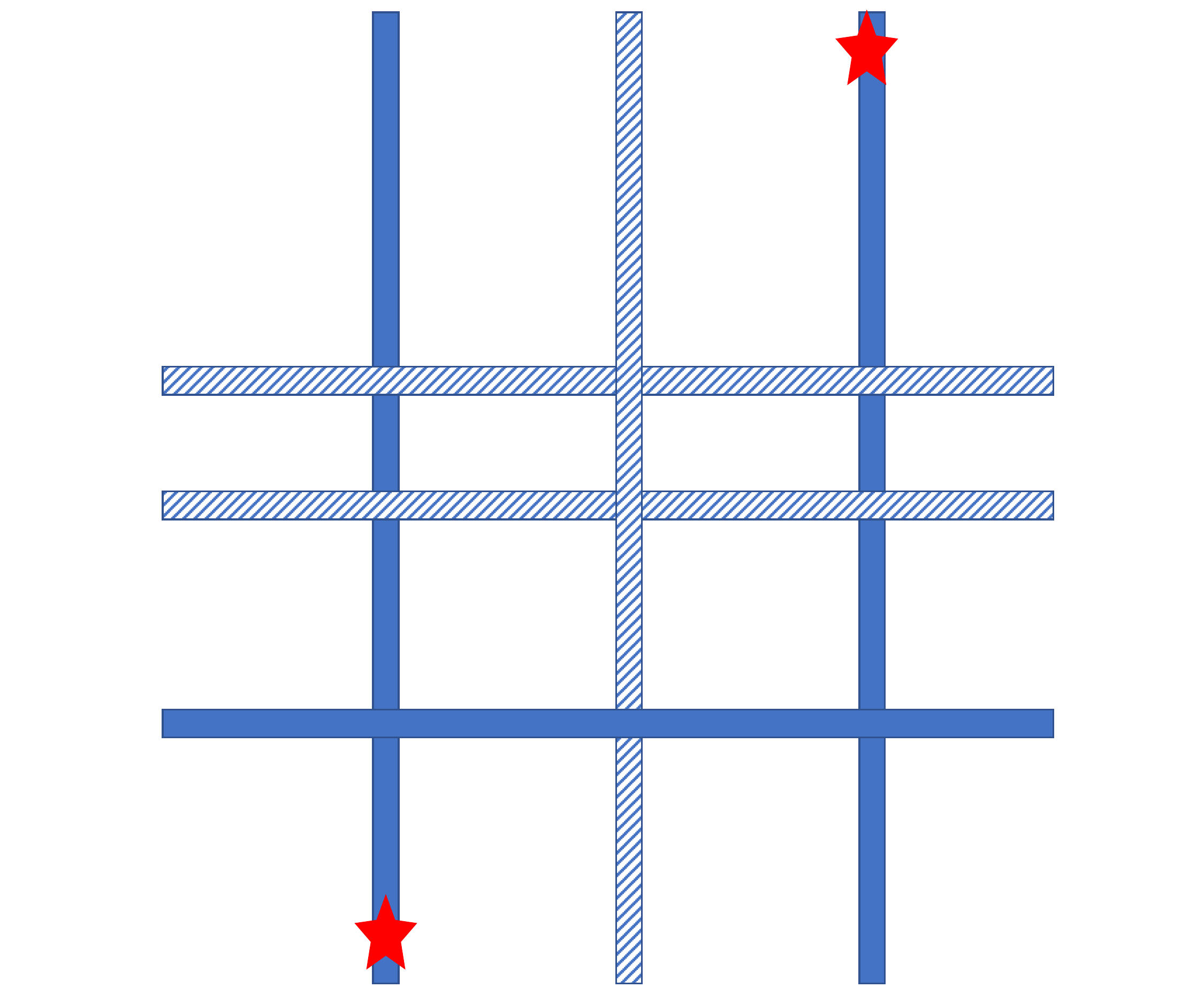}%
}\hfill
\subfloat[Event $E_{4,3,4}$\label{E434}]{%
  \includegraphics[width=0.33\columnwidth,keepaspectratio]{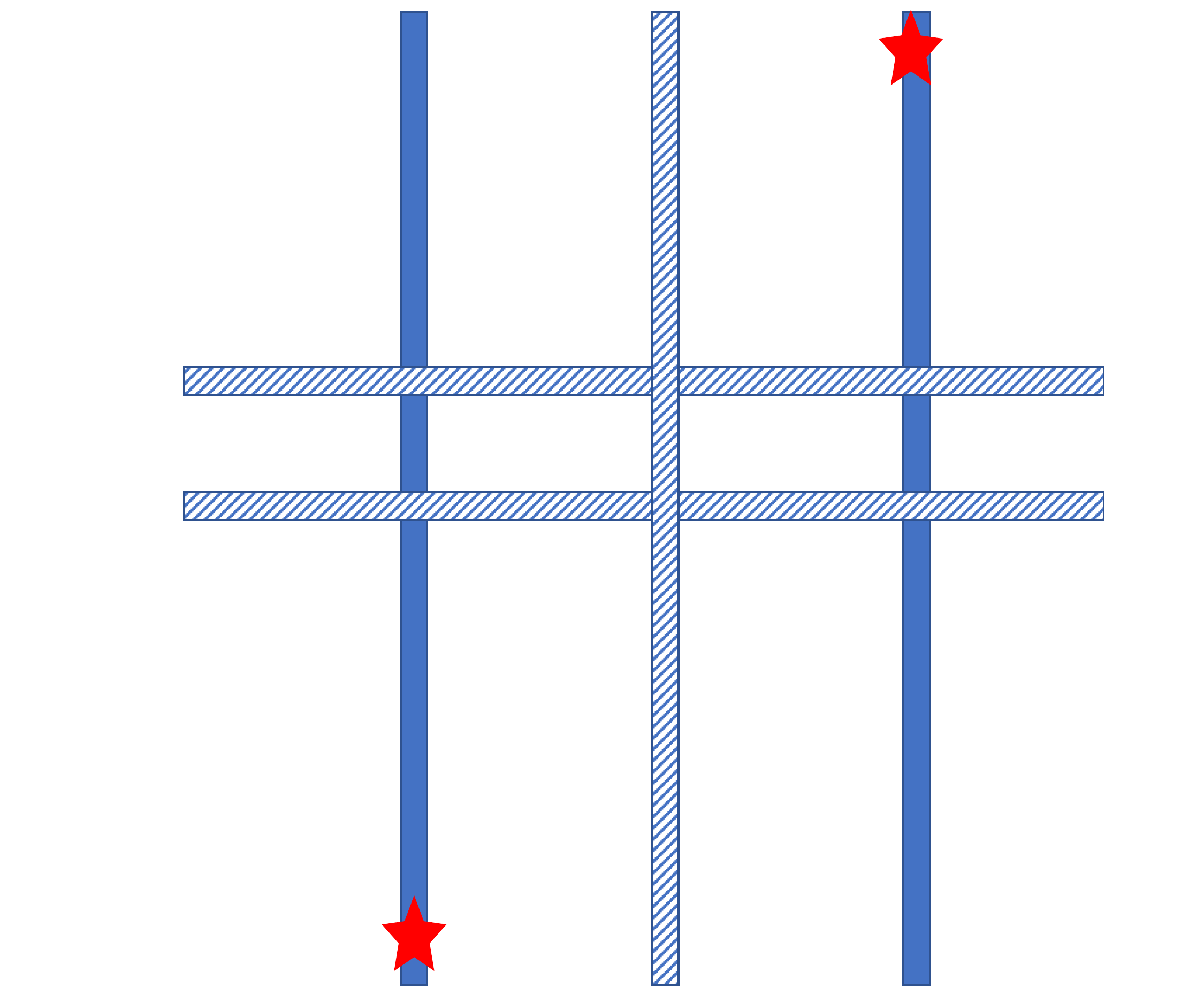}%
}\hfill
\subfloat{%
  \includegraphics[width=0.33\columnwidth,keepaspectratio]{dnx_legend_1.pdf}%
}

\caption{Subcases of tree $E_4$.}
\label{e4sub}
\end{figure}

\begin{itemize}[wide, labelwidth=!, labelindent=0pt]
    \item Event $E_{4,1,1}$: \\\textit{Description}: If there are no horizontal roads between S and D, as shown in Fig. \ref{E411}; 
    \\Event $L_{4,1} = E_{4,1,1} \cap E_4$;
    \\\textit{Probability}: $\mathbb{P}(E_{4,1,1}|E_4)=e^{-\lambda d_v}$,
    $\mathbb{P}(L_{4,1})=\mathbb{P}(E_{4,1,1}|E_4)\mathbb{P}(E_4)$; 
    \\\textit{Action}: We simply take the shortest path from S to D. 
    \begin{align}
    \label{411}
        &\mathbb{P}(D_n < x|L_{4,1})=\mathbb{P}(D_n < x|E_{4,1,1},E_4) 
        \nonumber\\&= \mathbb{P}(D_n < x |D_\mathrm{N-HC} > d_v, D_\mathrm{N-HNC} > d_v)
      	\nonumber\\& = {\textstyle \frac{p(g_1 + g_2 + g_3)}{f_1(0,\infty,d_v,t+d_v,y,d_v)+f_1(0,\infty,t+d_v,\infty,t+d_v,d_v)}}\times\nonumber\\&\mathbbm{1}\{x> d_v\}
      	\nonumber\\& + p \frac{f_2(d_v,\infty,0,{\rm min}(x,t-d_v),t,y+d_v)}{f_2(d_v,\infty,0,t-d_v,t,y+d_v)}
      	 \nonumber\\&+ p\frac{f_4(d_v,\infty,0,{\rm min}(x,t-d_v),t,y+d_v)}{f_4(d_v,\infty,0,t-d_v,t,y+d_v)}
    .
    \end{align}
    The proof for $\mathbb{P}(D_n < x|E_{4,1,1},E_{4})$ is similar to that of $\mathbb{P}(D_n < x|E_{3,1,1},E_{3})$ given in (\ref{311}).
    
    \item  Event $E_{4,1,2}$: \\\textit{Description}: If there is only one horizontal non-charging road and no horizontal charging road between S and D, as shown in Fig. \ref{E412}; 
    \\Event $L_{4,2} = E_{4,1,2} \cap E_4$;
    \\\textit{Probability}: $\mathbb{P}(E_{4,1,2}|E_4)=\lambda (1-p) d_v e^{-\lambda (1-p) d_v}e^{-\lambda p d_v}$,
    $\mathbb{P}(L_{4,2})=\mathbb{P}(E_{4,1,2}|E_4)\mathbb{P}(E_4)$;
    \\\textit{Action}: We take that horizontal non-charging road.  $$\mathbb{P}(D_n < x|L_{4,2})=\mathbb{P}(D_n < x |E_{4,1,2},E_4) = 0$$ 
    \item Event $E_{4,1,3}$: \\\textit{Description}: If there is only one horizontal charging road and no horizontal non-charging road between S and D, as shown in Fig. \ref{E413}; 
    \\Event $L_{4,3} = E_{4,1,3} \cap E_4$;
    \\\textit{Probability}: $\mathbb{P}(E_{4,1,3}|E_4)=\lambda p d_v e^{-\lambda p d_v}e^{-\lambda (1-p) d_v}$,
    $\mathbb{P}(L_{4,3})=\mathbb{P}(E_{4,1,3}|E_4)\mathbb{P}(E_4)$; 
    \\\textit{Action}: We take that horizontal charging road. 
    \begin{align}
    \label{413}
        &\mathbb{P}(D_n < x|L_{4,3})=\mathbb{P}(D_n < x|E_{4,1,3},E_4) 
        \nonumber\\&= \mathbb{P}(D_n < x |D_\mathrm{N-HC} < d_v, D_\mathrm{N-HNC} > d_v)\mathbbm{1}\{x < d_v\} \nonumber\\& + \mathbbm{1}\{x > d_v\}
        \nonumber\\&= \frac{\mathbb{P}(D_\mathrm{N-HC} < {\rm min}(x,d_v), D_\mathrm{N-HNC} > d_v)}{\mathbb{P}(D_\mathrm{N-HC} < d_v < D_\mathrm{N-HNC})}\mathbbm{1}\{x < d_v\} \nonumber\\& + \mathbbm{1}\{x > d_v\}
        \nonumber\\&= \frac{F_{D_\mathrm{N-HC}} (x)}{F_{D_\mathrm{N-HC}} (d_v)}\mathbbm{1}\{x < d_v\} + \mathbbm{1}\{x > d_v\}
    .
    \end{align}
    
    \item Event $E_{4,1,4}$: \\\textit{Description}: If there are no horizontal charging roads but at least two horizontal non-charging road between S and D; \\\textit{Probability}: $\mathbb{P}(E_{4,1,4}|E_4)=e^{-\lambda p d_v}(1-e^{-\lambda (1-p) d_v} - \lambda (1-p) d_v e^{- \lambda (1-p) d_v})$.
    \begin{itemize}[wide, labelwidth=!, labelindent=0pt]
        \item Event $E_{4,2,1}$: \\\textit{Description}: If there are no vertical charging roads between S and D, as shown in Fig. \ref{E421}; 
        \\Event $L_{4,4} = E_{4,2,1} \cap E_{4,1,4} \cap E_4$;
        \\\textit{Probability}: $\mathbb{P}(E_{4,2,1}|E_{4,1,4},E_4)=e^{-\lambda p d_h}$,
        $\mathbb{P}(L_{4,4})=\mathbb{P}(E_{4,2,1}|E_{4,1,4},E_4)\mathbb{P}(E_{4,1,4}|E_4)\mathbb{P}(E_4)$; 
        \\\textit{Action}: We take any horizontal non-charging road between S and D. $$\mathbb{P}(D_n < x|L_{4,4})=\mathbb{P}(D_n < x |E_{4,2,1},E_{4,1,4},E_4) = 0$$
        \item Event $E_{4,2,2}$: \\\textit{Description}: If there is at least one vertical charging road between S and D, as shown in Fig. \ref{E422}; 
        \\Event $L_{4,5} = E_{4,2,2} \cap E_{4,1,4} \cap E_4$;
        \\\textit{Probability}: $\mathbb{P}(E_{4,2,2}|E_{4,1,4},E_4)=1- e^{-\lambda p d_h}$,
        \\$\mathbb{P}(L_{4,5})=\mathbb{P}(E_{4,2,2}|E_{4,1,4},E_4)\mathbb{P}(E_{4,1,4}|E_4)\mathbb{P}(E_4)$; 
        \\\textit{Action}: We first go to the nearest horizontal non-charging road, then switch to the nearest vertical charging road, then switch to the furthest horizontal non-charging road. 
        \begin{align}
      \label{422}
      &\mathbb{P}(D_n < x|L_{4,5})=\mathbb{P}(D_n < x|E_{4,2,2},E_{4,1,4},E_4)
        \nonumber\\& = \biggl(\frac{\int_{{\rm max}(x-d_v,0)}^{{\rm min}(d_h,x)} F_{D_\mathrm{N-HNC}} (x-y) f_{D_\mathrm{N-VC}}(y) {\rm d}y}{F_{D_\mathrm{N-HNC}} (d_v)F_{D_\mathrm{N-VC}} (d_h)}
        \nonumber\\& + \frac{F_{D_\mathrm{N-VC}}({\rm min}(d_h,x-d_v)) \mathbbm{1}\{x>d_v\}}{F_{D_\mathrm{N-VC}} (d_h)}\biggr) \times  
       \nonumber\\& \mathbbm{1}\{x < d_h + d_v\}+ \mathbbm{1}\{x> d_h + d_v \}
      .
      \end{align}
      The proof for $\mathbb{P}(D_n < x|E_{4,2,2},E_{4,1,4},E_{4})$ is similar to that of $\mathbb{P}(D_n < x|E_{3,2,2},E_{3,1,3},E_{3})$ given in (\ref{322}).
        
    \end{itemize}
    \item Event $E_{4,1,5}$: \\\textit{Description}: If there is one horizontal charging road and at least one horizontal non-charging road between S and D; \\\textit{Probability}: $\mathbb{P}(E_{4,1,5}|E_4)=\lambda p d_v e^{-\lambda p d_v}(1-e^{-\lambda (1-p) d_v})$.
    \begin{itemize}[wide, labelwidth=!, labelindent=0pt]
        \item Event $E_{4,2,3}$: \\\textit{Description}: If there are no vertical charging roads between S and D, as shown in Fig. \ref{E423}; 
        \\Event $L_{4,6} = E_{4,2,3} \cap E_{4,1,5} \cap E_4$;
        \\\textit{Probability}: $\mathbb{P}(E_{4,2,3}|E_{4,1,5},E_4)=e^{-\lambda p d_h}$,
        $\mathbb{P}(L_{4,6})=\mathbb{P}(E_{4,2,3}|E_{4,1,5},E_4)\mathbb{P}(E_{4,1,5}|E_4)\mathbb{P}(E_4)$; 
        \\\textit{Action}: We take the horizontal charging road between S and D. 
        \begin{align}
    \label{423}
        &\mathbb{P}(D_n < x|L_{4,6})=\mathbb{P}(D_n < x|E_{4,2,3},E_{4,1,5},E_4)
        \nonumber\\&= \mathbb{P}(D_n < x |D_\mathrm{N-HC} < d_v, D_\mathrm{N-HNC} < d_v)\mathbbm{1}\{x < d_v\} \nonumber\\&+ \mathbbm{1}\{x > d_v\}
\nonumber\\&= \frac{F_{D_\mathrm{N-HC}} (x)}{F_{D_\mathrm{N-HC}} (d_v)}\mathbbm{1}\{x < d_v\} + \mathbbm{1}\{x > d_v\}
	  .
	  \end{align}
	  The proof of $\mathbb{P}(D_n < x|E_{4,2,3},E_{4,1,5},E_{4})$ is similar to that of $\mathbb{P}(D_n < x|E_{4,1,3},E_{4}) $ given in (\ref{413}).
        
        \item Event $E_{4,2,4}$: \\\textit{Description}: If there is at least one vertical charging road between S and D; \\\textit{Probability}: $\mathbb{P}(E_{4,2,4}|E_{4,1,5},E_4)=1- e^{-\lambda p d_h}$.
        \begin{itemize}[wide, labelwidth=!, labelindent=0pt]
            \item Event $E_{4,3,1}$: \\\textit{Description}: If there exists at least one horizontal non-charging road below the nearest horizontal charging road from S, as shown in Fig. \ref{E431}; \\\textit{Probability}: $\mathbb{P}(E_{4,3,1}|E_{4,2,4},E_{4,1,5},E_4)=1 - \frac{p - pe^{-\lambda d_v}}{1 -e^{-\lambda p d_v}}$; \\\textit{Action}: we compare (i) the distance between the nearest horizontal charging road and the nearest horizontal non-charging road, and (ii) the horizontal distance between the nearest vertical charging road and source, to take the longer one. 
            \begin{itemize}[wide, labelwidth=!, labelindent=0pt]
                \item Event $E_{4,4,1}$: \\\textit{Description}: If we take the nearest vertical charging road; 
                \\Event $L_{4,7} = E_{4,4,1} \cap E_{4,3,1} \cap E_{4,2,4} \cap E_{4,1,5} \cap E_4$;
                \\\textit{Probability}: $\mathbb{P}(E_{4,4,1}|E_{4,3,1},E_{4,2,4},E_{4,1,5},E_4)=1- \int_{0}^{d_h} F_{X_2}(x) f_{D_\mathrm{N-VC}}(x) {\rm d}x$; 
                \\$\mathbb{P}(L_{4,7})=\mathbb{P}(E_{4,4,1}|E_{4,3,1},E_{4,2,4},E_{4,1,5},E_4)\times \\\mathbb{P}(E_{4,3,1}|E_{4,2,4},E_{4,1,5},E_4)\times\\\mathbb{P}(E_{4,2,4}|E_{4,1,5},E_4)\mathbb{P}(E_{4,1,5}|E_4)\mathbb{P}(E_4)$;
                \begin{align}
        	  \label{441}
        	  &\mathbb{P}(D_n < x|L_{4,7})=\mathbb{P}(D_n < x|E_{4,4,1},E_{4,3,1},E_{4,2,4},E_{4,1,5},E_4)
        	  \nonumber\\& = \frac{f_8(0,x,x,d_v,x-t)+f_8(0,x,t,x,y-t)}{f_9(0,d_v,0,d_v-t,d_v,t+y)}\mathbbm{1}\{x < d_v\} \nonumber\\&+ \mathbbm{1}\{x>d_v \}
        	  .
        	  \end{align}
        	  The proof of $\mathbb{P}(D_n < x|E_{4,4,1},E_{4,3,1},E_{4,2,4},E_{4,1,5},E_{4})$ is similar to that of \\ $\mathbb{P}(D_n < x|E_{3,4,3},E_{3,3,3},E_{3,2,4},E_{3,1,4},E_{3})$ given in (\ref{343}).
                
                \item Event $E_{4,4,2}$: \\\textit{Description}: If we take the nearest horizontal charging road; \\Event $L_{4,8} = E_{4,4,2} \cap E_{4,3,1} \cap E_{4,2,4} \cap E_{4,1,5} \cap E_4$;
                \\\textit{Probability}: $\mathbb{P}(E_{4,4,2}|E_{4,3,1},E_{4,2,4},E_{4,1,5},E_4)=\int_{0}^{d_h} F_{X_2}(x) f_{D_\mathrm{N-VC}}(x) {\rm d}x$;
                \\$\mathbb{P}(L_{4,8})=\mathbb{P}(E_{4,4,2}|E_{4,3,1},E_{4,2,4},E_{4,1,5},E_4)\times
                \\\mathbb{P}(E_{4,3,1}|E_{4,2,4},E_{4,1,5},E_4)\times\\\mathbb{P}(E_{4,2,4}|E_{4,1,5},E_4)\mathbb{P}(E_{4,1,5}|E_4)\mathbb{P}(E_4)$;
                \begin{align}
	  \label{442}
	  &\mathbb{P}(D_n < x|L_{4,8})=\mathbb{P}(D_n < x|E_{4,4,2},E_{4,3,1},E_{4,2,4},E_{4,1,5},E_4)
	  \nonumber\\& = {\textstyle\frac{f_9 (0,x,0,{\rm max}(x-t,0),y+t,y)+f_9 (0,\infty, {\rm max}(x-t,0),x,x,y)}{f_9 (0,\infty,0,{\rm max}(d_v-t,0),y+t,y)+f_9 (0,\infty,{\rm max}(d_v-t,0),d_v,x,y)}}\times
	  \nonumber\\&\mathbbm{1}\{x < d_v\} + \mathbbm{1}\{x>d_v \}
	  .
	  \end{align}
	  The proof of $\mathbb{P}(D_n < x|E_{4,4,2},E_{4,3,1},E_{4,2,4},E_{4,1,5},E_{4})$ is similar to that of \\ $\mathbb{P}(D_n < x|E_{3,4,4},E_{3,3,3},E_{3,2,4},E_{3,1,4},E_{3})$ given in (\ref{344}).
                
            \end{itemize}
            \item Event $E_{4,3,2}$: \\\textit{Description}: If there does not exist horizontal non-charging roads below the nearest horizontal charging road from S, as shown in Fig. \ref{E432}; 
            \\Event $L_{4,9} =  E_{4,3,2} \cap E_{4,2,4} \cap E_{4,1,5} \cap E_4$;
            \\\textit{Probability}: $\mathbb{P}(E_{4,3,2}|E_{4,2,4},E_{4,1,5},E_4)=\frac{p - pe^{-\lambda d_v}}{1 -e^{-\lambda p d_v}}$,
            \\$\mathbb{P}(L_{4,9})=\mathbb{P}(E_{4,3,2}|E_{4,2,4},E_{4,1,5},E_4)\times
            \\\mathbb{P}(E_{4,2,4}|E_{4,1,5},E_4)\mathbb{P}(E_{4,1,5}|E_4)\mathbb{P}(E_4)$;
            \\\textit{Action}: we simply take the nearest horizontal charging road.  
            \begin{align}
	  \label{432}
	  &\mathbb{P}(L_{4,9})=\mathbb{P}(D_n < x|E_{4,3,2},E_{4,2,4},E_{4,1,5},E_4)
	   \nonumber\\&= \frac{f_7(0,x,y) + f_7(x,\infty,x)}{f_7(0,d_v,y)+f_7(d_v,\infty,d_v)}\mathbbm{1}\{x < d_v\} + \mathbbm{1}\{x>d_v \}
	  .
	  \end{align}
	  The proof of $\mathbb{P}(D_n < x|E_{4,3,2},E_{4,2,4},E_{4,1,5},E_{4})$ is similar to that of $\mathbb{P}(D_n < x|E_{3,3,4},E_{3,2,4},E_{3,1,4},E_{3}) $ given in (\ref{334}).
        \end{itemize}
    \end{itemize}
    \item Event $E_{4,1,6}$: \\\textit{Description}: If there are at least two horizontal charging roads between S and D; \\\textit{Probability}: $\mathbb{P}(E_{4,1,6}|E_4)=1-e^{-\lambda p d_v} - \lambda p d_v e^{-\lambda p d_v}$.
    \begin{itemize}[wide, labelwidth=!, labelindent=0pt]
        \item Event $E_{4,2,5}$: \\\textit{Description}: If there are no vertical charging roads between S and D, as shown in Fig. \ref{E425}; 
        \\Event $L_{4,10} =  E_{4,2,5} \cap E_{4,1,6} \cap E_4$;
        \\\textit{Probability}: $\mathbb{P}(E_{4,2,5}|E_{4,1,6},E_4)=e^{-\lambda p d_h}$,
        \\$\mathbb{P}(L_{4,10})=\mathbb{P}(E_{4,2,5}|E_{4,1,6},E_4)\mathbb{P}(E_{4,1,6}|E_4)\mathbb{P}(E_4)$;
        \\\textit{Action}: We take any horizontal charging road between S and D.  
        \begin{align}
	  \label{425}
	  &\mathbb{P}(D_n < x|L_{4,10})=\mathbb{P}(D_n < x|E_{4,2,5},E_{4,1,6},E_4)
	  \nonumber\\&= \frac{F_{D_\mathrm{N-HC}} (x)}{F_{D_\mathrm{N-HC}} (d_v)}\mathbbm{1}\{x < d_v\} + \mathbbm{1}\{x > d_v\}
	  .
	  \end{align}
	  The proof of $\mathbb{P}(D_n < x|E_{4,2,5},E_{4,1,6},E_{4})$ is similar to that of $\mathbb{P}(D_n < x|E_{4,1,3},E_{4}) $ given in (\ref{413}).
        
        \item Event $E_{4,2,6}$: \\\textit{Description}: If there is at least one vertical charging road between S and D; \\\textit{Probability}: $\mathbb{P}(E_{4,2,6}|E_{4,1,6},E_4)=1- e^{-\lambda p d_h}$.
        \begin{itemize}[wide, labelwidth=!, labelindent=0pt]
            \item Event $E_{4,3,3}$: \\\textit{Description}: If there exists at least one horizontal non-charging road below the nearest horizontal charging road from S, as shown in Fig. \ref{E433};
            \\\textit{Probability}: $\mathbb{P}(E_{4,3,3}|E_{4,2,6},E_{4,1,6},E_4)=1 - \frac{p - pe^{-\lambda d_v}}{1 -e^{-\lambda p d_v}}$; \\\textit{Action}: we compare (i) the distance between the nearest horizontal charging road and the nearest horizontal non-charging road, and (ii) the horizontal distance between the nearest vertical charging road and source, to take the longer one.
            \begin{itemize}[wide, labelwidth=!, labelindent=0pt]
                \item Event $E_{4,4,3}$: \\\textit{Description}: If we take the nearest vertical charging road; 
                \\Event $L_{4,11} = E_{4,4,3} \cap E_{4,3,3} \cap E_{4,2,6} \cap E_{4,1,6} \cap E_4$;
                \\\textit{Probability}: $\mathbb{P}(E_{4,4,3}|E_{4,3,3},E_{4,2,6},E_{4,1,6},E_4)=1- \int_{0}^{d_h} F_{X_2}(x) f_{D_\mathrm{N-VC}}(x) {\rm d}x$;
                \\$\mathbb{P}(L_{4,11})=\mathbb{P}(E_{4,4,3}|E_{4,3,3},E_{4,2,6},E_{4,1,6},E_4)\times
                \\\mathbb{P}(E_{4,3,3}|E_{4,2,6},E_{4,1,6},E_4)\times\\\mathbb{P}(E_{4,2,6}|E_{4,1,6},E_4)\mathbb{P}(E_{4,1,6}|E_4)\mathbb{P}(E_4)$;
                \begin{align}
	  \label{443}
	  &\mathbb{P}(D_n < x|L_{4,11})=\mathbb{P}(D_n < x|E_{4,4,3},E_{4,3,3},E_{4,2,6},E_{4,1,6},E_4)
	  \nonumber\\& = \frac{f_8(0,x,x,d_v,x-t)+f_8(0,x,t,x,y-t)}{f_9(0,d_v,0,d_v-t,d_v,t+y)}\mathbbm{1}\{x < d_v\} \nonumber\\&+ \mathbbm{1}\{x>d_v \}
	  .
	  \end{align}
	  The proof of $\mathbb{P}(D_n < x|E_{4,4,3},E_{4,3,3},E_{4,2,6},E_{4,1,6},E_{4})$ is similar to that of \\ $\mathbb{P}(D_n < x|E_{3,4,3},E_{3,3,3},E_{3,2,4},E_{3,1,4},E_{3})$ given in (\ref{343}).
                
                \item Event $E_{4,4,4}$: \\\textit{Description}: If we take the nearest horizontal charging road; \\Event $L_{4,12} = E_{4,4,4} \cap E_{4,3,3} \cap E_{4,2,6} \cap E_{4,1,6} \cap E_4$;
                \\\textit{Probability}: $\mathbb{P}(E_{4,4,4}|E_{4,3,3},E_{4,2,6},E_{4,1,6},E_4)=\int_{0}^{d_h} F_{X_2}(x) f_{D_\mathrm{N-VC}}(x) {\rm d}x$;
                \\$\mathbb{P}(L_{4,12})=\mathbb{P}(E_{4,4,4}|E_{4,3,3},E_{4,2,6},E_{4,1,6},E_4)\times\\\mathbb{P}(E_{4,3,3}|E_{4,2,6},E_{4,1,6},E_4)\times\\\mathbb{P}(E_{4,2,6}|E_{4,1,6},E_4)\mathbb{P}(E_{4,1,6}|E_4)\mathbb{P}(E_4)$;
                \begin{align}
	  \label{444}
		&\mathbb{P}(D_n < x|L_{4,12})=\mathbb{P}(D_n < x|E_{4,4,4},E_{4,3,3},E_{4,2,6},E_{4,1,6},E_4)
		\nonumber\\& = {\textstyle\frac{f_9 (0,x,0,{\rm max}(x-t,0),y+t,y)+f_9 (0,\infty, {\rm max}(x-t,0),x,x,y)}{f_9 (0,\infty,0,{\rm max}(d_v-t,0),y+t,y)+f_9 (0,\infty,{\rm max}(d_v-t,0),d_v,x,y)}}\times
	    \nonumber\\&\mathbbm{1}\{x < d_v\} + \mathbbm{1}\{x>d_v \}
	  .
	  \end{align}
	  The proof of $\mathbb{P}(D_n < x|E_{4,4,4},E_{4,3,3},E_{4,2,6},E_{4,1,6},E_{4})$ is similar to that of \\ $\mathbb{P}(D_n < x|E_{3,4,4},E_{3,3,3},E_{3,2,4},E_{3,1,4},E_{3})$ given in (\ref{344}).
            \end{itemize}
            \item Event $E_{4,3,4}$: \\\textit{Description}: If there does not exist horizontal non-charging roads below the nearest horizontal charging road from S, as shown in Fig. \ref{E434};
            \\Event $L_{4,13} = E_{4,3,4} \cap E_{4,2,6} \cap E_{4,1,6} \cap E_4$;
            \\\textit{Probability}: $\mathbb{P}(E_{4,3,4}|E_{4,2,6},E_{4,1,6},E_4)=\frac{p - pe^{-\lambda d_v}}{1 -e^{-\lambda p d_v}}$,
            \\$\mathbb{P}(L_{4,13})=\mathbb{P}(E_{4,3,4}|E_{4,2,6},E_{4,1,6},E_4)\times
            \\\mathbb{P}(E_{4,2,6}|E_{4,1,6},E_4)\mathbb{P}(E_{4,1,6}|E_4)\mathbb{P}(E_4)$;
            \\\textit{Action}: we simply take the nearest horizontal charging road. 
            \begin{align}
	  \label{434}
	  &\mathbb{P}(D_n < x|L_{4,13})=\mathbb{P}(D_n < x|E_{4,3,4},E_{4,2,6},E_{4,1,6},E_4)
	  \nonumber\\&= \frac{f_7(0,x,y) + f_7(x,\infty,x)}{f_7(0,d_v,y)+f_7(d_v,\infty,d_v)}\mathbbm{1}\{x < d_v\} + \mathbbm{1}\{x>d_v \}
	  .
	  \end{align}
	  The proof of $\mathbb{P}(D_n < x|E_{4,3,4},E_{4,2,6},E_{4,1,6},E_{4})$ is similar to that of $\mathbb{P}(D_n < x|E_{3,3,4},E_{3,2,4},E_{3,1,4},E_{3}) $ given in (\ref{334}). 
            
        \end{itemize}
    \end{itemize}
\end{itemize}
\end{IEEEproof}

\section{proof of Lemma \ref{DnE6}}
\label{appendxe6}
\begin{figure}[!h]
\centering
\includegraphics[width=0.9\columnwidth]{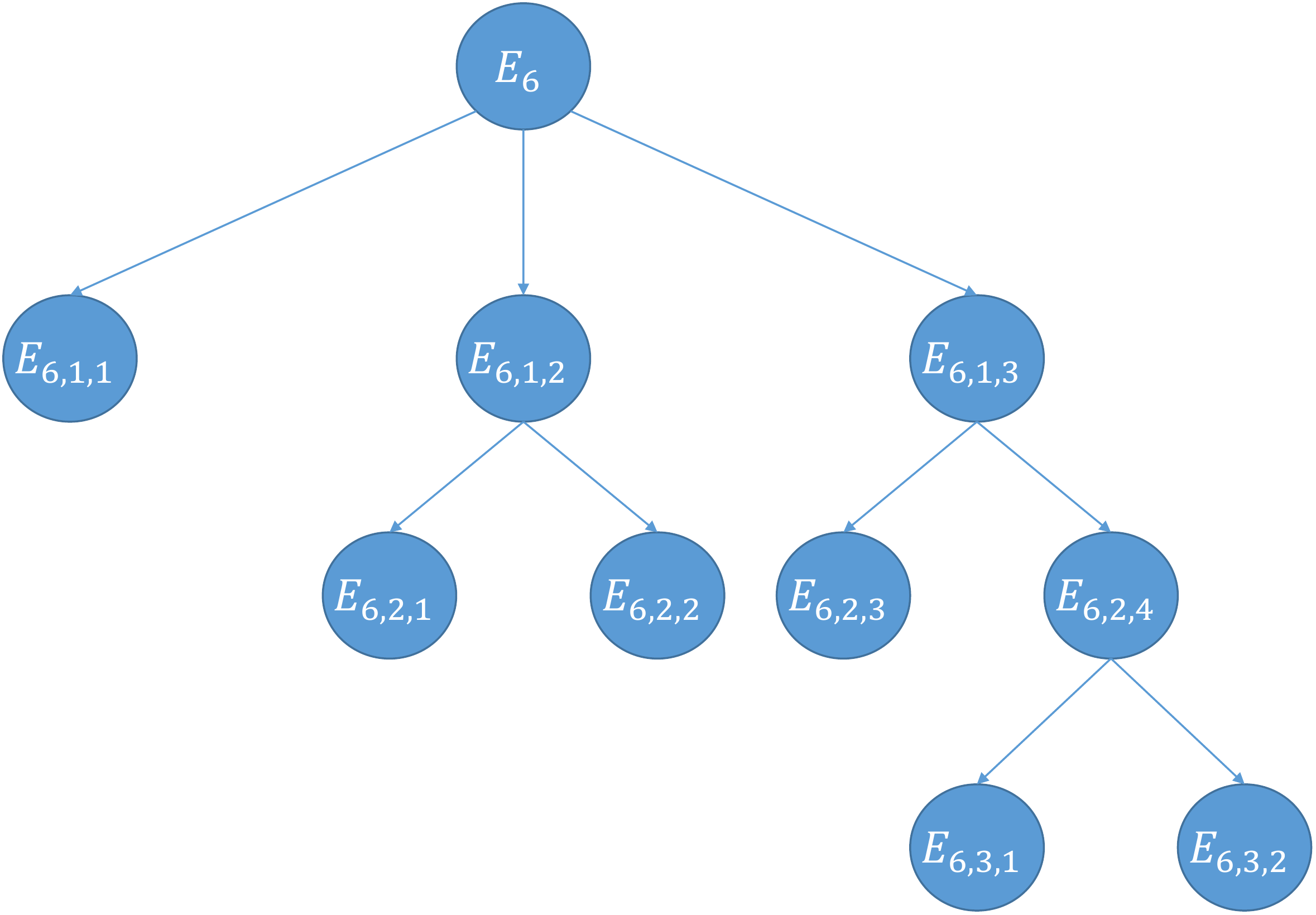}
\caption{Tree $E_6$: source and destination roads are on two perpendicular roads and only the source road is charging.}
\label{E6}
\end{figure}

\begin{figure}[h]
\centering
\captionsetup[subfigure]{font=scriptsize,labelfont=normalsize}
\subfloat[Event $E_{6,1,1}$\label{E611}]{%
  \includegraphics[width=0.33\columnwidth,keepaspectratio]{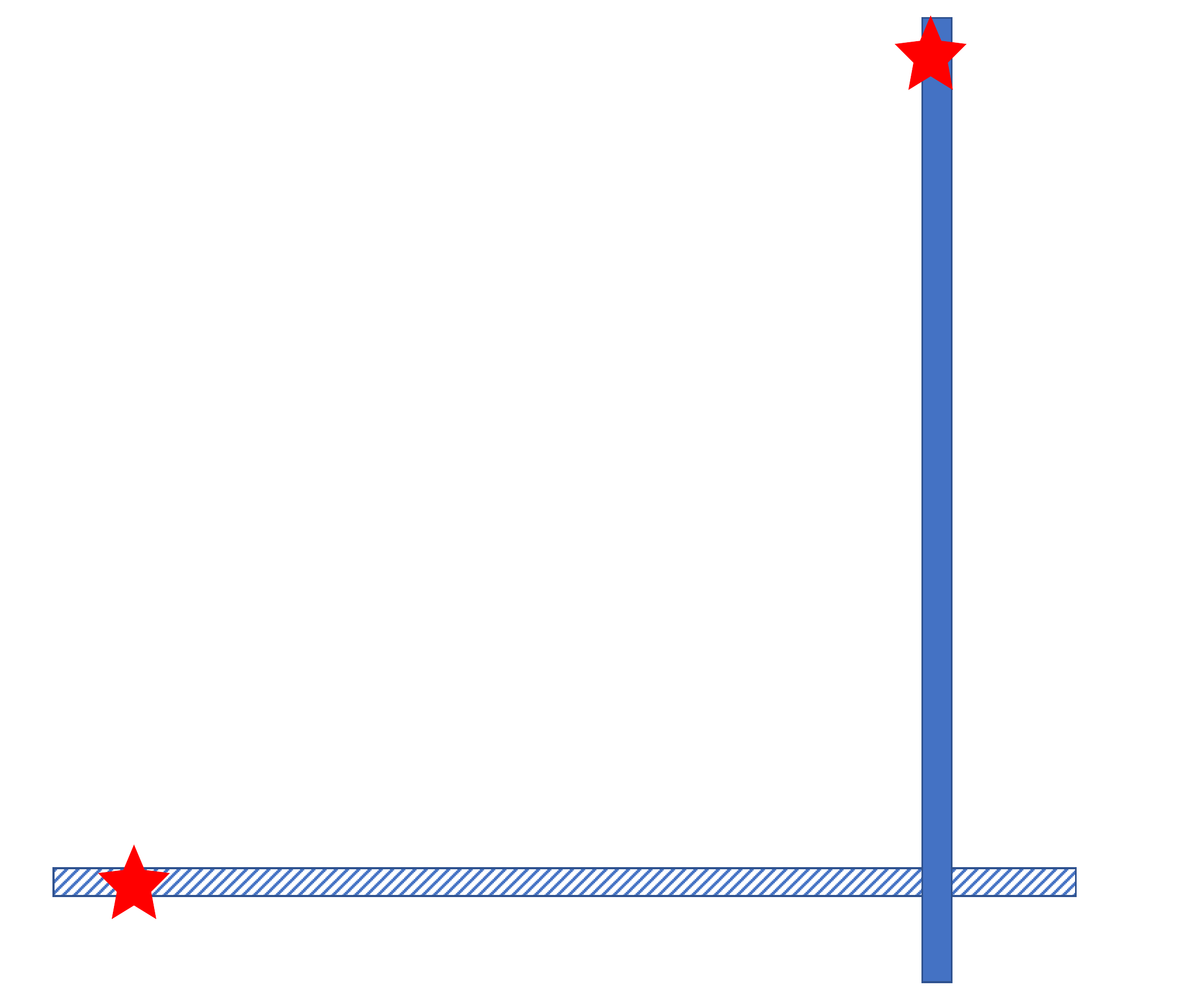}%
}\hfill
\subfloat[Event $E_{6,2,1}$\label{E621}]{%
  \includegraphics[width=0.33\columnwidth,keepaspectratio]{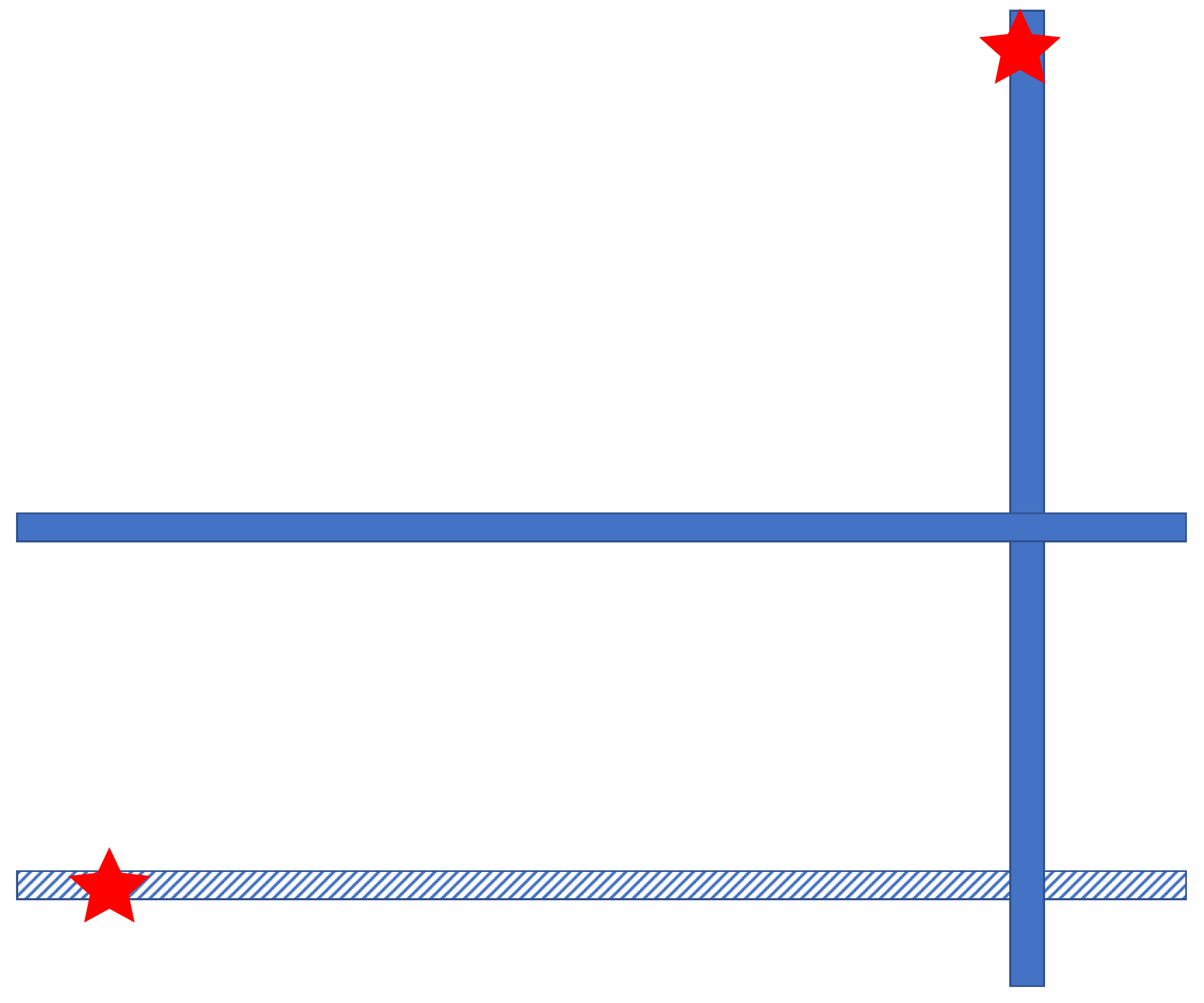}%
}
\hfill
\subfloat[Event $E_{6,2,2}$\label{E622}]{%
  \includegraphics[width=0.33\columnwidth,keepaspectratio]{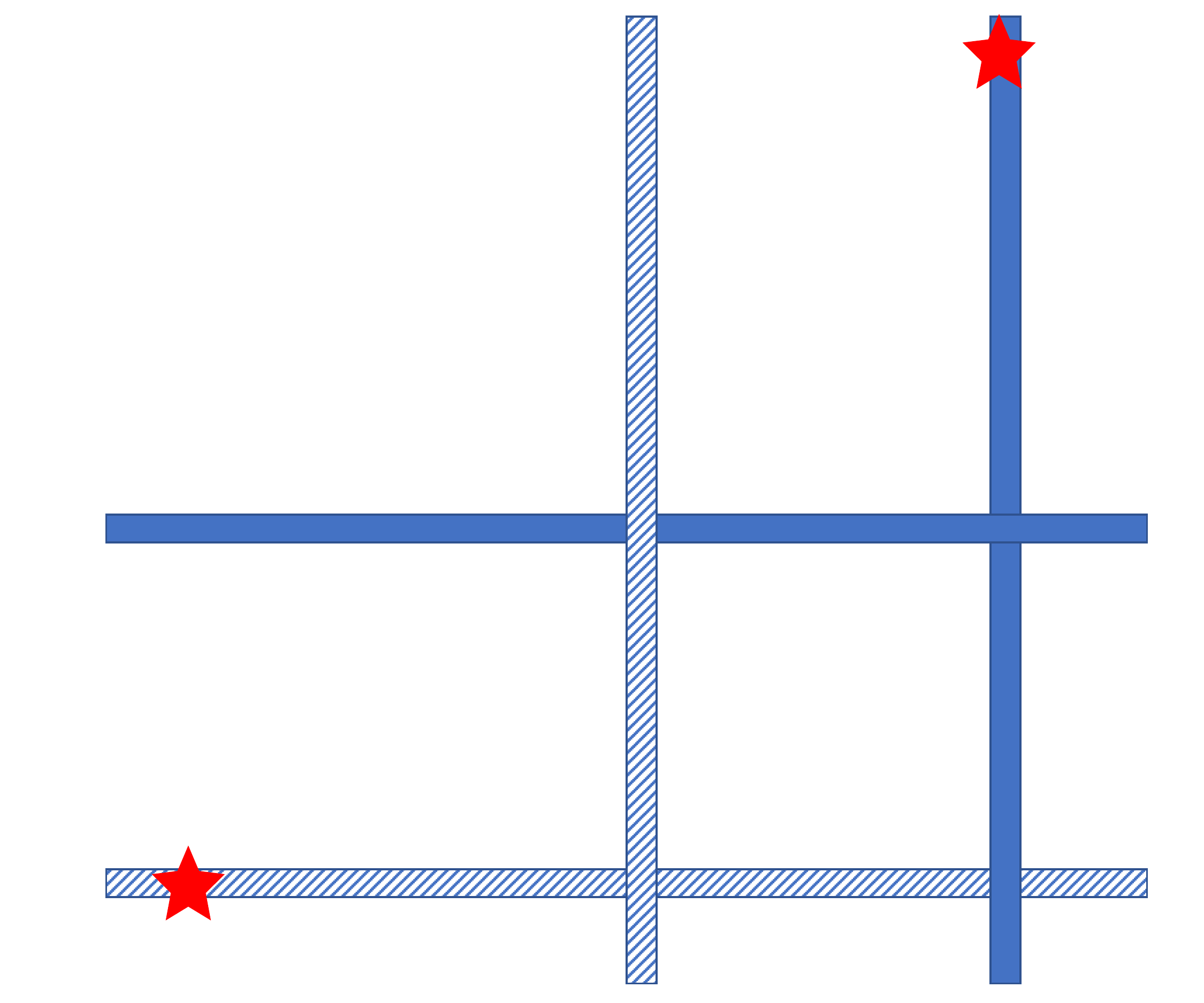}%
}

\subfloat[Event $E_{6,2,3}$\label{E623}]{%
  \includegraphics[width=0.33\columnwidth,keepaspectratio]{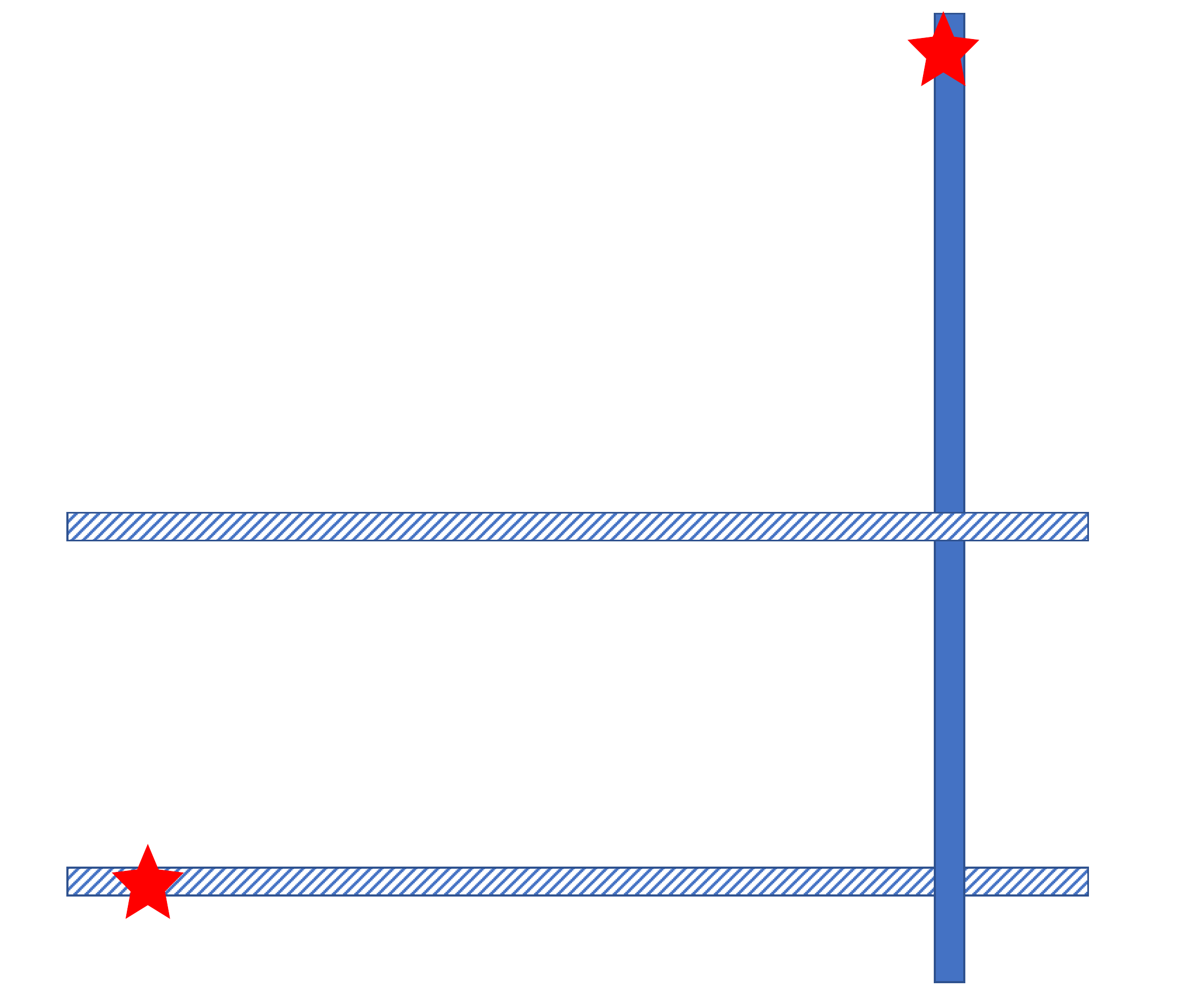}%
}\hfill
\subfloat[Event $E_{6,3,1}$\label{E631}]{%
  \includegraphics[width=0.33\columnwidth,keepaspectratio]{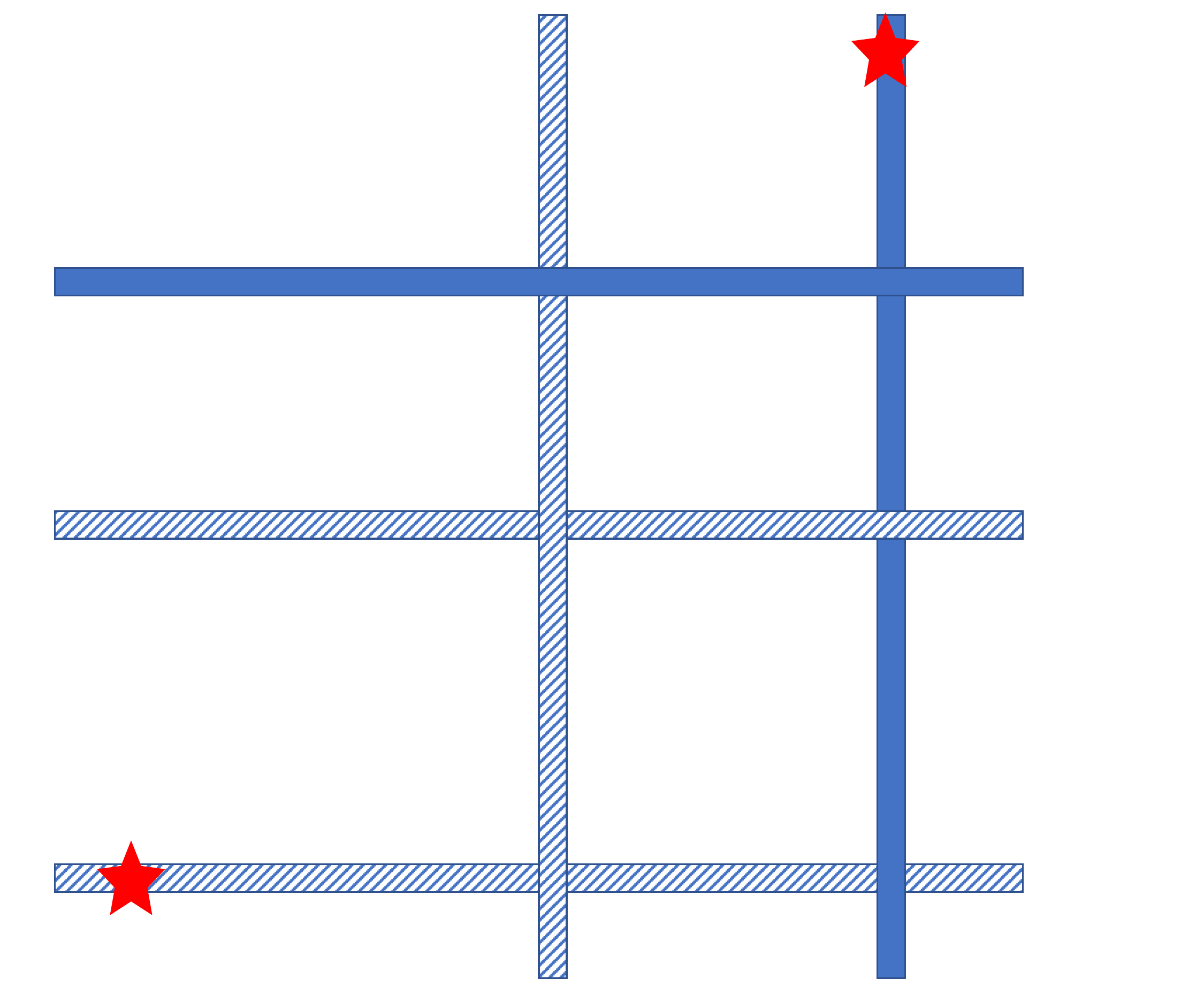}%
}
\hfill
\subfloat[Event $E_{6,3,2}$\label{E632}]{%
  \includegraphics[width=0.33\columnwidth,keepaspectratio]{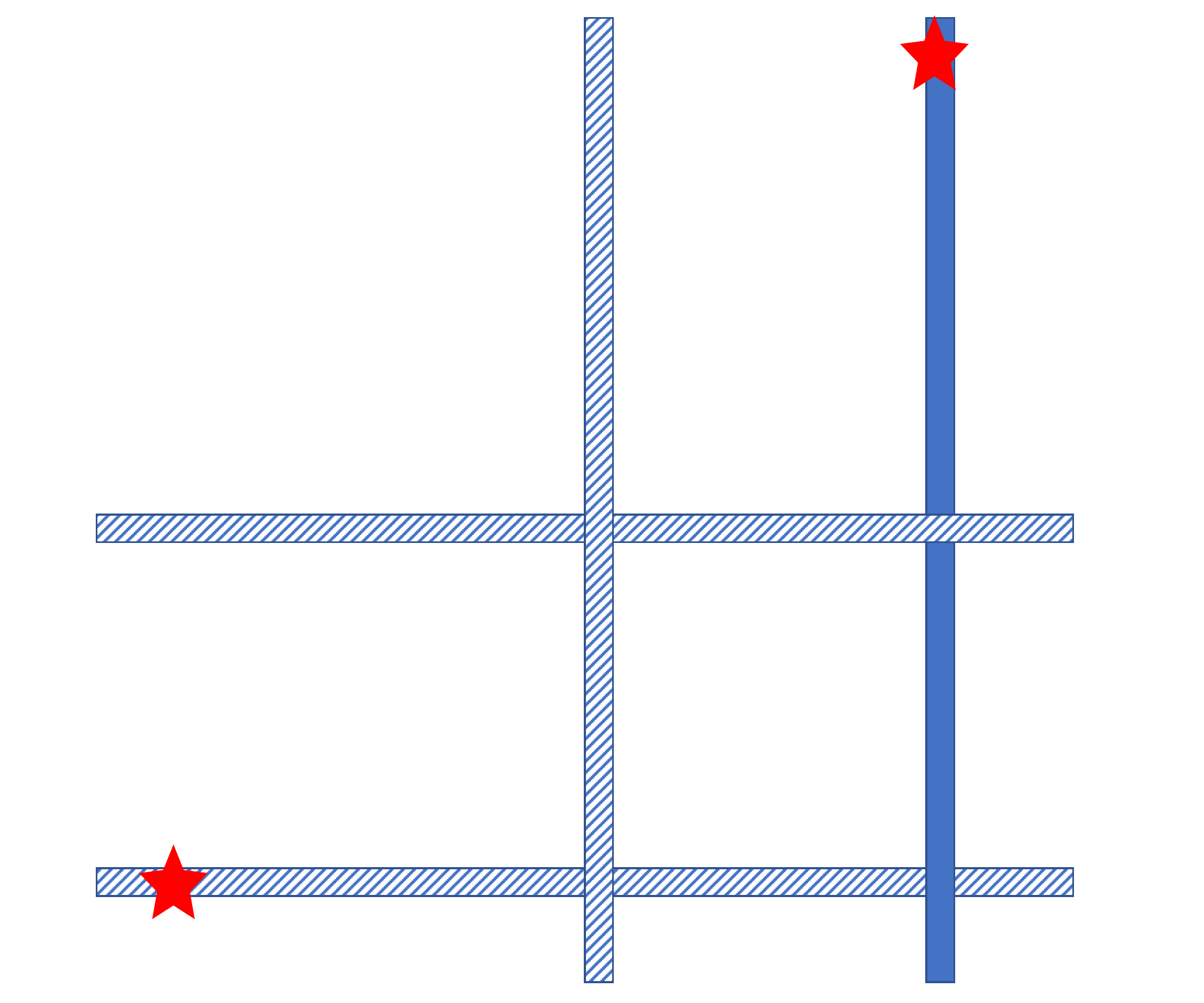}%
}

\subfloat{%
  \includegraphics[width=0.8\columnwidth,keepaspectratio]{dnx_legend_2.pdf}%
}

\caption{Subcases of tree $E_6$.}
\label{e6sub}
\end{figure}

In this appendix, we outline the proof for the probability that the distance to the nearest charging road is less than a positive real number $x$ given the event $E_6$, i.e., $\mathbb{P}(D_n < x|E_6)$. As shown in Fig.~\ref{E6}, we hereby denote subevents as $E_{6,i,j}$, in which $i$ is the level of depth of the event in the probability tree and $j$ is the index of the event at that level. Representative figures for $E_6$ are shown in Fig.~\ref{e6sub}. The distribution of $D_n$ given $E_6$ can be derived as follows:

\begin{align*}
\mathbb{P}(D_n < x | E_6)\mathbb{P}(E_6) = \sum_{i=1}^{N_6} \mathbb{P}(D_n < x | L_{6,i})\mathbb{P}(L_{6,i}),
\end{align*} 
where $N_6$ denotes the number of leaves of tree $E_6$, i.e., $N_6=6$, and $L_{6,i}$'s are successive events ending at the leaves of tree $E_6$ as shown in Fig. \ref{E6}. The definition for each event $L_{6,i}$ will be given in more details as we visit each leaf of the tree.  

%

\begin{itemize}[wide, labelwidth=!, labelindent=0pt]
    \item Event $E_{6,1,1}$: \\\textit{Description}: If there are no horizontal roads between S and D, as shown in Fig. \ref{E611}; 
    \\Event $L_{6,1} = E_{6,1,1} \cap E_6$;
    \\\textit{Probability}: $\mathbb{P}(E_{6,1,1}|E_6)=e^{-\lambda d_v}$,
    $\mathbb{P}(L_{6,1})=\mathbb{P}(E_{6,1,1}|E_6)\mathbb{P}(E_6)$;
    \\\textit{Action}: We simply take the source road then the destination road.
    \item Event $E_{6,1,2}$: \\\textit{Description}: If there are no horizontal charging roads but at least one horizontal non-charging road between S and D; 
    \\\textit{Probability}: $\mathbb{P}(E_{6,1,2}|E_6)=e^{-\lambda p d_v}(1-e^{-\lambda (1-p) d_v})$.
    \begin{itemize}[wide, labelwidth=!, labelindent=0pt]
        \item Event $E_{6,2,1}$: \\\textit{Description}: If there are no vertical charging roads between S and D, as shown in Fig. \ref{E621}; 
        \\Event $L_{6,2} = E_{6,2,1} \cap E_{6,1,2} \cap E_6$;
        \\\textit{Probability}: $\mathbb{P}(E_{6,2,1}|E_{6,1,2},E_6)=e^{-\lambda p d_h}$,
        \\$\mathbb{P}(L_{6,2})=\mathbb{P}(E_{6,2,1}|E_{6,1,2},E_6)\mathbb{P}(E_{6,1,2}|E_6)\mathbb{P}(E_6)$;
        \\\textit{Action}: We take the source road then the destination road. 
        \item Event $E_{6,2,2}$: \\\textit{Description}: If there is at least one vertical charging road between S and D, as shown in Fig. \ref{E622};
        \\Event $L_{6,3} = E_{6,2,2} \cap E_{6,1,2} \cap E_6$;
        \\\textit{Probability}: $\mathbb{P}(E_{6,2,2}|E_{6,1,2},E_6)=1- e^{-\lambda p d_h}$,
        \\$\mathbb{P}(L_{6,3})=\mathbb{P}(E_{6,2,2}|E_{6,1,2},E_6)\mathbb{P}(E_{6,1,2}|E_6)\mathbb{P}(E_6)$;
        \\\textit{Action}: we compare (i) the distance between source and the furthest horizontal non-charging road and (ii) the distance between the furthest vertical charging road to destination, to take the longer one.
    \end{itemize}
    \item Event $E_{6,1,3}$: \\\textit{Description}: If there is at least one horizontal charging road between S and D; \\\textit{Probability}: $\mathbb{P}(E_{6,1,3}|E_6)=1-e^{-\lambda p d_v}$.
    \begin{itemize}[wide, labelwidth=!, labelindent=0pt]
        \item Event $E_{6,2,3}$: \\\textit{Description}: If there are no vertical charging roads between S and D, as shown in Fig. \ref{E623}; 
        \\Event $L_{6,4} = E_{6,2,3} \cap E_{6,1,3} \cap E_6$;
        \\\textit{Probability}: $\mathbb{P}(E_{6,2,3}|E_{6,1,3},E_6)=e^{-\lambda p d_h}$,
        \\$\mathbb{P}(L_{6,4})=\mathbb{P}(E_{6,2,3}|E_{6,1,3},E_6)\mathbb{P}(E_{6,1,3}|E_6)\mathbb{P}(E_6)$;
        \\\textit{Action}: We take the source road then the destination road. 
        \item Event $E_{6,2,4}$: \\\textit{Description}: If there is at least one vertical charging road between S and D; \\\textit{Probability}: $\mathbb{P}(E_{6,2,4}|E_{6,1,3},E_6)=1- e^{-\lambda p d_h}$.
        \begin{itemize}[wide, labelwidth=!, labelindent=0pt]
            \item Event $E_{6,3,1}$: \\\textit{Description}: If there exists at least one horizontal non-charging road above the furthest horizontal charging road from S, as shown in Fig. \ref{E631};
            \\Event $L_{6,5} = E_{6,3,1} \cap E_{6,2,4} \cap E_{6,1,3} \cap E_6$;
            \\\textit{Probability}: $\mathbb{P}(E_{6,3,1}|E_{6,2,4},E_{6,1,3},E_6)=1 - \frac{p - pe^{-\lambda d_v}}{1 -e^{-\lambda p d_v}}$,
            \\$\mathbb{P}(L_{6,5})=\mathbb{P}(E_{6,3,1}|E_{6,2,4},E_{6,1,3},E_6)\times
            \\\mathbb{P}(E_{6,2,4}|E_{6,1,3},E_6)\mathbb{P}(E_{6,1,3}|E_6)\mathbb{P}(E_6)$;
            \\\textit{Action}: we compare (i) the distance between the furthest horizontal charging road and the furthest horizontal non-charging road, and (ii) the horizontal distance between the furthest vertical charging road and destination, to take the longer one.
            \item Event $E_{6,3,2}$: \\\textit{Description}: If there does not exist  one horizontal non-charging road above the furthest horizontal charging road from S, as shown in Fig. \ref{E632};
            \\Event $L_{6,6} = E_{6,3,2} \cap E_{6,2,4} \cap E_{6,1,3} \cap E_6$;
            \\\textit{Probability}: $\mathbb{P}(E_{6,3,2}|E_{6,2,4},E_{6,1,3},E_6)= \frac{p - pe^{-\lambda d_v}}{1 -e^{-\lambda p d_v}}$,
            \\$\mathbb{P}(L_{6,6})=\mathbb{P}(E_{6,3,2}|E_{6,2,4},E_{6,1,3},E_6)\times
            \\\mathbb{P}(E_{6,2,4}|E_{6,1,3},E_6)\mathbb{P}(E_{6,1,3}|E_6)\mathbb{P}(E_6)$;
            \\\textit{Action}: we simply go with the furthest horizontal charging road.
        \end{itemize}
    \end{itemize}
\end{itemize}

Since the source road is already a charging road, $\mathbb{P}(D_n < x | L_{6,i}) = 1$ for all $i$.

\section{proof of Lemma \ref{DnE7}}
\label{appendxd}	
In this appendix, we first provide the complete form of Lemma~\ref{DnE7}, i.e., 
the distribution of $D_n$ given $E_7$, as follows:
\begin{align}
\label{lemmae7eqn}
&\mathbb{P}(D_n < x | E_7)\mathbb{P}(E_7) = \sum_{i=1}^{9} C_i,
\end{align}
where
\begin{align*}
&C_1 = e^{-\lambda d_v}\mathbbm{1}\{x > d_h\}\frac{p(1-p)}{2},
\\&C_2 = e^{-\lambda p d_v}(1-e^{-\lambda (1-p) d_v})e^{-\lambda p d_h}\mathbbm{1}\{x > d_h\}\frac{p(1-p)}{2},
\\&C_3 = 
\nonumber\\& \biggl(\frac{F_{D_\mathrm{N-VC}} (x)}{F_{D_\mathrm{N-VC}} (d_h)}\mathbbm{1}\{x < d_h\} + \mathbbm{1}\{x > d_h\}\biggr)(1 - e^{-\lambda p d_h})\times\\&e^{-\lambda p d_v}(1-e^{-\lambda (1-p) d_v})\frac{p(1-p)}{2},
\\&C_4 = (1-e^{-\lambda p d_v})\mathbbm{1}\{x > d_h\}\frac{p(1-p)}{2},
\\&C_5 = 
\nonumber\\& \biggl({\textstyle\frac{f_{11}({\rm max}(x-d_v,0),x,x-y)+f_{11}(0,x-d_v,d_v)\mathbbm{1}\{x > d_v\}}{f_{11}({\rm max}(d_h-d_v,0),d_h,d_h-y)+f_{11}(0,d_h-d_v,d_v)\mathbbm{1}\{d_h > d_v\}}}\times
\nonumber\\&\mathbbm{1}\{x < d_h\} + \mathbbm{1}\{x > d_h\}\biggr)\biggl(\int_{0}^{d_h} \frac{1-e^{-\lambda p (d_h-w)}}{1-e^{-\lambda p d_v}}\times\\&\frac{\lambda (1-p) e^{-\lambda (1-p) w}}{1-e^{-\lambda p d_v}} dw\biggr) e^{-\lambda p d_h}(1-e^{-\lambda (1-p) d_h})\\&\times(1-e^{-\lambda p d_v})\frac{p(1-p)}{2},
\\&C_6 = \mathbbm{1}\{x > d_h\}\biggl(1-\int_{0}^{d_h} \frac{1-e^{-\lambda p (d_h-w)}}{1-e^{-\lambda p d_v}}\times\\&\frac{\lambda (1-p) e^{-\lambda (1-p) w}}{1-e^{-\lambda p d_v}} dw\biggr) e^{-\lambda p d_h}(1-e^{-\lambda (1-p) d_h})\\&\times(1-e^{-\lambda p d_v})\frac{p(1-p)}{2},
\\&C_7 = 
\nonumber\\& \biggl(\frac{f_{12}(0,x,x,d_h,x-t)+f_{12}(0,x,0,x,y-t)}{f_{13}(0,d_h,0,d_h-t,d_h,t+y)}\mathbbm{1}\{x < d_h\} \nonumber\\&+ \mathbbm{1}\{x >  d_h\}\biggr)\biggl(1- \int_{0}^{d_v} F_{X_1}(x) f_{D_\mathrm{N-HC}(x)} {\rm d}x\biggr)\times\\&\biggl(1 - \frac{p - pe^{-\lambda d_h}}{1 -e^{-\lambda p d_h}}\biggr)(1-e^{-\lambda p d_h})(1-e^{-\lambda p d_v})\frac{p(1-p)}{2},
\\&C_8 = 
\nonumber\\& \biggl({\textstyle\frac{f_{13}(0,\infty,0,{\rm max}(x-t,0),y+t,y)-f_{13}(0,\infty,{\rm max}(x-t,0),x,x,y)}{f_{13}(0,\infty,0,{\rm max}(d_h-t,0),y+t,y) - f_{13}(0,\infty,{\rm max}(d_h-t,0),d_h,x,y)}}\times
\nonumber\\& \mathbbm{1}\{x < d_h\} + \mathbbm{1}\{x >  d_h\}\biggr)\biggl(\int_{0}^{d_v} F_{X_1}(x) f_{D_\mathrm{N-HC}(x)} {\rm d}x\biggr)\\&\times\biggl(1 - \frac{p - pe^{-\lambda d_h}}{1 -e^{-\lambda p d_h}}\biggr)(1-e^{-\lambda p d_h})(1-e^{-\lambda p d_v})\frac{p(1-p)}{2},
\\&C_9 = 
\nonumber\\& \biggl(\frac{f_{11}(0,x,y)+f_{11}(x,\infty,x)}{f_{11}(0,d_h,y)+f_{11}(d_h,\infty,d_h)}\mathbbm{1}\{x < d_h\} + \mathbbm{1}\{x > d_h\}\biggr)\\&\times\biggl(\frac{p - pe^{-\lambda d_h}}{1 -e^{-\lambda p d_h}}\biggr)(1-e^{-\lambda p d_h})(1-e^{-\lambda p d_v})\frac{p(1-p)}{2}.
\end{align*}

\begin{IEEEproof}
We prove (\ref{lemmae7eqn}) by further dividing it into subevents, as shown in Fig.~\ref{E7}, we hereby denote subevents as $E_{7,i,j}$, in which $i$ is the level of depth of the event in the probability tree and $j$ is the index of the event at that level. Representative figures for $E_7$ are shown in Fig.~\ref{e7sub}. The distribution of $D_n$ given $E_7$ can be derived as follows:
\begin{align*}
\mathbb{P}(D_n < x | E_7)\mathbb{P}(E_7) = \sum_{i=1}^{N_7} \mathbb{P}(D_n < x | L_{7,i})\mathbb{P}(L_{7,i}),
\end{align*} 
where $N_7$ denotes the number of leaves of tree $E_7$, i.e., $N_7=9$, and $L_{7,i}$'s are successive events ending at the leaves of tree $E_7$ as shown in Fig. \ref{E7}. The definition for each event $L_{7,i}$ will be given in more details as we visit each leaf of the tree.

\begin{figure}[!h]
\centering
\includegraphics[width=1\columnwidth]{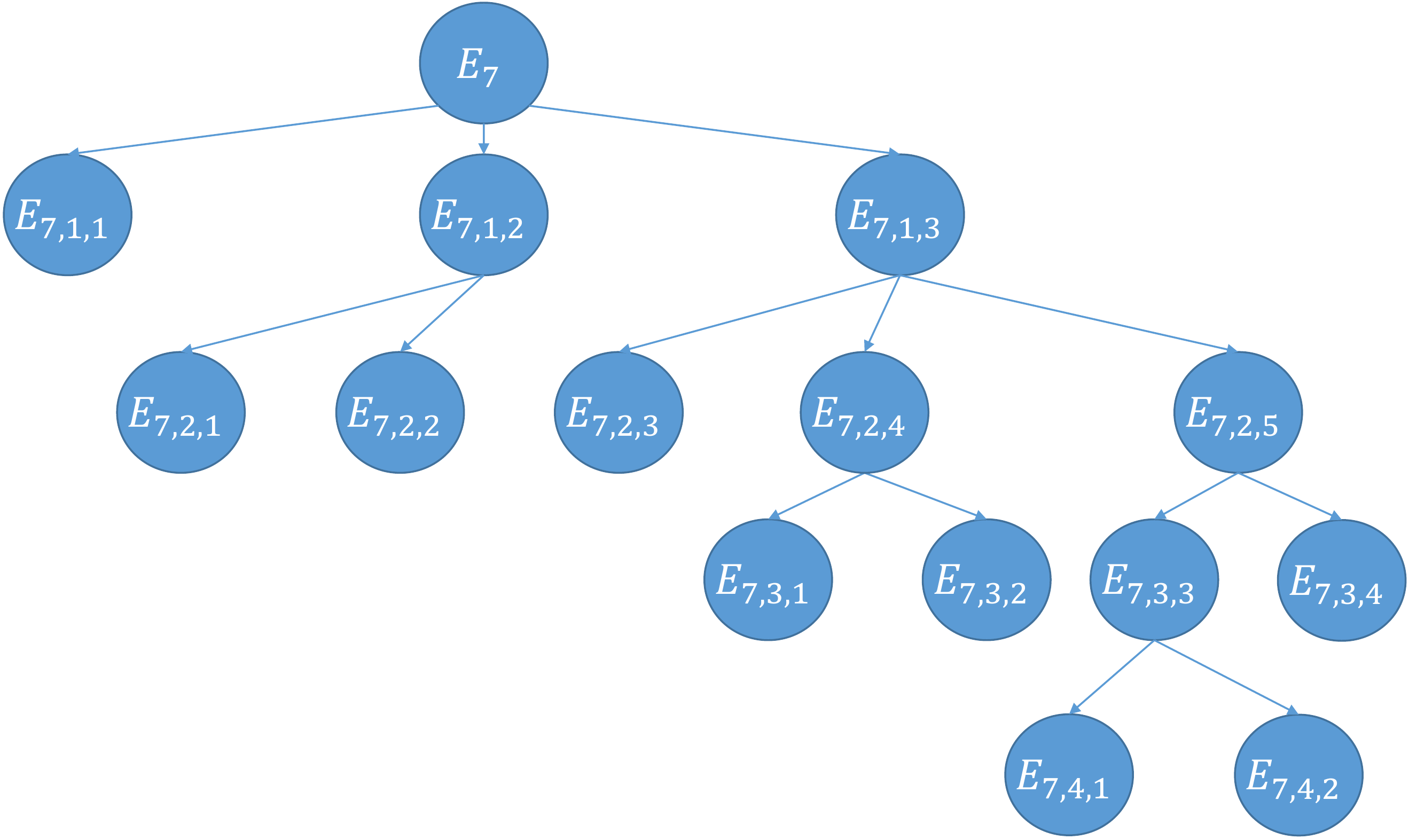}
\caption{Tree $E_7$: source and destination roads are on two perpendicular roads and the destination road is charging.}
\label{E7}
\end{figure}

%
%

\begin{figure}[h]
\centering
\captionsetup[subfigure]{font=scriptsize,labelfont=normalsize}
\subfloat[Event $E_{7,1,1}$\label{E711}]{%
  \includegraphics[width=0.33\columnwidth,keepaspectratio]{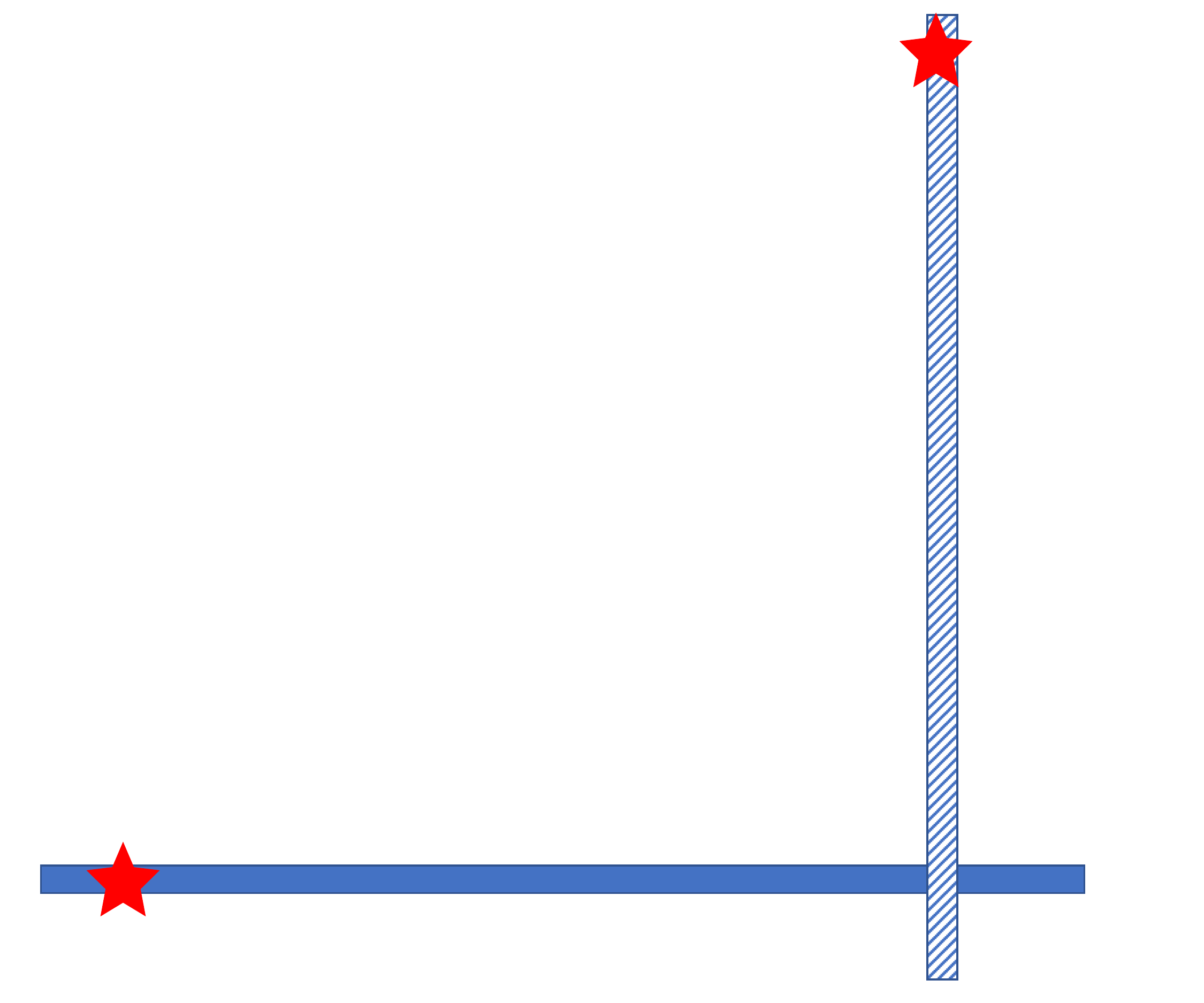}%
}\hfill
\subfloat[Event $E_{7,2,1}$\label{E721}]{%
  \includegraphics[width=0.33\columnwidth,keepaspectratio]{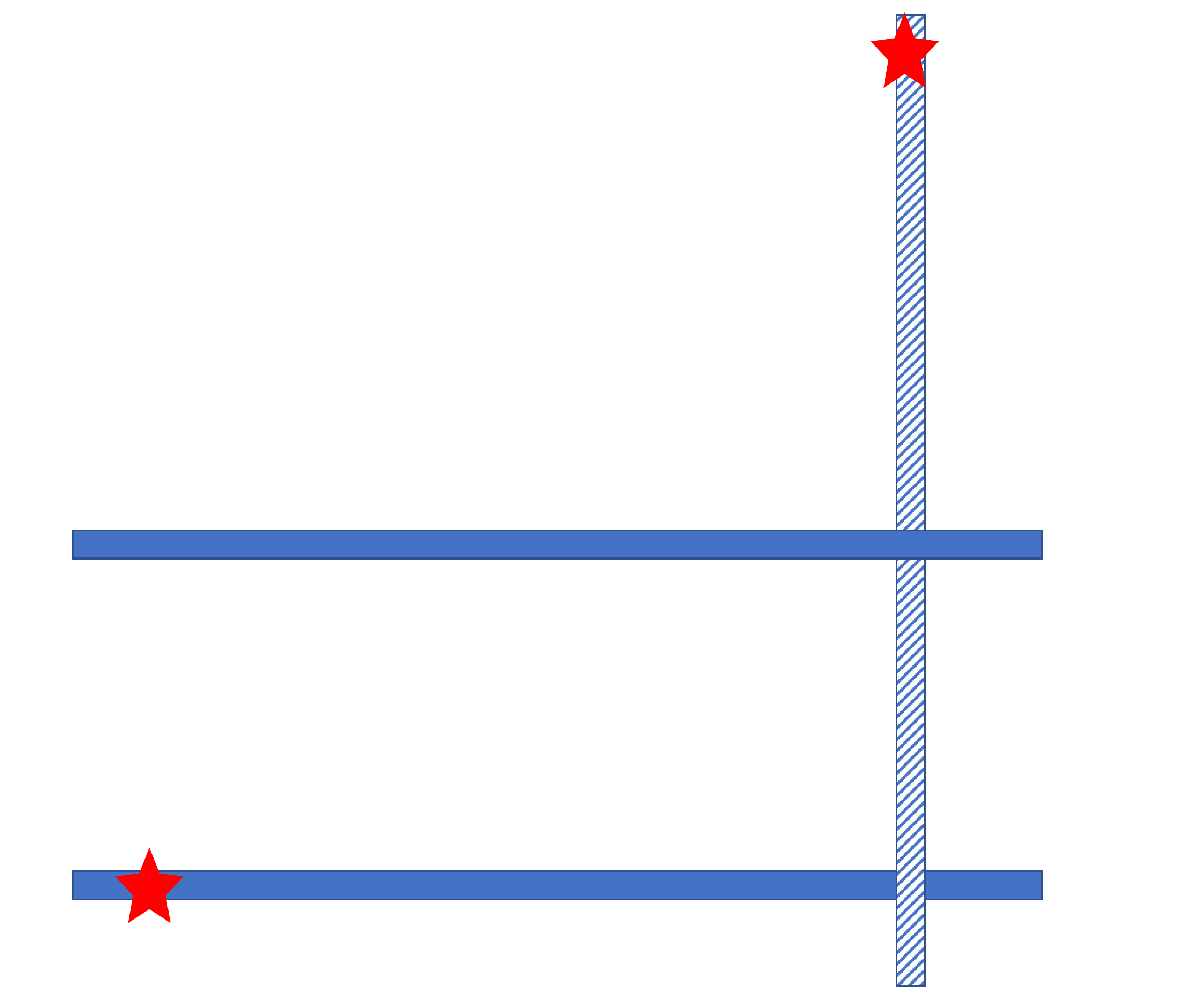}%
}
\hfill
\subfloat[Event $E_{7,2,2}$\label{E722}]{%
  \includegraphics[width=0.33\columnwidth,keepaspectratio]{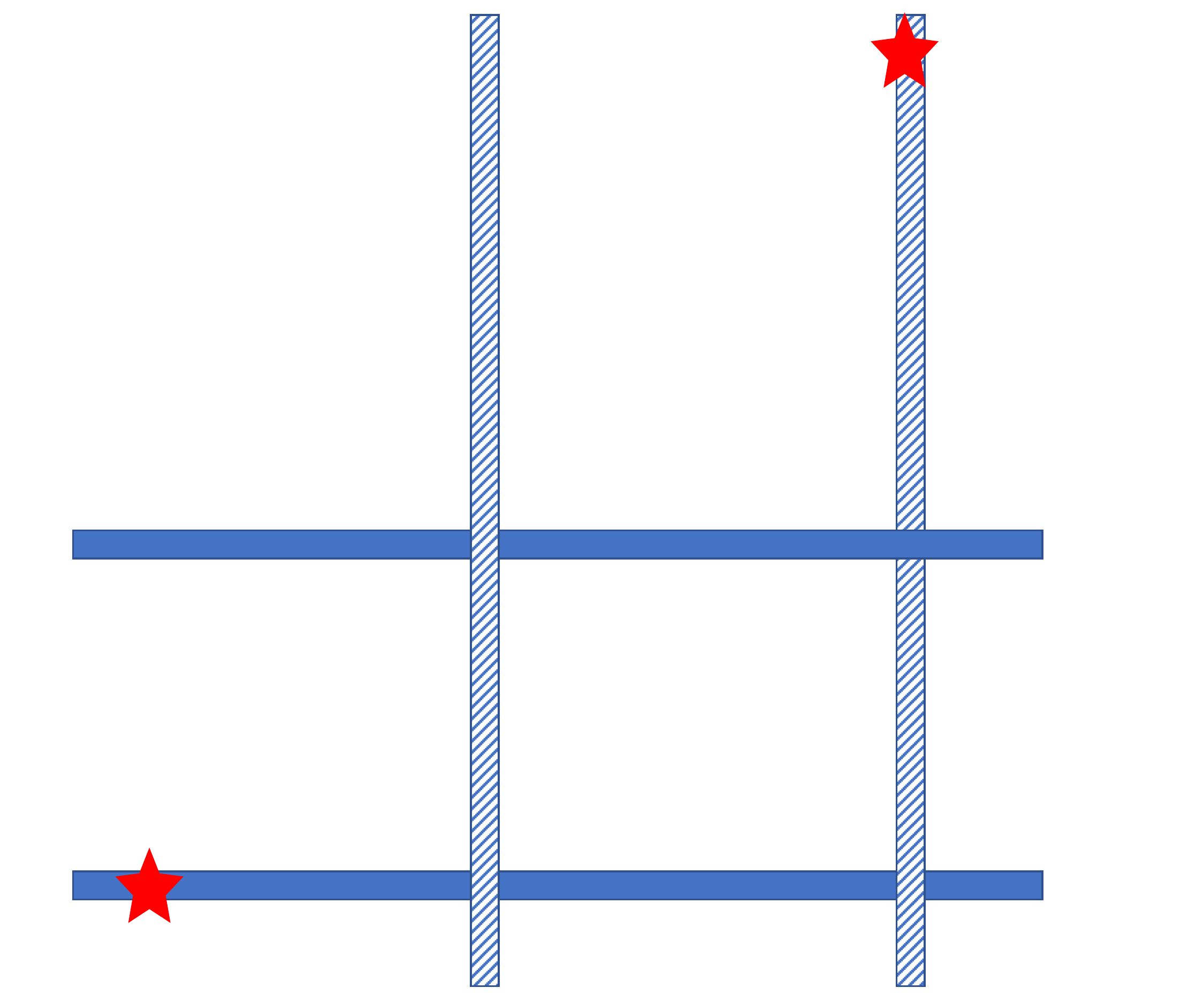}%
}

\subfloat[Event $E_{7,2,3}$\label{E723}]{%
  \includegraphics[width=0.33\columnwidth,keepaspectratio]{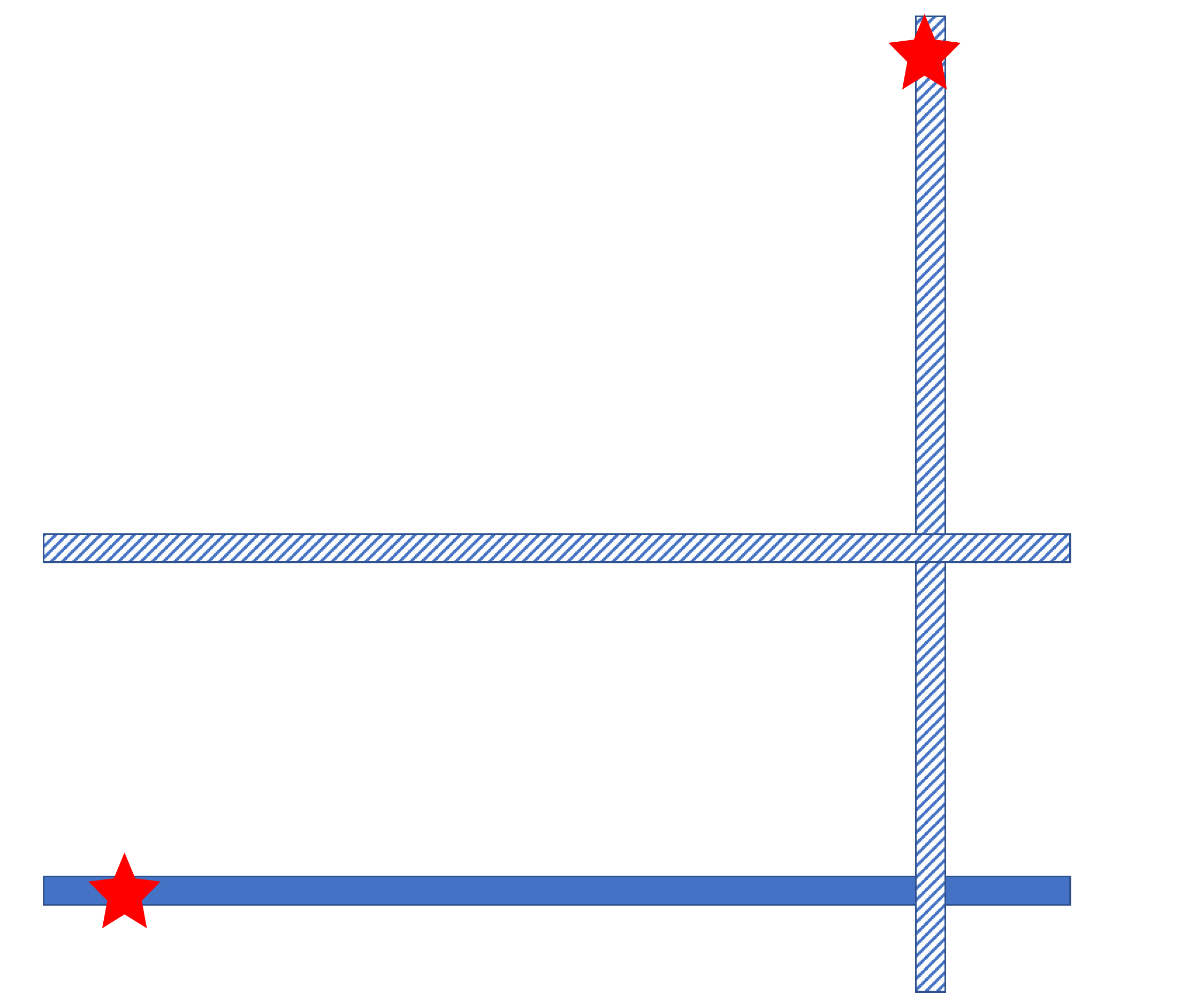}%
}\hfill
\subfloat[Event $E_{7,2,4}$\label{E724}]{%
  \includegraphics[width=0.33\columnwidth,keepaspectratio]{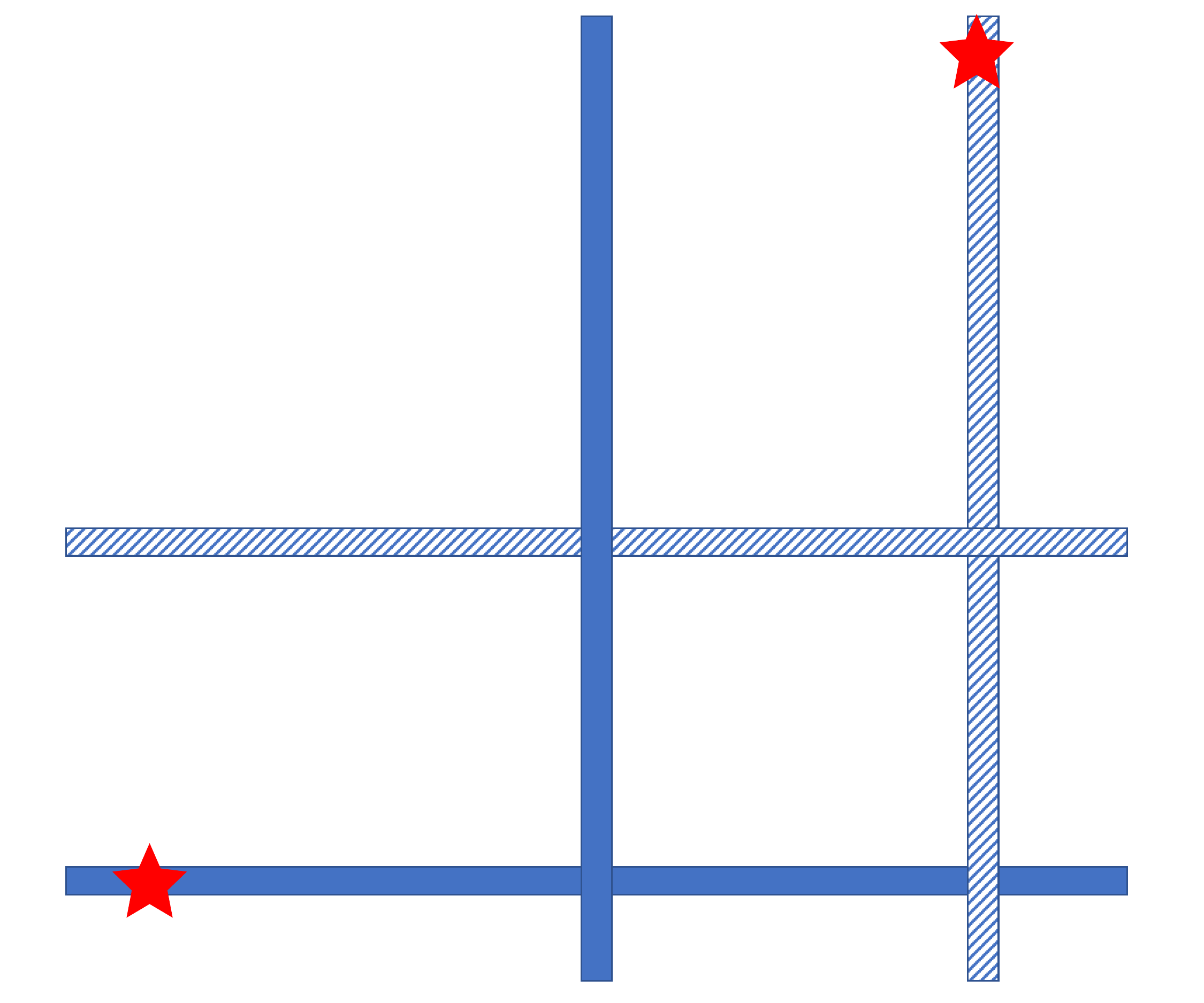}%
}
\hfill
\subfloat[Event $E_{7,3,3}$\label{E733}]{%
  \includegraphics[width=0.33\columnwidth,keepaspectratio]{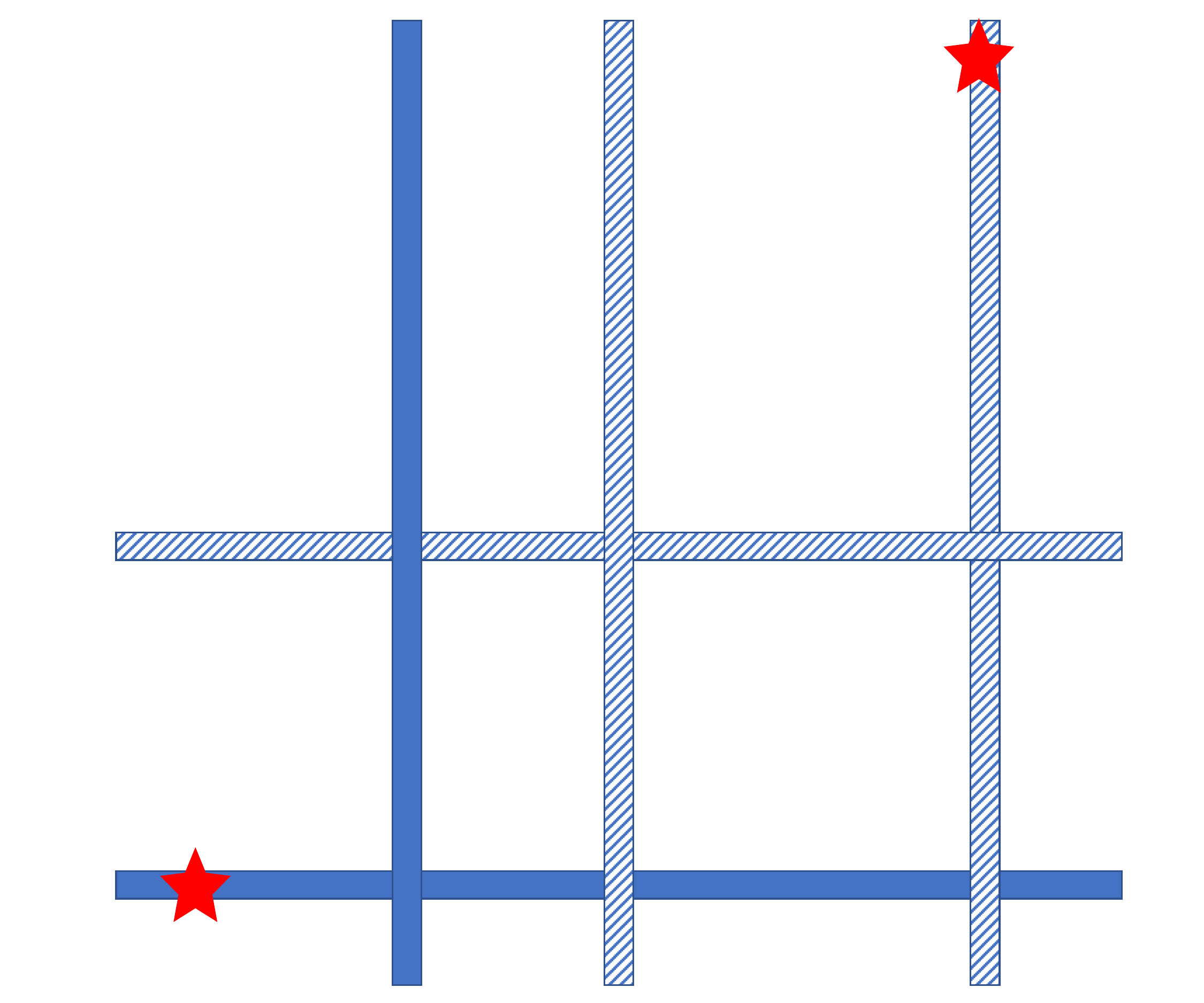}%
}

\subfloat[Event $E_{7,3,4}$\label{E734}]{%
  \includegraphics[width=0.33\columnwidth,keepaspectratio]{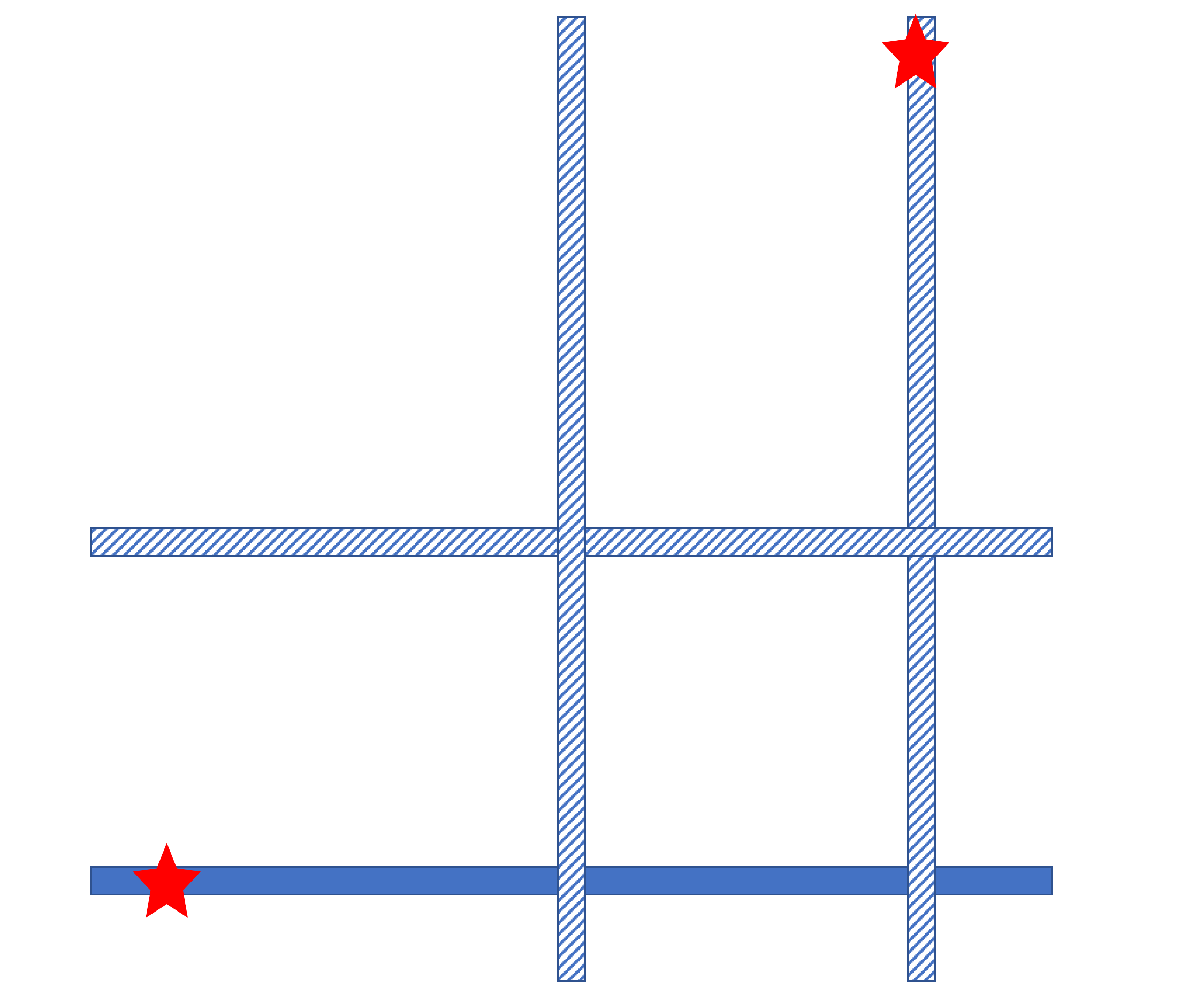}%
}
\subfloat{%
  \includegraphics[width=0.33\columnwidth,keepaspectratio]{dnx_legend_1.pdf}%
}

\caption{Subcases of tree $E_7$.}
\label{e7sub}
\end{figure}

\begin{itemize}[wide, labelwidth=!, labelindent=0pt]
\item Event $E_{7,1,1}$: \\\textit{Description}: If there are no horizontal roads between S and D, as shown in Fig. \ref{E711}; 
\\Event $L_{7,1} = E_{7,1,1} \cap E_7$;
\\\textit{Probability}: $\mathbb{P}(E_{7,1,1}|E_7)=e^{-\lambda d_v}$, $\mathbb{P}(L_{7,1})=\mathbb{P}(E_{7,1,1}|E_7)\mathbb{P}(E_7)$; 
\\\textit{Action}: We simply take the source road then the destination road. $$\mathbb{P}(D_n < x|L_{7,1})=\mathbb{P}(D_n < x|E_{7,1,1},E_7) = \mathbbm{1}\{x > d_h\}.$$
\item Event $E_{7,1,2}$: \\\textit{Description}: If there are no horizontal charging roads but at least one horizontal non-charging road between S and D; \\\textit{Probability}: $\mathbb{P}(E_{7,1,2}|E_7)=e^{-\lambda p d_v}(1-e^{-\lambda (1-p) d_v})$.
	\begin{itemize}[wide, labelwidth=!, labelindent=0pt]
	\item Event $E_{7,2,1}$: \\\textit{Description}: If there are no vertical charging roads between S and D, as shown in Fig. \ref{E721}; 
	\\Event $L_{7,2} = E_{7,2,1} \cap E_{7,1,2} \cap E_7$;
	\\\textit{Probability}: $\mathbb{P}(E_{7,2,1}|E_{7,1,2},E_7)=e^{-\lambda p d_h}$,
	$\mathbb{P}(L_{7,2})=\mathbb{P}(E_{7,2,1}|E_{7,1,2},E_7)\mathbb{P}(E_{7,1,2}|E_7)\mathbb{P}(E_7)$; 
	\\\textit{Action}: We simply take the source road then the destination road. $\mathbb{P}(D_n < x|L_{7,2})=\mathbb{P}(D_n < x|E_{7,2,1},E_{7,1,2},E_7) = \mathbbm{1}\{x > d_h\}.$
	\item Event $E_{7,2,2}$: \\\textit{Description}: If there is at least one vertical charging road between S and D, as shown in Fig. \ref{E722}; 
	\\Event $L_{7,3} = E_{7,2,2} \cap E_{7,1,2} \cap E_7$;
	\\\textit{Probability}: $\mathbb{P}(E_{7,2,2}|E_{7,1,2},E_7)=1 - e^{-\lambda p d_h}$,
	\\$\mathbb{P}(L_{7,3})=\mathbb{P}(E_{7,2,2}|E_{7,1,2},E_7)\mathbb{P}(E_{7,1,2}|E_7)\mathbb{P}(E_7)$;
	\\\textit{Action}: We take the nearest vertical charging road.
	    \begin{align}
    \label{722}
        &\mathbb{P}(D_n < x|L_{7,3})=\mathbb{P}(D_n < x|E_{7,2,2},E_{7,1,2},E_7) 
        \nonumber\\&= \mathbb{P}(D_n < x |D_\mathrm{N-HNC} < d_v, D_\mathrm{N-HC} > d_v, D_\mathrm{N-VC} < d_h)
        \nonumber\\&= \mathbb{P}(D_\mathrm{N-VC} < x |D_\mathrm{N-HNC} < d_v, D_\mathrm{N-HC} > d_v,\nonumber\\& D_\mathrm{N-VC} < d_h)\mathbbm{1}\{x < d_h\} + \mathbbm{1}\{x > d_h\}
        \nonumber\\&= \frac{F_{D_\mathrm{N-VC}} (x)}{F_{D_\mathrm{N-VC}} (d_h)}\mathbbm{1}\{x < d_h\} + \mathbbm{1}\{x > d_h\}
    .
    \end{align} 
	\end{itemize}

\item Event $E_{7,1,3}$: \\\textit{Description}: If there is at least one horizontal charging road between S and D; \\\textit{Probability}: $\mathbb{P}(E_{7,1,3}|E_7)=1-e^{-\lambda p d_v}$. 
\begin{itemize}[wide, labelwidth=!, labelindent=0pt]
    \item Event $E_{7,2,3}$: \\\textit{Description}: If there are no vertical roads between S and D, as shown in Fig. \ref{E723}; 
    \\Event $L_{7,4} = E_{7,2,3} \cap E_{7,1,3} \cap E_7$;
    \\\textit{Probability}: $\mathbb{P}(E_{7,2,3}|E_{7,1,3},E_7)=e^{-\lambda d_h}$,
    \\$\mathbb{P}(L_{7,4})=\mathbb{P}(E_{7,2,3}|E_{7,1,3},E_7)\mathbb{P}(E_{7,1,3}|E_7)\mathbb{P}(E_7)$;
    \\\textit{Action}: We simply take the source road then the destination road. 
    $\mathbb{P}(D_n < x|L_{7,4})=\mathbb{P}(D_n < x|E_{7,2,3},E_{7,1,3},E_7) = \mathbbm{1}\{x > d_h\}.$
    \item Event $E_{7,2,4}$: \\\textit{Description}: If there are no vertical charging roads but at least one vertical non-charging road between S and D, as shown in Fig. \ref{E724}; \\\textit{Probability}: $\mathbb{P}(E_{7,2,4}|E_{7,1,3},E_7)=e^{-\lambda p d_h}(1-e^{-\lambda (1-p) d_h})$; \\\textit{Action}: We compare (i) the distance from the nearest vertical non-charging road to destination and (ii) the distance from source to the nearest horizontal charging road, to take the longer one.
    \begin{itemize}[wide, labelwidth=!, labelindent=0pt]
        \item Event $E_{7,3,1}$: \\\textit{Description}: If we take the horizontal charging road; 
        \\Event $L_{7,5} = E_{7,3,1} \cap E_{7,2,4} \cap E_{7,1,3} \cap E_7$;
        \\\textit{Probability}: $\mathbb{P}(E_{7,3,1}|E_{7,2,4},E_{7,1,3},E_7)=\int_{0}^{d_h} \frac{1-e^{-\lambda p (d_h-w)}}{1-e^{-\lambda p d_v}}\frac{\lambda (1-p) e^{-\lambda (1-p) w}}{1-e^{-\lambda p d_v}} dw$,
        \\$\mathbb{P}(L_{7,5})=\mathbb{P}(E_{7,3,1}|E_{7,2,4},E_{7,1,3},E_7)\times
        \\\mathbb{P}(E_{7,2,4}|E_{7,1,3},E_7)\mathbb{P}(E_{7,1,3}|E_7)\mathbb{P}(E_7)$;
            \begin{align}
    \label{731}
        &\mathbb{P}(D_n < x|L_{7,5})=\mathbb{P}(D_n < x|E_{7,3,1},E_{7,2,4},E_{7,1,3},E_7) 
        \nonumber\\&= \mathbb{P}(D_n < x |D_\mathrm{N-HC} < d_v, D_\mathrm{N-VNC} < d_h, D_\mathrm{N-VC} > d_h,\nonumber\\& d_h - D_\mathrm{N-VNC} > D_\mathrm{N-HC})
        \nonumber\\&= \mathbb{P}(D_\mathrm{N-VNC} + D_\mathrm{N-HC} < x |D_\mathrm{N-HC} < d_v, D_\mathrm{N-VNC} < d_h, \nonumber\\& D_\mathrm{N-VC} > d_h, d_h - D_\mathrm{N-VNC} > D_\mathrm{N-HC})=
        \nonumber\\& {\scriptstyle\frac{\mathbb{P}(D_\mathrm{N-HC} < {\rm min}(x - D_\mathrm{N-VNC}, d_v, d_h - D_\mathrm{N-VNC}), D_\mathrm{N-VNC} < d_h, D_\mathrm{N-VC} > d_h)}{\mathbb{P}(D_\mathrm{N-HC} < {\rm min}(d_v, d_h - D_\mathrm{N-VNC}), D_\mathrm{N-VNC} < d_h, D_\mathrm{N-VC} > d_h)}}
		\nonumber\\& = {\textstyle\frac{f_{11}({\rm max}(x-d_v,0),x,x-y)+f_{11}(0,x-d_v,d_v)\mathbbm{1}\{x > d_v\}}{f_{11}({\rm max}(d_h-d_v,0),d_h,d_h-y)+f_{11}(0,d_h-d_v,d_v)\mathbbm{1}\{d_h > d_v\}}}\times
		\nonumber\\&\mathbbm{1}\{x < d_h\} + \mathbbm{1}\{x > d_h\}
    .
    \end{align} 
        
        \item Event $E_{7,3,2}$: \\\textit{Description}: If we take the destination vertical road; 
        \\Event $L_{7,6} = E_{7,3,2} \cap E_{7,2,4} \cap E_{7,1,3} \cap E_7$;
        \\\textit{Probability}: $\mathbb{P}(E_{7,3,2}|E_{7,2,4},E_{7,1,3},E_7)=1 - \int_{0}^{d_h} \frac{1-e^{-\lambda p (d_h-w)}}{1-e^{-\lambda p d_v}}\frac{\lambda (1-p) e^{-\lambda (1-p) w}}{1-e^{-\lambda p d_v}} dw$,
        \\$\mathbb{P}(L_{7,6})=\mathbb{P}(E_{7,3,2}|E_{7,2,4},E_{7,1,3},E_7)\times
        \\\mathbb{P}(E_{7,2,4}|E_{7,1,3},E_7)\mathbb{P}(E_{7,1,3}|E_7)\mathbb{P}(E_7)$;
        $\mathbb{P}(D_n < x|L_{7,6})=\mathbb{P}(D_n < x|E_{7,3,2},E_{7,2,4},E_{7,1,3},E_7) = \mathbbm{1}\{x > d_h\}.$
    \end{itemize}
    \item Event $E_{7,2,5}$: \\\textit{Description}: If there is at least one vertical charging road between S and D; \\\textit{Probability}: $\mathbb{P}(E_{7,2,5}|E_{7,1,3},E_7)=1-e^{-\lambda p d_h}$. 
    \begin{itemize}[wide, labelwidth=!, labelindent=0pt]
        \item Event $E_{7,3,3}$: \\\textit{Description}: If there exists at least one vertical non-charging road before the nearest vertical charging road from source, as shown in Fig. \ref{E733}; \\\textit{Probability}: $\mathbb{P}(E_{7,3,3}|E_{7,2,5},E_{7,1,3},E_7)=1 - \frac{p - pe^{-\lambda d_h}}{1 -e^{-\lambda p d_h}}$; \\\textit{Action}: we compare (i) the distance between the nearest vertical charging road and the nearest vertical non-charging road, and (ii) the vertical distance between the nearest horizontal charging road and source, to take the longer one. 
        \begin{itemize}[wide, labelwidth=!, labelindent=0pt]
            \item Event $E_{7,4,1}$: \\\textit{Description}: If we take the nearest horizontal charging road; 
            \\Event $L_{7,7} = E_{7,4,1} \cap E_{7,3,3} \cap E_{7,2,5} \cap E_{7,1,3} \cap E_7$;
            \\\textit{Probability}: $\mathbb{P}(E_{7,4,1}|E_{7,3,3},E_{7,2,5},E_{7,1,3},E_7)=1- \int_{0}^{d_v} F_{X_1}(x) f_{D_\mathrm{N-HC}(x)} {\rm d}x$;
            \\$\mathbb{P}(L_{7,7})=\mathbb{P}(E_{7,4,1}|E_{7,3,3},E_{7,2,5},E_{7,1,3},E_7)\times
            \\\mathbb{P}(E_{7,3,3}|E_{7,2,5},E_{7,1,3},E_7)\times\\\mathbb{P}(E_{7,2,5}|E_{7,1,3},E_7)\mathbb{P}(E_{7,1,3}|E_7)\mathbb{P}(E_7)$;
            	\begin{align}
	\label{741}
                &\mathbb{P}(D_n < x|L_{7,7})=\mathbb{P}(D_n < x|E_{7,4,1},E_{7,3,3},E_{7,2,5},E_{7,1,3},E_7) 
                \nonumber\\&= \mathbb{P}(D_\mathrm{N-VNC} + D_\mathrm{N-HC} < x|D_\mathrm{N-VC} < d_h, \nonumber\\& D_\mathrm{N-VNC} < D_\mathrm{N-VC}, D_\mathrm{N-VC} - D_\mathrm{N-VNC} > D_\mathrm{N-HC})\times\nonumber\\&\mathbbm{1}\{x < d_h\} + \mathbbm{1}\{x >  d_h\}
				\nonumber\\& = \frac{f_{12}(0,x,x,d_h,x-t)+f_{12}(0,x,0,x,y-t)}{f_{13}(0,d_h,0,d_h-t,d_h,t+y)}\mathbbm{1}\{x < d_h\} \nonumber\\&+ \mathbbm{1}\{x >  d_h\}
                .
                \end{align}    
    The proof for $\mathbb{P}(D_n < x|E_{7,4,1},E_{7,3,3},E_{7,2,5},E_{7,1,3},E_{7})$ is similar to that of \\ $\mathbb{P}(D_n < x|E_{3,4,3},E_{3,3,3},E_{3,2,4},E_{3,1,4},E_{3})$ given in (\ref{343}).
            
            \item Event $E_{7,4,2}$: \\\textit{Description}: If we take the nearest vertical charging road; 
            \\Event $L_{7,8} = E_{7,4,2} \cap E_{7,3,3} \cap E_{7,2,5} \cap E_{7,1,3} \cap E_7$;
            \\\textit{Probability}: $\mathbb{P}(E_{7,4,2}|E_{7,3,3},E_{7,2,5},E_{7,1,3},E_7)= \int_{0}^{d_v} F_{X_1}(x) f_{D_\mathrm{N-HC}(x)} {\rm d}x$;
            \\$\mathbb{P}(L_{7,8})=\mathbb{P}(E_{7,4,2}|E_{7,3,3},E_{7,2,5},E_{7,1,3},E_7)\times
            \\\mathbb{P}(E_{7,3,3}|E_{7,2,5},E_{7,1,3},E_7)\times\\\mathbb{P}(E_{7,2,5}|E_{7,1,3},E_7)\mathbb{P}(E_{7,1,3}|E_7)\mathbb{P}(E_7)$;
                \begin{align}
    \label{742}
                    &\mathbb{P}(D_n < x|L_{7,8})=\mathbb{P}(D_n < x|E_{7,4,2},E_{7,3,3},E_{7,2,5},E_{7,1,3},E_7) 
                    \nonumber\\&= \mathbb{P}(D_\mathrm{N-VC} < x|D_\mathrm{N-VC} < d_h, D_\mathrm{N-VNC} < D_\mathrm{N-VC}, \nonumber\\& D_\mathrm{N-VC} - D_\mathrm{N-VNC} < D_\mathrm{N-HC})\mathbbm{1}\{x < d_h\} + \mathbbm{1}\{x >  d_h\}
					\nonumber\\& = {\textstyle\frac{f_{13}(0,\infty,0,{\rm max}(x-t,0),y+t,y)-f_{13}(0,\infty,{\rm max}(x-t,0),x,x,y)}{f_{13}(0,\infty,0,{\rm max}(d_h-t,0),y+t,y) - f_{13}(0,\infty,{\rm max}(d_h-t,0),d_h,x,y)}}\times
					\nonumber\\& \mathbbm{1}\{x < d_h\} + \mathbbm{1}\{x >  d_h\}
                .
                \end{align}
    The proof for $\mathbb{P}(D_n < x|E_{7,4,2},E_{7,3,3},E_{7,2,5},E_{7,1,3},E_{7})$ is similar to that of \\ $\mathbb{P}(D_n < x|E_{3,4,4},E_{3,3,3},E_{3,2,4},E_{3,1,4},E_{3})$ given in (\ref{344}). 
            
        \end{itemize}
        \item Event $E_{7,3,4}$: \\\textit{Description}: If there exists no vertical non-charging road before the nearest vertical charging road from source, as shown in Fig. \ref{E734}; 
        \\Event $L_{7,9} =  E_{7,3,4} \cap E_{7,2,5} \cap E_{7,1,3} \cap E_7$;
        \\\textit{Probability}: $\mathbb{P}(E_{7,3,4}|E_{7,2,5},E_{7,1,3},E_7)=\frac{p - pe^{-\lambda d_h}}{1 -e^{-\lambda p d_v}}$,
        \\$\mathbb{P}(L_{7,9})=\mathbb{P}(E_{7,3,4}|E_{7,2,5},E_{7,1,3},E_7)\times
        \\\mathbb{P}(E_{7,2,5}|E_{7,1,3},E_7)\mathbb{P}(E_{7,1,3}|E_7)\mathbb{P}(E_7)$;
        \\\textit{Action}: we simply go with the nearest vertical charging road. 
        	\begin{align}
	\label{734}
                    &\mathbb{P}(D_n < x|L_{7,9})=\mathbb{P}(D_n < x|E_{7,3,4},E_{7,2,5},E_{7,1,3},E_7) 
                    \nonumber\\&= \mathbb{P}(D_\mathrm{N-VC} < x|D_\mathrm{N-VC} < D_\mathrm{N-VNC}, D_\mathrm{N-VC} < d_h) 
                    \nonumber\\&= \frac{\mathbb{P}(D_\mathrm{N-VC} <{\rm min}(x, D_\mathrm{N-VNC})}{\mathbb{P}(D_\mathrm{N-VC}<{\rm min}(D_\mathrm{N-VNC}, d_h))}\mathbbm{1}\{x < d_h\} + \nonumber\\& \mathbbm{1}\{x > d_h\}
					\nonumber\\& = \frac{f_{11}(0,x,y)+f_{11}(x,\infty,x)}{f_{11}(0,d_h,y)+f_{11}(d_h,\infty,d_h)}\mathbbm{1}\{x < d_h\} + \mathbbm{1}\{x > d_h\}
                .
            \end{align}    
    The proof for $\mathbb{P}(D_n < x|E_{7,3,4},E_{7,2,5},E_{7,1,3},E_{7})$ is similar to that of $\mathbb{P}(D_n < x|E_{3,3,4},E_{3,2,4},E_{3,1,4},E_{3})$ given in (\ref{334}).
        
    \end{itemize}
\end{itemize}

\end{itemize}
\end{IEEEproof}
 
\section{proof of Lemma \ref{DnE8}}
\label{appendxe}  
In this appendix, we first provide the complete form of Lemma~\ref{DnE8}, i.e., 
the distribution of $D_n$ given $E_8$, as follows:  
\begin{align}
\label{lemmae8eqn}
&\mathbb{P}(D_n < x | E_8)\mathbb{P}(E_8) = \sum_{i=1}^{5} C_i,
\end{align}
where
\begin{align*}
&C_1 = 
\nonumber\\& \biggl(\frac{F_{D_\mathrm{N-VC}} (x)}{F_{D_\mathrm{N-VC}} (d_h)}\mathbbm{1}\{x < d_h\} + \mathbbm{1}\{x > d_h\}\biggr)(1 - e^{-\lambda p d_h})\times\\&e^{-\lambda p d_v}(1-e^{-\lambda (1-p) d_v})\frac{(1-p)^2}{2},
\\&C_2 = 
\nonumber\\& \biggl({\textstyle\frac{f_{11}({\rm max}(x-d_v,0),{\rm min}(x,d_h),x-o)+f_{11}(0,{\rm min}(x-d_v,d_h),d_v)\mathbbm{1}\{x > d_v\}}{F_{D_\mathrm{N-HC}} (d_v)F_{D_\mathrm{N-VNC}} (d_h)}} \nonumber\\& \times\mathbbm{1}\{x < d_h + d_v\} + \mathbbm{1}\{x > d_h + d_v\}\biggr) e^{-\lambda p d_h}\times \\&(1-e^{-\lambda (1-p) d_h})(1-e^{-\lambda p d_v})\frac{(1-p)^2}{2},
\\&C_3 = 
\nonumber\\& \biggl(\frac{f_{12}(0,x,x,d_h,x-t)+f_{12}(0,x,0,x,y-t)}{f_{13}(0,d_h,0,d_h-t,d_h,t+y)}\mathbbm{1}\{x < d_h\} \nonumber\\&+ \mathbbm{1}\{x >  d_h\}\biggr)\biggl(1- \int_{0}^{d_v} F_{X_1}(x) f_{D_\mathrm{N-HC}(x)} {\rm d}x\biggr)\times\\&\biggl(1 - \frac{p - pe^{-\lambda d_h}}{1 -e^{-\lambda p d_h}}\biggr)(1-e^{-\lambda p d_h})(1-e^{-\lambda p d_v})\frac{(1-p)^2}{2},
\\&C_4 = 
\nonumber\\& \biggl({\textstyle\frac{f_{13}(0,\infty,0,{\rm max}(x-t,0),y+t,y)-f_{13}(0,\infty,{\rm max}(x-t,0),x,x,y)}{f_{13}(0,\infty,0,{\rm max}(d_h-t,0),y+t,y) - f_{13}(0,\infty,{\rm max}(d_h-t,0),d_h,x,y)}}\times
\nonumber\\& \mathbbm{1}\{x < d_h\} + \mathbbm{1}\{x >  d_h\}\biggr)\biggl(\int_{0}^{d_v} F_{X_1}(x) f_{D_\mathrm{N-HC}(x)} {\rm d}x\biggr)\\&\times\biggl(1 - \frac{p - pe^{-\lambda d_h}}{1 -e^{-\lambda p d_h}}\biggr)(1-e^{-\lambda p d_h})(1-e^{-\lambda p d_v})\frac{(1-p)^2}{2},
\\&C_5 = 
\nonumber\\& \biggl(\frac{f_{11}(0,x,y)+f_{11}(x,\infty,x)}{f_{11}(0,d_h,y)+f_{11}(d_h,\infty,d_h)}\mathbbm{1}\{x < d_h\} + \mathbbm{1}\{x > d_h\}\biggr)\\&\times\biggl(\frac{p - pe^{-\lambda d_h}}{1 -e^{-\lambda p d_h}}\biggr)(1-e^{-\lambda p d_h})(1-e^{-\lambda p d_v})\frac{(1-p)^2}{2}.
\end{align*}

\begin{IEEEproof}
We prove (\ref{lemmae8eqn}) by further dividing it into subevents, as shown in Fig.~\ref{E8}, we hereby denote subevents as $E_{8,i,j}$, in which $i$ is the level of depth of the event in the probability tree and $j$ is the index of the event at that level. Representative figures for $E_8$ are shown in Fig.~\ref{e8sub}. The distribution of $D_n$ given $E_8$ can be derived as follows:

\begin{align*}
\mathbb{P}(D_n < x | E_8)\mathbb{P}(E_8) = \sum_{i=1}^{N_8} \mathbb{P}(D_n < x | L_{8,i})\mathbb{P}(L_{8,i}),
\end{align*} 
where $N_8$ denotes the number of leaves of tree $E_8$, i.e., $N_8=8$, and $L_{8,i}$'s are successive events ending at the leaves of tree $E_8$ as shown in Fig. \ref{E8}. The definition for each event $L_{8,i}$ will be given in more details as we visit each leaf of the tree.

\begin{figure}[!h]
\centering
\includegraphics[width=1\columnwidth]{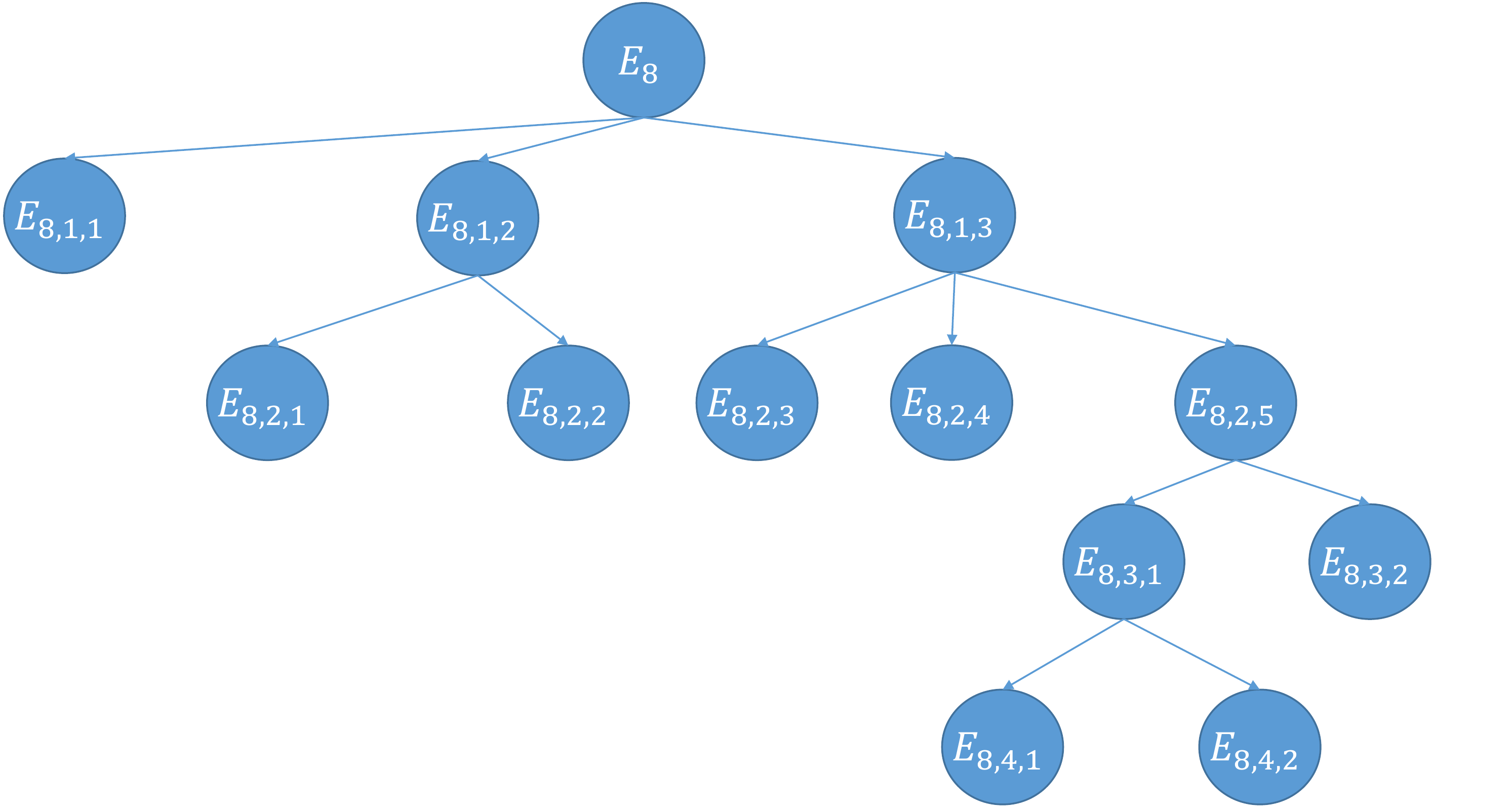}
\caption{Tree $E_8$: source and destination roads are on two perpendicular roads and both are not charging.}
\label{E8}
\end{figure}

\begin{figure}[h]
\centering
\captionsetup[subfigure]{font=scriptsize,labelfont=normalsize}
\subfloat[Event $E_{8,1,1}$\label{E811}]{%
  \includegraphics[width=0.33\columnwidth,keepaspectratio]{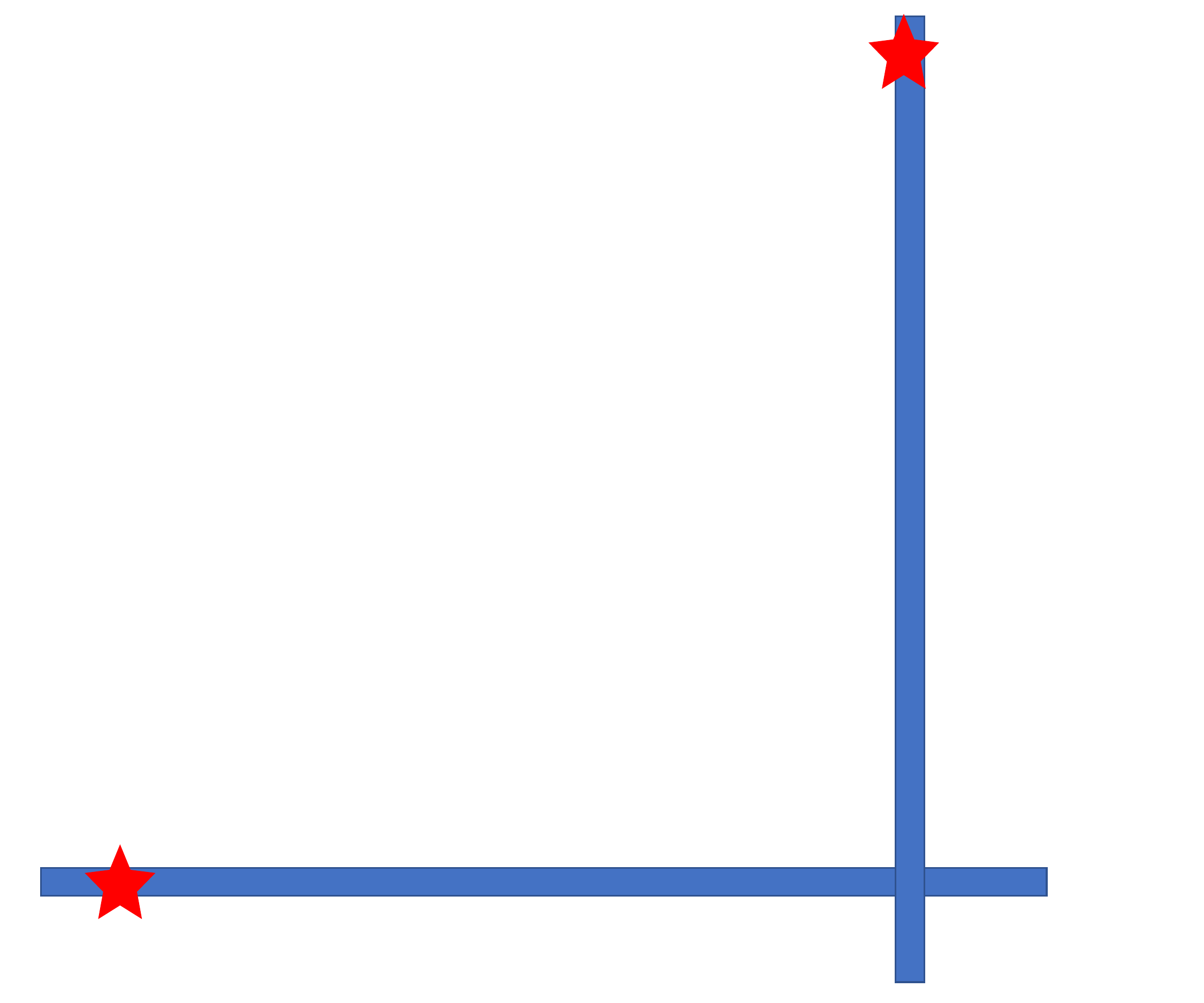}%
}\hfill
\subfloat[Event $E_{8,2,1}$\label{E821}]{%
  \includegraphics[width=0.33\columnwidth,keepaspectratio]{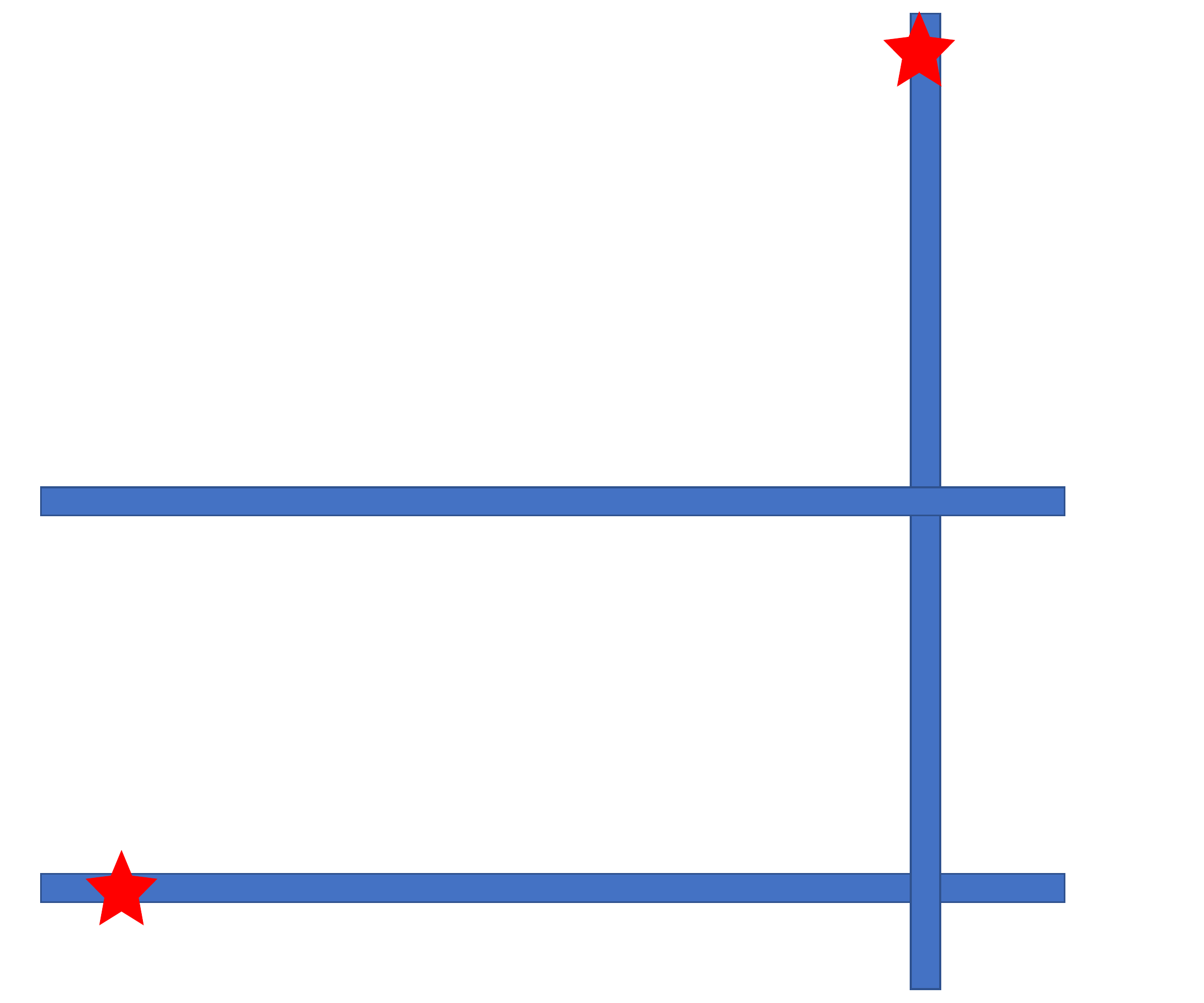}%
}
\hfill
\subfloat[Event $E_{8,2,2}$\label{E822}]{%
  \includegraphics[width=0.33\columnwidth,keepaspectratio]{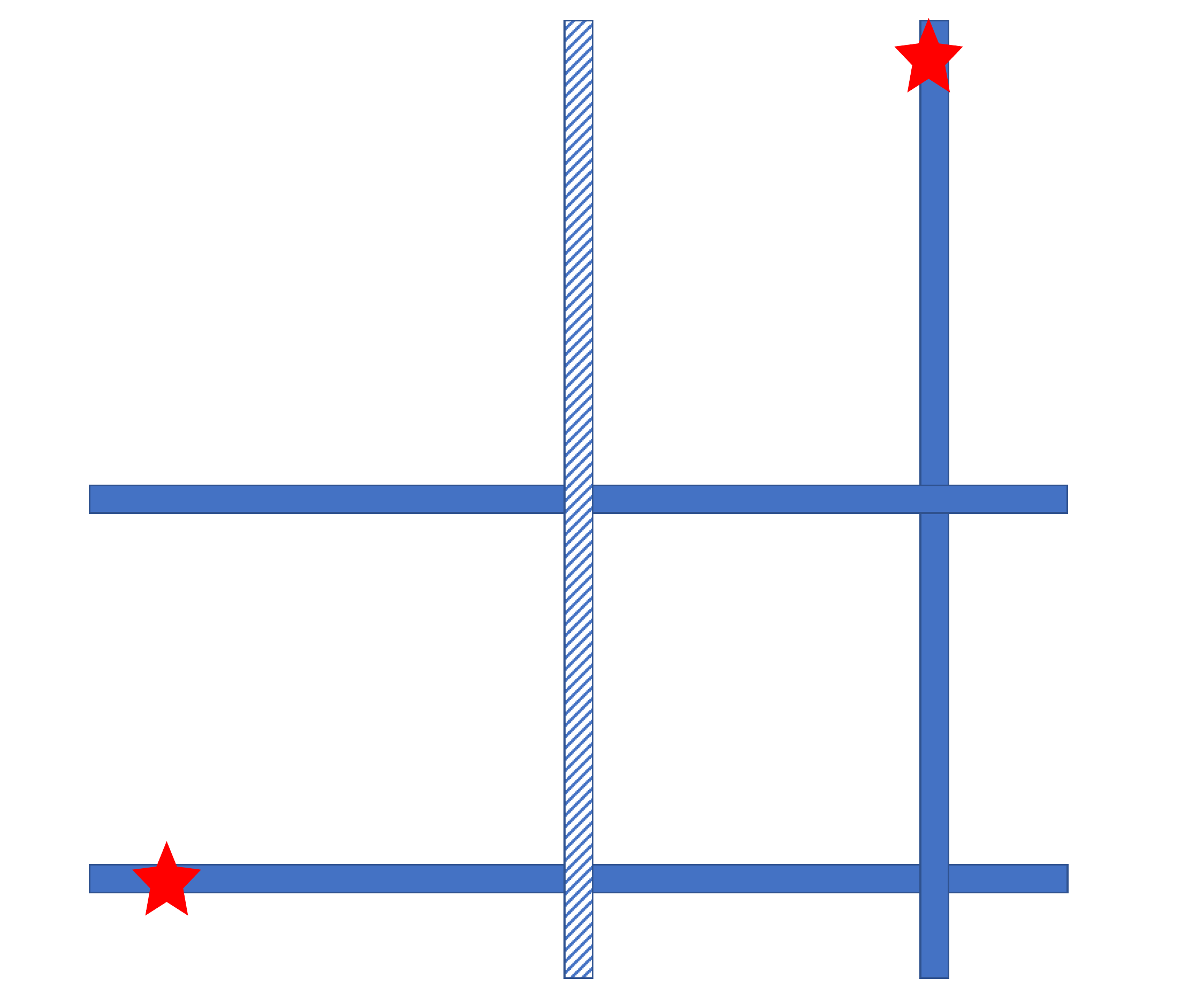}%
}

\subfloat[Event $E_{8,2,3}$\label{E823}]{%
  \includegraphics[width=0.33\columnwidth,keepaspectratio]{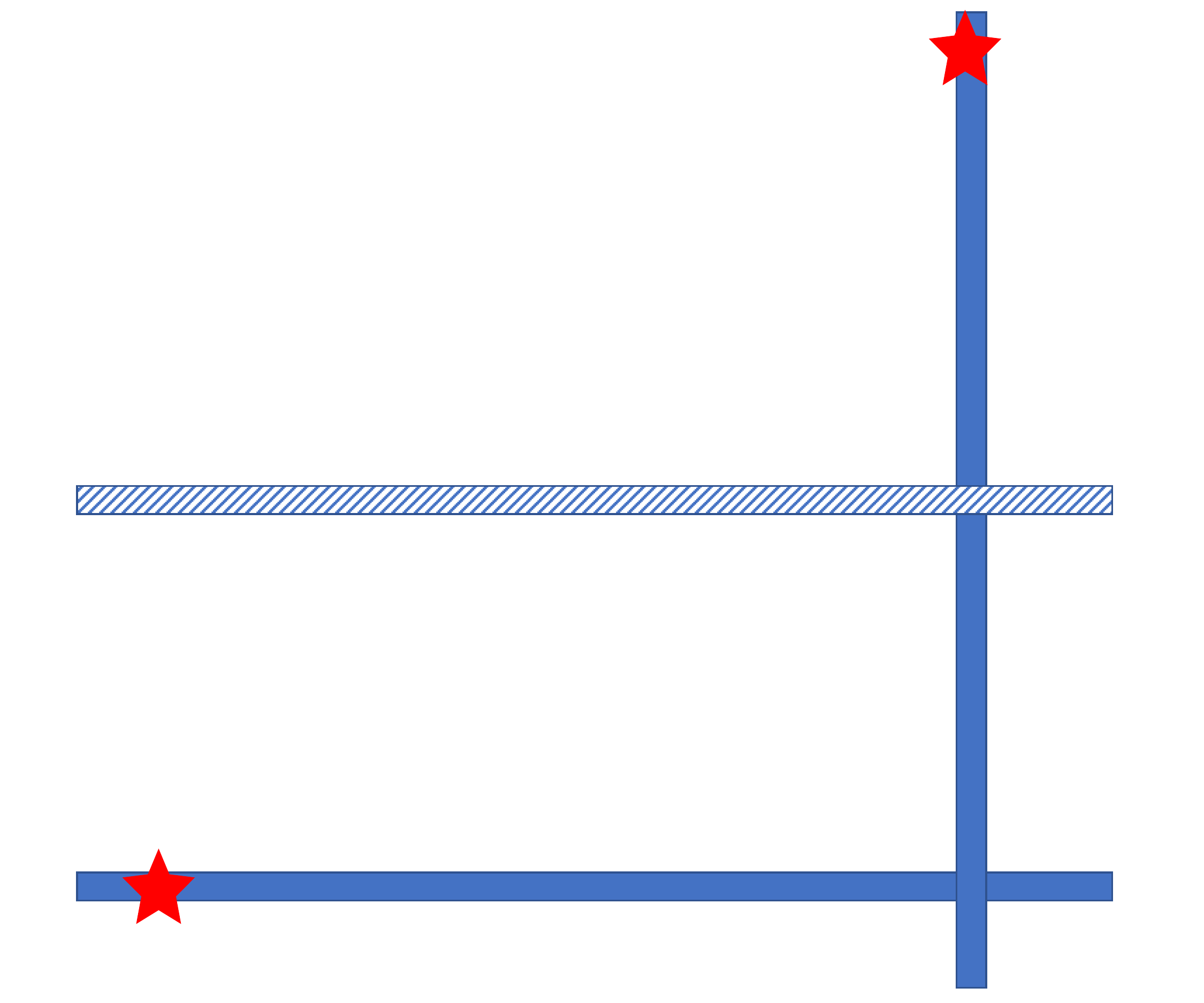}%
}\hfill
\subfloat[Event $E_{8,2,4}$\label{E824}]{%
  \includegraphics[width=0.33\columnwidth,keepaspectratio]{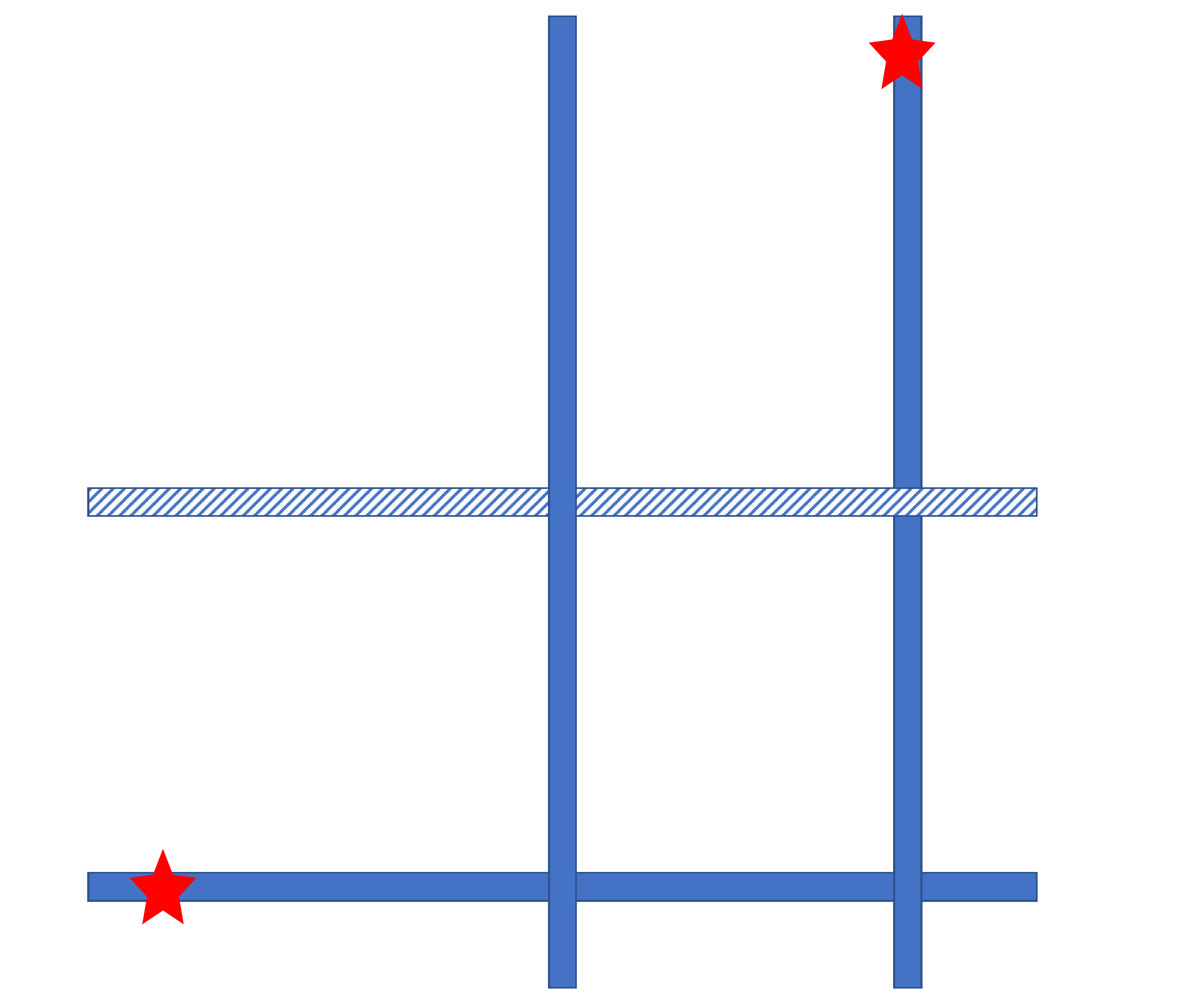}%
}
\hfill
\subfloat[Event $E_{8,3,1}$\label{E831}]{%
  \includegraphics[width=0.33\columnwidth,keepaspectratio]{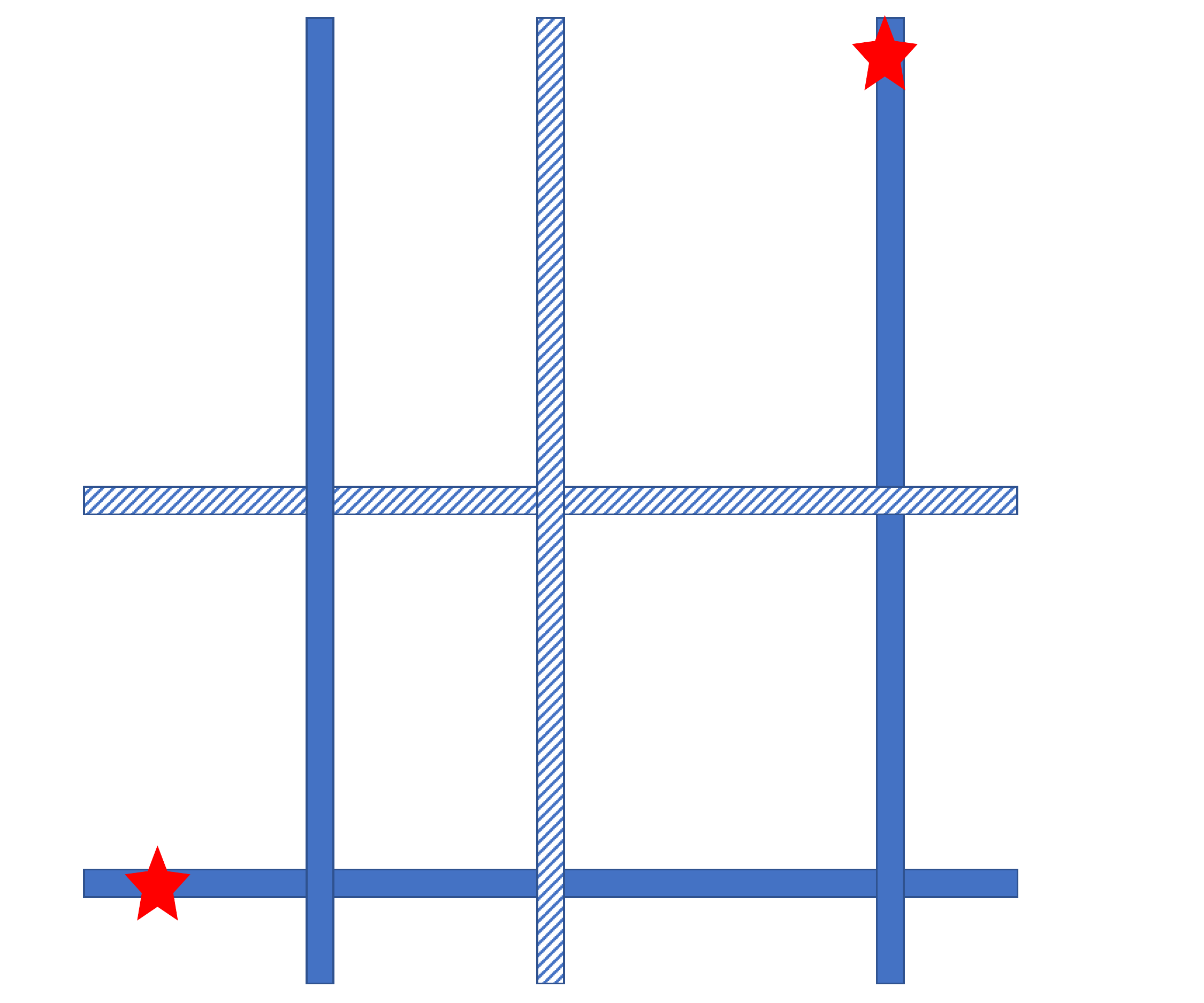}%
}

\subfloat[Event $E_{8,3,2}$\label{E832}]{%
  \includegraphics[width=0.33\columnwidth,keepaspectratio]{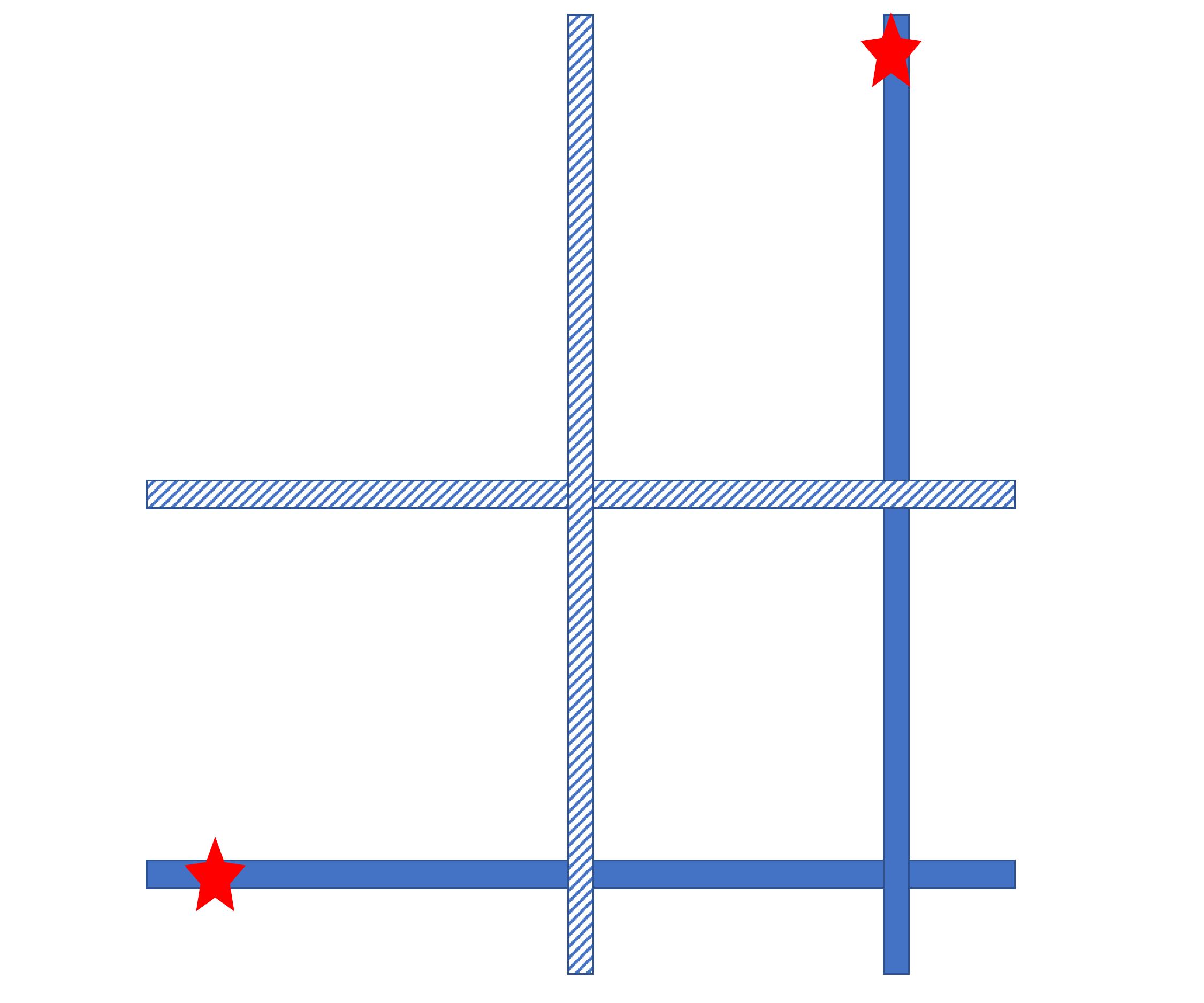}%
}
\subfloat{%
  \includegraphics[width=0.33\columnwidth,keepaspectratio]{dnx_legend_1.pdf}%
}

\caption{Subcases of tree $E_8$.}
\label{e8sub}
\end{figure}

\begin{itemize}[wide, labelwidth=!, labelindent=0pt]
    \item Event $E_{8,1,1}$: \\\textit{Description}: If there are no horizontal roads between S and D, as shown in Fig. \ref{E811}; 
    \\Event $L_{8,1} =  E_{8,1,1} \cap E_8$;
    \\\textit{Probability}: $\mathbb{P}(E_{8,1,1}|E_8)=e^{-\lambda d_v}$, $\mathbb{P}(L_{8,1})=\mathbb{P}(E_{8,1,1}|E_8)\mathbb{P}(E_8)$; 
    \\\textit{Action}: We simply take the source road then the destination road. $$\mathbb{P}(D_n < x|L_{8,1})=\mathbb{P}(D_n < x|E_{8,1,1},E_8) = 0.$$
    \item Event $E_{8,1,2}$: \\\textit{Description}: If there are no horizontal charging roads but at least one horizontal non-charging road between S and D; \\\textit{Probability}: $\mathbb{P}(E_{8,1,2}|E_8)=e^{-\lambda p d_v}(1-e^{-\lambda (1-p) d_v})$. 
    \begin{itemize}[wide, labelwidth=!, labelindent=0pt]
        \item Event $E_{8,2,1}$: \\\textit{Description}: If there are no vertical charging roads between S and D, as shown in Fig. \ref{E821}; 
        \\Event $L_{8,2} =  E_{8,2,1} \cap E_{8,1,2} \cap E_8$;
        \\\textit{Probability}: $\mathbb{P}(E_{8,2,1}|E_{8,1,2},E_8)=e^{-\lambda p d_h}$,
        \\$\mathbb{P}(L_{8,2})=\mathbb{P}(E_{8,2,1}|E_{8,1,2},E_8)\mathbb{P}(E_{8,1,2}|E_8)\mathbb{P}(E_8)$; \\\textit{Action}: We take the source road then the destination road. $$\mathbb{P}(D_n < x|L_{8,2})=\mathbb{P}(D_n < x|E_{8,2,1},E_{8,1,2},E_8) = 0.$$
        \item Event $E_{8,2,2}$: \\\textit{Description}: If there is at least one vertical charging road between S and D, as shown in Fig. \ref{E822};
        \\Event $L_{8,3} =  E_{8,2,2} \cap E_{8,1,2} \cap E_8$;
        \\\textit{Probability}: $\mathbb{P}(E_{8,2,2}|E_{8,1,2},E_8)=1- e^{-\lambda p d_h}$,
        \\$\mathbb{P}(L_{8,3})=\mathbb{P}(E_{8,2,2}|E_{8,1,2},E_8)\mathbb{P}(E_{8,1,2}|E_8)\mathbb{P}(E_8)$; 
        \\\textit{Action}: We go to the nearest vertical charging road from source, then switch to the furthest horizontal non-charging road from source. 
        	\begin{align}
	\label{822}
	&\mathbb{P}(D_n < x|L_{8,3})=\mathbb{P}(D_n < x|E_{8,2,2},E_{8,1,2},E_8)
	\nonumber\\&= \frac{F_{D_\mathrm{N-VC}} (x)}{F_{D_\mathrm{N-VC}} (d_h)}\mathbbm{1}\{x < d_h\} + \mathbbm{1}\{x > d_h\}
	.
	\end{align}
	The proof for $\mathbb{P}(D_n < x|E_{8,2,2},E_{8,1,2},E_{8})$ is similar to that of $\mathbb{P}(D_n < x|E_{7,2,2},E_{7,1,2},E_{7})$ given in (\ref{722}).
        
    \end{itemize}
    \item Event $E_{8,1,3}$: \\\textit{Description}: If there is at least one horizontal charging road between S and D; \\\textit{Probability}: $\mathbb{P}(E_{8,1,3}|E_8)=1-e^{-\lambda p d_v}$.
    \begin{itemize}[wide, labelwidth=!, labelindent=0pt]
        \item Event $E_{8,2,3}$: \\\textit{Description}: If there are no vertical roads between S and D, as shown in Fig. \ref{E823}; 
        \\Event $L_{8,4} =  E_{8,2,3} \cap E_{8,1,3} \cap E_8$;
        \\\textit{Probability}: $\mathbb{P}(E_{8,2,3}|E_{8,1,3},E_8)=e^{-\lambda d_h}$,
        \\$\mathbb{P}(L_{8,4})=\mathbb{P}(E_{8,2,3}|E_{8,1,3},E_8)\mathbb{P}(E_{8,1,3}|E_8)\mathbb{P}(E_8)$; \\\textit{Action}: We take the source road then the destination road. $$\mathbb{P}(D_n < x|L_{8,4})=\mathbb{P}(D_n < x|E_{8,2,3},E_{8,1,3},E_8) = 0.$$  
        \item Event $E_{8,2,4}$: \\\textit{Description}: If there are no vertical charging roads but at least one vertical non-charging road between S and D, as shown in Fig. \ref{E824}; 
        \\Event $L_{8,5} =  E_{8,2,4} \cap E_{8,1,3} \cap E_8$;
        \\\textit{Probability}: $\mathbb{P}(E_{8,2,4}|E_{8,1,3},E_8)=e^{-\lambda p d_h}(1-e^{-\lambda (1-p) d_h)})$,
        \\$\mathbb{P}(L_{8,5})=\mathbb{P}(E_{8,2,4}|E_{8,1,3},E_8)\mathbb{P}(E_{8,1,3}|E_8)\mathbb{P}(E_8)$; 
        \\\textit{Action}: We go to the nearest vertical non-charging road, then switch to any horizontal charging road between S and D.  
        \begin{align}
    \label{824}
            &\mathbb{P}(D_n < x|L_{8,5})=\mathbb{P}(D_n < x|E_{8,2,4},E_{8,1,3},E_8) 
            \nonumber\\&= \mathbb{P}(D_\mathrm{N-VNC} + D_\mathrm{N-HC} < x| D_\mathrm{N-HC} < d_v, D_\mathrm{N-VC} > d_h, \nonumber\\& D_\mathrm{N-VNC} < d_h)
            \nonumber\\&= {\textstyle\frac{\mathbb{P}( D_\mathrm{N-HC} < {\rm min}(x-D_\mathrm{N-VNC}, d_v), D_\mathrm{N-VC} > d_h, D_\mathrm{N-VNC} < d_h)}{\mathbb{P}( D_\mathrm{N-HC} < d_v, D_\mathrm{N-VNC} < d_h <D_\mathrm{N-VC})}}
			\nonumber\\& = {\textstyle\frac{f_{11}({\rm max}(x-d_v,0),{\rm min}(x,d_h),x-o)+f_{11}(0,{\rm min}(x-d_v,d_h),d_v)\mathbbm{1}\{x > d_v\}}{F_{D_\mathrm{N-HC}} (d_v)F_{D_\mathrm{N-VNC}} (d_h)}} \nonumber\\& \times\mathbbm{1}\{x < d_h + d_v\} + \mathbbm{1}\{x > d_h + d_v\}
        .
        \end{align}   
        
        \item Event $E_{8,2,5}$: \\\textit{Description}: If there is at least one vertical charging road between S and D; \\\textit{Probability}: $\mathbb{P}(E_{8,2,5}|E_{8,1,3},E_8)=1-e^{-\lambda p d_h}$.
        \begin{itemize}[wide, labelwidth=!, labelindent=0pt]
            \item Event $E_{8,3,1}$: \\\textit{Description}: If there exists at least one vertical non-charging road before the nearest vertical charging road from S, as shown in Fig. \ref{E831}; \\\textit{Probability}: $\mathbb{P}(E_{8,3,1}|E_{8,2,5},E_{8,1,3},E_8)=1 - \frac{p - pe^{-\lambda d_h}}{1 -e^{-\lambda p d_h}}$; \\\textit{Action}: we compare (i) the distance between the nearest vertical charging road and the nearest vertical non-charging road, and (ii) the vertical distance between the nearest horizontal charging road and source, to take the longer one. 
            \begin{itemize}[wide, labelwidth=!, labelindent=0pt]
                \item Event $E_{8,4,1}$: \\\textit{Description}: If we take the nearest horizontal charging road; \\Event $L_{8,6} =  E_{8,4,1} \cap E_{8,3,1} \cap E_{8,2,5} \cap E_{8,1,3} \cap E_8$;
                \\\textit{Probability}: $\mathbb{P}(E_{8,4,1}|E_{8,3,1},E_{8,2,5},E_{8,1,3},E_8)=1- \int_{0}^{d_v} F_{X_1}(x) f_{D_\mathrm{N-HC}(x)} {\rm d}x$;
                \\$\mathbb{P}(L_{8,6})=\mathbb{P}(E_{8,4,1}|E_{8,3,1},E_{8,2,5},E_{8,1,3},E_8)\times
                \\\mathbb{P}(E_{8,3,1}|E_{8,2,5},E_{8,1,3},E_8)\times\\\mathbb{P}(E_{8,2,5}|E_{8,1,3},E_8)\mathbb{P}(E_{8,1,3}|E_8)\mathbb{P}(E_8)$;
                      \begin{align}
		\label{841}
		&\mathbb{P}(D_n < x|L_{8,6})=\mathbb{P}(D_n < x|E_{8,4,1},E_{8,3,1},E_{8,2,5},E_{8,1,3},E_8)
		\nonumber\\& = \frac{f_{12}(0,x,x,d_h,x-t)+f_{12}(0,x,0,x,y-t)}{f_{13}(0,d_h,0,d_h-t,d_h,t+y)}\mathbbm{1}\{x < d_h\} \nonumber\\&+ \mathbbm{1}\{x >  d_h\}	
		\end{align}
      The proof for $\mathbb{P}(D_n < x|E_{8,4,1},E_{8,3,1},E_{8,2,5},E_{8,1,3},E_{8})$ is similar to that of \\ $\mathbb{P}(D_n < x|E_{7,4,1},E_{7,3,3},E_{7,2,5},E_{7,1,3},E_{7})$ given in (\ref{741}).
            
                \item Event $E_{8,4,2}$: \\\textit{Description}: If we take the nearest vertical charging road;
                \\Event $L_{8,7} =  E_{8,4,2} \cap E_{8,3,1} \cap E_{8,2,5} \cap E_{8,1,3} \cap E_8$;
                \\\textit{Probability}: $\mathbb{P}(E_{8,4,2}|E_{8,3,1}|E_{8,2,5},E_{8,1,3},E_8)= \int_{0}^{d_v} F_{X_1}(x) f_{D_\mathrm{N-HC}(x)} {\rm d}x$;
                \\$\mathbb{P}(L_{8,7})=\mathbb{P}(E_{8,4,2}|E_{8,3,1},E_{8,2,5},E_{8,1,3},E_8)\times
                \\\mathbb{P}(E_{8,3,1}|E_{8,2,5},E_{8,1,3},E_8)\times\\\mathbb{P}(E_{8,2,5}|E_{8,1,3},E_8)\mathbb{P}(E_{8,1,3}|E_8)\mathbb{P}(E_8)$;
                      \begin{align}
      \label{842}
      & \mathbb{P}(D_n < x|L_{8,7})=\mathbb{P}(D_n < x|E_{8,4,2},E_{8,3,1},E_{8,2,5},E_{8,1,3},E_8)
      \nonumber\\& = {\textstyle\frac{f_{13}(0,\infty,0,{\rm max}(x-t,0),y+t,y)-f_{13}(0,\infty,{\rm max}(x-t,0),x,x,y)}{f_{13}(0,\infty,0,{\rm max}(d_h-t,0),y+t,y) - f_{13}(0,\infty,{\rm max}(d_h-t,0),d_h,x,y)}}\times
      \nonumber\\&\mathbbm{1}\{x < d_h\} + \mathbbm{1}\{x >  d_h\}
      .
      \end{align}
      The proof for $\mathbb{P}(D_n < x|E_{8,4,2},E_{8,3,1},E_{8,2,5},E_{8,1,3},E_{8})$ is similar to that of \\ $\mathbb{P}(D_n < x|E_{7,4,2},E_{7,3,3},E_{7,2,5},E_{7,1,3},E_{7})$ given in (\ref{742}).
                
            \end{itemize}
            \item Event $E_{8,3,2}$: \\\textit{Description}: If there exists no vertical non-charging road before the nearest vertical charging road from source, as shown in Fig. \ref{E832}; 
            \\Event $L_{8,8} = E_{8,3,2} \cap E_{8,2,5} \cap E_{8,1,3} \cap E_8$;
            \\\textit{Probability}: $\mathbb{P}(E_{8,3,2}|E_{8,2,5},E_{8,1,3},E_8)=\frac{p - pe^{-\lambda d_h}}{1 -e^{-\lambda p d_h}}$,
            \\$\mathbb{P}(L_{8,8})=\mathbb{P}(E_{8,3,2}|E_{8,2,5},E_{8,1,3},E_8)\times
            \\\mathbb{P}(E_{8,2,5}|E_{8,1,3},E_8)\mathbb{P}(E_{8,1,3}|E_8)\mathbb{P}(E_8)$;
            \\\textit{Action}: we simply go with the nearest vertical charging road. 
            \begin{align}
      \label{832}
      & \mathbb{P}(D_n < x|L_{8,8})=\mathbb{P}(D_n < x|E_{8,3,2},E_{8,2,5},E_{8,1,3},E_8)
       \nonumber\\&= \frac{f_{11}(0,x,y)+f_{11}(x,\infty,x)}{f_{11}(0,d_h,y)+f_{11}(d_h,\infty,d_h)}\mathbbm{1}\{x < d_h\} + \mathbbm{1}\{x > d_h\}
      .
      \end{align}
      The proof for $\mathbb{P}(D_n < x|E_{8,3,2},E_{8,2,5},E_{8,1,3},E_{8})$ is similar to that of $\mathbb{P}(D_n < x|E_{7,3,4},E_{7,2,5},E_{7,1,3},E_{7})$ given in (\ref{734}).
            
        \end{itemize}
    \end{itemize}
\end{itemize}
\end{IEEEproof}
      
\section{Proof of Theorem \ref{theorem2}}
\label{appendxf}  
In this appendix, we outline the proof for the probability that any given trip passes through at least one charging road, i.e., $\mathbb{P}(T_c)$.
    \newline$\mathbb{P}(\overline{T_c} | E_4)\mathbb{P}(E_4)$:
    \begin{align}
  	\label{tce4}
    &\mathbb{P}(\overline{T_c} | E_4)\mathbb{P}(E_4) = [\mathbb{P}(E_{4,1,1})(1-p) + \mathbb{P}(E_{4,1,2}) \nonumber\\& + \mathbb{P}(E_{4,1,4}) \mathbb{P}(E_{4,2,1})]\mathbb{P}(E_4) 
    \nonumber\\&=[e^{-\lambda d_v}(1-p) + \lambda (1-p) d_v e^{-\lambda (1-p) d_v} e^{-\lambda p d_v} \nonumber\\& + e^{-\lambda p d_v}(1-e^{-\lambda (1-p) d_v} - \lambda (1-p) d_v e^{- \lambda (1-p) d_v}) e^{-\lambda p d_h}] \nonumber\\&\times(1-p)^2
.
\end{align}
      $\mathbb{P}(\overline{T_c} | E_8)\mathbb{P}(E_8)$:
    \begin{align}
    \label{tce8}
    &\mathbb{P}(\overline{T_c} | E_8)\mathbb{P}(E_8) = [\mathbb{P}(E_{8,1,1}) + \mathbb{P}(E_{8,1,2})\mathbb{P}(E_{8,2,1}) \nonumber\\&+ \mathbb{P}(E_{8,1,3}) \mathbb{P}(E_{8,2,3})]\mathbb{P}(E_8) 
    \nonumber\\&=[e^{-\lambda d_v} + e^{-\lambda p d_v}(1-e^{-\lambda (1-p) d_v})e^{-\lambda p d_h} \nonumber\\&+ (1-e^{-\lambda p d_v})e^{-\lambda d_h}](1-p)^2
.
\end{align}

\ifCLASSOPTIONcaptionsoff
   
\fi
\balance
\bibliographystyle{IEEEtran}
\bibliography{hokie-HD}
\begin{IEEEbiography}{Duc Minh Nguyen}
was born in Hanoi, Vietnam. He received the M.S. degree in Electrical and Computer Engineering from King Abdullah University of Science and Technology (KAUST), Saudi Arabia, and the B.Eng. degree in Mobile Systems Engineering from Dankook University, Republic of Korea, in 2020 and 2018, respectively. He is currently pursuing the Ph.D. degree with the Computer, Electrical and Mathematical Science and Engineering Division, KAUST, Saudi Arabia. His research interests include social/vehicular networks analysis, data mining, and machine learning.
\end{IEEEbiography}
\begin{IEEEbiography}{Mustafa A. Kishk}
[S’16, M’18] is a postdoctoral
research fellow in the communication theory lab at King Abdullah University of Science and Technology (KAUST). He received his B.Sc. and M.Sc. degree from Cairo University in 2013 and 2015, respectively, and his Ph.D. degree from Virginia Tech in 2018. His current research interests include stochastic geometry, energy harvesting wireless networks, UAV-enabled communication systems, and satellite communications.
\end{IEEEbiography}
\begin{IEEEbiography}{Mohamed-Slim Alouini}
[S’94-M’98-SM’03-F’09] was born in Tunis, Tunisia. He received the Ph.D. degree in Electrical Engineering from the California
Institute of Technology (Caltech), Pasadena, CA, USA, in 1998. He served as a faculty member in the University of Minnesota, Minneapolis, MN, USA, then in the Texas A\&M University at Qatar, Education City, Doha, Qatar before joining King Abdullah University of Science and Technology (KAUST), Thuwal, Makkah Province, Saudi Arabia as a Professor of Electrical Engineering in 2009. His current research interests include the modeling, design, and performance analysis of wireless communication systems.
\end{IEEEbiography}
\end{document}